\RequirePackage{fix-cm}
\documentclass[smallextended]{svjour3}

\smartqed

\usepackage{graphicx}
\usepackage{amssymb}    
\usepackage{epstopdf}
\usepackage{color}

\usepackage{lscape}
\usepackage{epsfig}
\usepackage{fullpage}
\usepackage{graphicx}
\usepackage[titletoc]{appendix}
\usepackage{fancyhdr}
\usepackage{amsmath,amssymb,xspace}
\usepackage{tabularx}
\usepackage[caption=false]{subfig}

\newfont{\bbb}{msbm10 scaled\magstep1}

\newtheorem{prop}{Proposition}[section]

\renewcommand{\thesection}{\arabic{section}}

\let \leq \leqslant
\let \geq \geqslant

\let \epsilon \varepsilon
\let \hat \widehat
\newcommand{\norm}[2]{\left\lVert#1\right\rVert_{#2}}
{ \par \medskip \par
  \noindent \textit{\textbf{Demonstration\/}} : }{\null \hfill $\Box$ \par }
 
\newcommand{\R} {\ensuremath{\mathbb{R}}}
\newcommand{\Z} {\ensuremath{\mathbb{Z}}}
\newcommand{\N} {\ensuremath{\mathbb{N}}}

\usepackage{mathrsfs}

\begin{document}

\title{Simple digital quantum algorithm for symmetric first order linear hyperbolic systems}


\author{F. Fillion-Gourdeau  \and E. Lorin}

\institute{F. Fillion-Gourdeau \at Universit\'{e} du Qu\'{e}bec, INRS-\'{E}nergie, Mat\'{e}riaux et T\'{e}l\'{e}communications, Varennes, Canada, J3X 1S2 \\
Institute for Quantum Computing, University of Waterloo, Waterloo,	Ontario, Canada, N2L 3G1\\
\email{francois.fillion@emt.inrs.ca}
\and E. Lorin \at 
Centre de Recherches Math\'{e}matiques, Universit\'{e} de Montr\'{e}al, Montr\'{e}al, Canada, H3T~1J4.\\
School of Mathematics and Statistics, Carleton University, Ottawa, Canada, K1S 5B6.\\
\email{elorin@math.carleton.ca}}


\maketitle

\begin{abstract}
This paper is devoted to the derivation of a digital quantum algorithm for the Cauchy problem for symmetric first order linear hyperbolic systems, thanks to the reservoir technique. The reservoir technique is a method designed to avoid artificial diffusion generated by first order finite volume methods approximating hyperbolic systems of conservation laws. For some class of hyperbolic systems, namely those with constant matrices in several dimensions, we show that the combination of i) the reservoir method and ii) the alternate direction iteration operator splitting approximation,  allows for the derivation of algorithms only based on simple unitary transformations, thus perfectly suitable for an implementation on a quantum computer. The same approach can also be adapted to scalar one-dimensional systems with non-constant velocity by combining with a non-uniform mesh. The asymptotic computational complexity for the time evolution is determined and it is demonstrated that the quantum algorithm is more efficient than the classical version. However, in the quantum case, the solution is encoded in probability amplitudes of the quantum register. As a consequence, as with other similar quantum algorithms, a post-processing mechanism has to be used to obtain general properties of the solution because a direct reading cannot be performed as efficiently as the time evolution.  

\end{abstract}

\section{Introduction}

Quantum computing is a new paradigm in information science which benefits from quantum mechanics to perform some computational tasks. In the last few decades, it has attracted a lot of attention because it promises efficient solutions to a large class of problems deemed unsolvable on classical computers. Shor's algorithm, for the prime number factorization of integers, is the foremost example of the strength of quantum computing \cite{shor1998}. This algorithm runs in polynomial time, i.e. the computation time scales like a  polynomial function of the input size, while the same task runs in (sub-)exponential time on a classical computer. This quantum speedup has motivated the development of many other algorithms for the solution of problems in the BQP complexity class but outside the P class, i.e. problems with bounded error for which the amount of quantum resources is a polynomial function and having an exponential speedup over classical computations \cite{nielsen2010quantum,0034-4885-61-2-002}. 

One of the promising applications of quantum computing is the simulation of quantum systems. Inspired from Feynman's quantum simulator \cite{feynman1982simulating}, it has been demonstrated that \textit{universal quantum computers} \cite{Deutsch97} can simulate efficiently the dynamics of any local quantum Hamiltonian with a number of quantum operations scaling polynomially with the system size \cite{SLoyd}. Following these seminal results, other algorithms have been developed for other types of quantum physical systems. Some examples include algorithms for the simulation of non-relativistic single-particle quantum mechanics \cite{OPPROP:PR877,Zalka08011998,Strini2008}, relativistic mechanics \cite{fillion}, many-body physics \cite{boghosian1998simulating,OPPROP:PR877,Zalka08011998,wiesner1996simulations,PhysRevA.65.042323}, quantum field theory \cite{jordan2012quantum}, quantum chemistry \cite{PMID:21166541,yung2014,kassal2008polynomial,Aspuru-Guzik1704}  and many others \cite{e12112268,RevModPhys.86.153}. 

Generally the simulation of quantum systems on a quantum computer is based on two main ingredients: (i) the encoding of the quantum state on the quantum register, and (ii) the existence of a set of operations that modify a quantum register according to the dynamics of the physical system under study.
When both are available, it is possible to use the quantum computer to emulate another quantum physical system.

In contrast to analogue quantum simulators \cite{RevModPhys.86.153}, based on a direct analogy between two Hamiltonians and thus restricted to a certain category of systems, the digital quantum computers (DQC) considered in this article are built from a set of entangled two-level quantum systems, the qubits, forming the quantum register. Similar to bits in classical computing, qubits are discrete entities calling for the discretization of the physical system under consideration. However, due to their inherent quantum nature allowing them to be in a superposition of states, the quantum register is characterized by $N=2^{n} \in \mathbb{N}^*$ complex coefficients,  
where $n$ is the number of qubits. Then, a possible strategy consists in mapping the coefficients of the physical system wave function expressed in some suitable basis into the probability amplitudes of the quantum register, as in the aforementioned algorithms. This process is called amplitude encoding. Thus, DQC allows for the simulation of quantum systems in their discretized form, analogously to classical computers.

Furthermore, the dynamics of the quantum computer and the simulated system proceed by unitary operations. In a quantum computer, the state of the quantum register is modified by simple unitary operations: the quantum logic gates \cite{nielsen2010quantum}. In turn, a quantum system evolves according to unitary operations given by the evolution operator, according to the laws of quantum mechanics. A mapping of the evolution operator onto quantum gates can be implemented via a unitary decomposition, often performed using Trotter(splitting)-like approximations \cite{1751-8121-43-6-065203}. However, these mappings are not unique, each one corresponding to a different numerical method: the encoding is determined by the basis choice while the chosen unitary decomposition sets the numerical scheme used to evolve the system in time. 

Using these two mappings, i.e. the wave function on the quantum register and the evolution operator on quantum gates, it is possible to simulate quantum systems. Moreover, for many cases of interest, these simulations are more efficient than their classical counterparts, in the sense that computing resources are scaling polynomially with the size of the system.

For classical systems, such as fluids, plasmas and electromagnetic fields, the analogy described above between the quantum computer and the physical system is not as explicit and may even be non-existing as the time evolution may be non-unitary. Mathematically, this implies that in contrast to quantum mechanics, the $L^{2}$-norm is not always preserved by the dynamics, complicating the mapping on quantum computers which are based on unitary operations. Nevertheless, it is possible to devise algorithms for the quantum simulation of some classical systems. 
For instance, fluid-like mechanics equations have been considered in \cite{mezzacapo2015quantum}, but they required the implementation of non-unitary operations. The latter can be implemented on a quantum computer, but the algorithm becomes nondeterministic and the probability of success depends on the time step. For these reasons, algorithms for the simulation of classical systems are scarce, with some notable exceptions where classical thermal state \cite{PhysRevA.82.060302}, classical diffusion \cite{Meyer395}, electromagnetism \cite{Sinha2010} and the Poisson equation \cite{1367-2630-15-1-013021} have been considered. In addition, there exists algorithms for the solution of ordinary differential equations, which may be relevant for many physical applications \cite{1751-8121-47-10-105301}.  

In this article, we consider the quantum simulation of an important class of classical partial differential equations: linear hyperbolic systems. Our approach is similar to a scheme developed for the Dirac equation where an analogy between the split operator method and quantum walks was explicitly constructed \cite{fillion}. The Dirac equation is in fact a particular hyperbolic system, where the mass and the electromagnetic potential are local source terms. Therefore, it was expected that techniques for the quantum Dirac equation can be adapted to more general hyperbolic systems. This was noticed in \cite{1751-8121-47-46-465302}, where a quantum algorithm for constant linear hyperbolic systems with rational eigenvalues have been investigated. Here, we are proposing a quantum implementation of the reservoir numerical scheme \cite{res0,res1,res2,res3,res4}, extending the results in \cite{1751-8121-47-46-465302} to more general hyperbolic systems. The reservoir method was developed to avoid spurious diffusion generated by low order finite volume methods, by adding a reservoir and a Courant-Friedrichs-Lewy (CFL \cite{strikwerda}) counter at the interface between discretization volumes. This numerical scheme is then  particularly well-suited for a quantum implementation because in some cases, it reduces to simple streaming steps which preserve the $L^{2}$-norm and also, can be implemented efficiently on a quantum computer. Notice that the assumption that matrices in the hyperbolic system are symmetric is not indispensable in the reservoir method, but is required in order to have unitary operations. At the same time, it guarantees that the system of equations is hyperbolic. 

The main result of this paper, stated precisely in Propositions \ref{prop:quant_speed} and \ref{prop:quant_speed_eps}, is that by combining these ideas and assuming that the quantum register can be initialized in $O(\mathrm{polylog}\; N)$ operations, it is possible to solve a symmetric hyperbolic system on a DQC with a quantum speedup 
\begin{eqnarray*}
S =
 O\left( \cfrac{m^{2}}{\log^{2} m} \cfrac{N^{d}}{\log^{2} N} \right), 
\end{eqnarray*}
for a large number of grid points $N$,
where $d$ is the number of dimensions and $m$ is the number of characteristic fields. This corresponds to an exponential speedup, i.e. the number of operations in the quantum algorithm increases logarithmically with the number of grid points while it increases linearly in the classical implementation of the same algorithm. This is clearly an interesting advantage of the quantum approach, although there is an important caveat: as the solution is encoded into probability amplitudes, the measurement/reading of the solution on the DQC requires $O(N)$ repetitions of the algorithm, potentially eliminating the exponential speedup. This is a typical problem in the field of quantum simulations (see \cite{1751-8121-47-10-105301} for instance) which is usually solved partially by post-processing the solution to obtain some observables or some general properties of the solution. To preserve the exponential speedup, one requires that this post-processing is performed using $O(\mathrm{polylog}\; N)$ operations, which is not possible if one wants the whole solution. Of course, this is an important but standard limitation of the quantum approach.

Although the problem under consideration, with constant symmetric matrices, seems relatively simple from a mathematical point of view, there still exist some complex open problems in physics involving this type of systems. For instance, the advection equation for transport problem, linear acoustic equations for sound waves, Maxwell's equations in electrodynamics, linear elasticity equation and the wave equation are some examples of homogeneous symmetric hyperbolic systems with considerable importance in physics. As mentioned above, the Dirac equation is a first order hyperbolic system with constant coefficients. In quantum electrodynamics, it is well known that accurate computations of electron-positron pair production from Schwinger's effect \cite{schwinger} requires the solution to a given Dirac equation with many initial conditions, which is not tractable on classical computers.  Such calculations require fundamental breakthrough from the computing point of view, which could be provided by efficient quantum algorithms. In addition, our work can be considered as a first step towards the development of quantum algorithms for more general hyperbolic equations. We give possible directions to generalize our approach at the end of this article.

This article is organized as follows. In Section \ref{survival}, we first give some basic notions on quantum computing. In Section \ref{sec:res_meth}, we review  the reservoir method for linear systems. We then show in Section \ref{sec:general_case}, how to derive a quantum algorithm for the reservoir method for one-dimensional first order hyperbolic systems and then, for multidimensional first order systems. The computational complexity and quantum speedup are discussed in Section \ref{sec:eff}.  In Section \ref{sec:extension}, we propose some possible avenues to extend the ideas presented in this paper, first to scalar first order hyperbolic equations with non-constant velocity, then using quantum algorithms based on the method of lines. We conclude in Section \ref{sec:conclusion}.
   
\section{Survival kit on quantum computing}\label{survival}
In this section, we recall some basic definitions in quantum computing,  and quantum simulation theory for readers without or with a very limited knowledge of these fields. The objective is to present the mathematical objects and tools which are necessary to construct or understand a quantum algorithm. The reader who already has some advanced knowledge in quantum computing can skip this section. More information on this topic can be found in Ref. \cite{nielsen2010quantum}.

First, we recall that the qubit in QC is the analogue of the bit on digital computers, that is the basic unit of information. More specifically, a qubit is a quantum mechanical two-level system. In principle, it can be realized by any quantum physical systems having two degrees of freedom, such as single photon polarization, electron spin, superconducting qubits, and many others. In practice, some physical systems are more suitable for quantum computing because they can be controlled more easily and have a larger decoherence time owing to a weaker interaction with the environment.

The state for all two-level systems is described quantum mechanically by a unit vector, denoted here by $|u\rangle$, defined in a two-dimensional Hilbert space $\mathcal{H}$. It can then be written in the form : $|u \rangle  = \alpha_{0} |0\rangle + \alpha_{1} |1 \rangle$, where $|0\rangle$, $|1\rangle$ denote the computational basis vectors of $\mathcal{H}$ and $\alpha_0$, $\alpha_1$ are complex numbers representing the quantum amplitudes normalized as $|\alpha_{0}|^2+|\alpha_{1}|^2=1$.  

A quantum register is a set of $\ell$ two-level entangled systems. In this case, the quantum state of the total system $|u_{\ell}\rangle $, according to quantum mechanics principles, is a vector in the Hilbert space $\mathcal{H}_{\ell} = \otimes_{n=1}^{\ell} \mathcal{H}$, the tensor product of $\ell$ two-dimensional spaces.
Then, any $|u_{\ell}\rangle \in \mathcal{H}_{\ell}$ reads
\begin{eqnarray}
\label{eq:qu_register}
|u_{\ell}\rangle = \sum_{s_1=0}^1\cdots \sum_{s_{\ell}=0}^1\alpha_{s_1\cdots s_{\ell}}\otimes_{i=1}^{\ell}|s_i\rangle,
\end{eqnarray}
where $\alpha_{s_1\cdots s_{\ell}} $ (for all $s_1\cdots s_{\ell}$) are complex numbers representing the coefficients or amplitudes of the quantum state, and $\{|s_i\rangle\}_{1 \leq i \leq \ell}$ are the $\ell$ qubit basis functions of the $\ell$ two-dimensional spaces $\{\mathcal{H}_i\}_{1 \leq i \leq \ell}$. We note that $|u_{\ell}\rangle$ is a unit vector ($\langle u_{\ell}| u_{\ell} \rangle =1$), implying that the amplitudes should be normalized as
\begin{eqnarray*}
\sum_{s_1=0}^1\cdots \sum_{s_{\ell}=0}^1 |\alpha_{s_1\cdots s_{\ell}}|^{2} = 1 .
\end{eqnarray*}
This normalization is introduced to have a probability interpretation of the quantum state.

Here, it is important to note that although we only have $\ell$ qubits, the number of amplitude coefficients is $2^{\ell}$ owing to the tensorial structure of the vector space. Information can be stored into these coefficients. However, reading all of them is challenging because it requires $O(2^{\ell})$ measurements. This occurs because  each amplitude $\alpha_{s_1\cdots s_{\ell}}$ is related to the probability of finding the system in some state $\otimes_{i=1}^{\ell}|s_i\rangle$ as $P_{s_1\cdots s_{\ell}} = |\alpha_{s_1\cdots s_{\ell}}|^{2}$. Then, according to the measurement postulate in quantum mechanics, a Von Neumann measurement on a quantum system characterized by the Hilbert space $\mathcal{H}_{\ell}$ outputs a classical value $s_1\cdots s_{\ell}$ with probability $P_{s_1\cdots s_{\ell}}$. After such measurement, the system has collapsed to the state $\otimes_{i=1}^{\ell}|s_i\rangle$, i.e. any subsequent measurement will obtain $s_1\cdots s_{\ell}$ with probability 1. Then, measuring all coefficients $\alpha_{s_1\cdots s_{\ell}}$ entails reconstructing the probability distribution for all possible states, a process called quantum tomography, which usually requires $O(2^{\ell})$ measurements.

Realizing physically a quantum register by entangling a certain number of qubits is a very challenging experimental task. Nevertheless, this has been achieved using several physical systems such as superconducting circuits \cite{kelly2015state,barends2015digital,barends2016digitized}, trapped ions \cite{Lanyon57}, nuclear magnetic resonance \cite{PhysRevLett.96.170501} photons \cite{PhysRevLett.117.210502} and cavity quantum electrodynamics \cite{PhysRevX.5.021027}. The typical size for quantum registers varies presently up to $\sim 50$ qubits, but this number is likely to increase in the future. 

\subsection{Quantum logic gates}
Quantum logic gates are analogues to logical gates in classical computing. More precisely, the quantum gates are operators acting on a quantum register, modifying its state according to the laws of quantum mechanics. Thus, they must be unitary reversible operators. One of the most simple, but also important quantum gate is the {\it Hadamard gate}. The latter acts on a single qubit and corresponds mathematically to one rotation of $\pi$ around the $x$-axis and another rotation of $\pi/2$ around the $y$-axis, so that
\begin{eqnarray*}
 H = \frac{1}{\sqrt{2}} \begin{bmatrix} 1 & 1 \\ 1 & -1 \end{bmatrix}.
\end{eqnarray*}
Hence, any qubit $|u \rangle =  \alpha_0 |0\rangle + \alpha_1 |1 \rangle$, is transformed by $H$ as 
\begin{eqnarray*}
H|u\rangle = \cfrac{\alpha_0+\alpha_1}{\sqrt{2}}|0\rangle + \cfrac{\alpha_0-\alpha_1}{\sqrt{2}}|1\rangle.
\end{eqnarray*}
Another elementary  quantum gate, is the NOT-gate which is the analogue of digital NOT-gate (or inverter), which reads
\begin{eqnarray*}
\textrm{NOT} = \begin{bmatrix} 0 & 1 \\ 1 & 0 \end{bmatrix}.
\end{eqnarray*}
There exists an infinite number of possible $2^{\ell} \times 2^{\ell}$-dimensional unitary operations $U_{\ell}$, each of them corresponding to a quantum logic gate. However, it can be demonstrated that any unitary operations applied on a quantum register with $\ell$-qubits can be approximated to a certain accuracy $\epsilon$ by a finite sequence of elementary quantum gates taken from a \textit{universal set} \cite{nielsen2010quantum}. As any real quantum device used for computation will be able to implement a certain universal set, any quantum algorithm needs to be decomposed as a sequence of these elementary gates.

A typical example is the \textit{standard universal set} formed of two single qubit gates: Hadamard and $\pi/8$-gates ($=R_{\pi/4}$), and one two-qubit gate: the controlled-NOT-gate. They are explicitly given by
\begin{eqnarray*}
	R_{\phi} = \begin{bmatrix}
		1 & 0 \\
		0 & e^{i\phi}
	\end{bmatrix}\;,\;
	\mbox{CNOT}=\begin{bmatrix}1&0&0&0\\0&1&0&0\\0&0&0&1\\0&0&1&0\end{bmatrix},
\end{eqnarray*}
along with the Hadamard gate defined previously.
The controlled-gates will be important in the design of our algorithm. They operate on at least two qubits, where one of the qubit serves as a control: if the control qubit is in a certain state ($|0\rangle$ or $|1\rangle$), then some operation $X$ is applied on other qubits. For example, if $X$ is an arbitrary gate acting on a single qubit, that is
\begin{eqnarray*}
{\displaystyle X={\begin{bmatrix}X_{00}&X_{01}\\X_{10}&X_{11}\end{bmatrix}}},
\end{eqnarray*}
the corresponding $C(X)$-gate which acts of two qubits and which perform the operation when the first qubit is in the state $|1\rangle$, reads
\begin{eqnarray*}
{\displaystyle {\mbox{C}}(X)={\begin{bmatrix}1&0&0&0\\0&1&0&0\\0&0&X_{00}&X_{01}\\0&0&X_{10}&X_{11}\end{bmatrix}}}.
\end{eqnarray*}
We refer again to \cite{nielsen2010quantum} for more details about quantum gates. 

One of the main challenges in quantum computation is to determine efficient decompositions of unitary transformations in terms of elementary gates defined above. A given decomposition has an exponential speedup when the ratio of the cost (number of operations) for the classical algorithm and the cost for the quantum algorithm is an exponential function in the amount of resources. In a quantum algorithm, such sequence of operations forms a quantum circuit, discussed in more details in the next section. The purpose of this paper is to decompose a classical numerical solver for hyperbolic systems into quantum logic gates which are applied to a quantum register.

\subsection{Quantum circuits} 
\begin{figure}
\begin{center}
\subfloat[]{\includegraphics[scale=1.5,keepaspectratio]{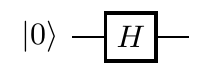}}
\subfloat[]{\includegraphics[scale=1.5, keepaspectratio]{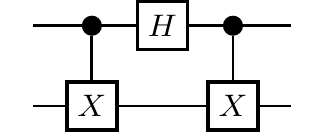}}
\caption{(a) Hadamard gate applied to a qubit initialized to the state $|u\rangle = |0\rangle$. (b) Diagram of rotation operator in $x$-direction, with qubits initialized in arbitrary states.}
\label{fig:had}
\end{center}
\end{figure}

As mentioned above, the quantum algorithm has to be written as a sequence of elementary quantum gates. A convenient way to represent this decomposition visually is to use what is commonly called a quantum circuit or quantum diagram. These circuits allow for a visual representation of operations on $\ell-$qubits, where $\ell$ is called the width of circuit. The most simple quantum circuit is depicted in  Fig. \ref{fig:had}(a). The line represents the qubit under consideration $|u\rangle$, and $H$ is the Hadamard matrix$/$operator. The circuit is read from the left to the right: first, the qubit is initialized to $|u\rangle = |0\rangle$ and then the Hadamard gate is applied, yielding the qubit in the state $|u \rangle = \cfrac{1}{\sqrt{2}}|0\rangle + \cfrac{1}{\sqrt{2}}|1\rangle $, at the right of the circuit.

A more complex example for $2$-qubits is the one shown in Fig. \ref{fig:had}(b), where the two qubits are initialized to arbitrary states. The top (resp. bottom) line represents the first (resp. second) qubit. The first part of the circuit represents $C(X)$-gate applied to the 2-qubits, then a Hadamard gate on the first qubit, then a $C(X)$-gate again. The vertical lines represent the control: here, if the first qubit is in the state $|1\rangle$ (represented by a black dot), then the operation $X$ is performed on the second qubit. A white dot in the controlled operation can also be used when one needs the control qubit to be in the state $|0\rangle$ to perform the operation $X$.  

\begin{figure}
\begin{center}
\subfloat[]{\includegraphics[scale=0.4,keepaspectratio]{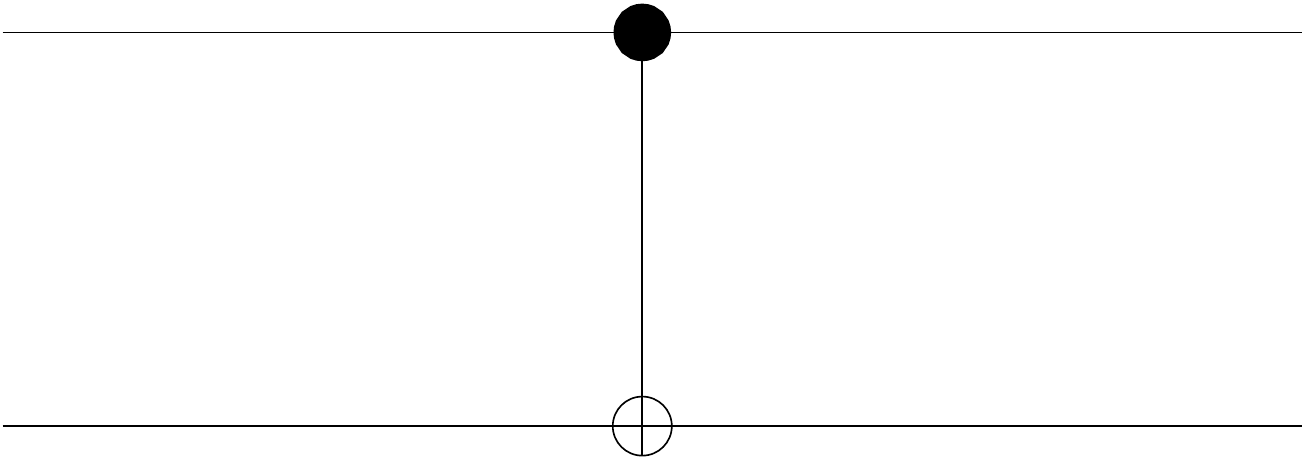}} \hspace{5.0mm}
\subfloat[]{\includegraphics[scale=0.4,keepaspectratio]{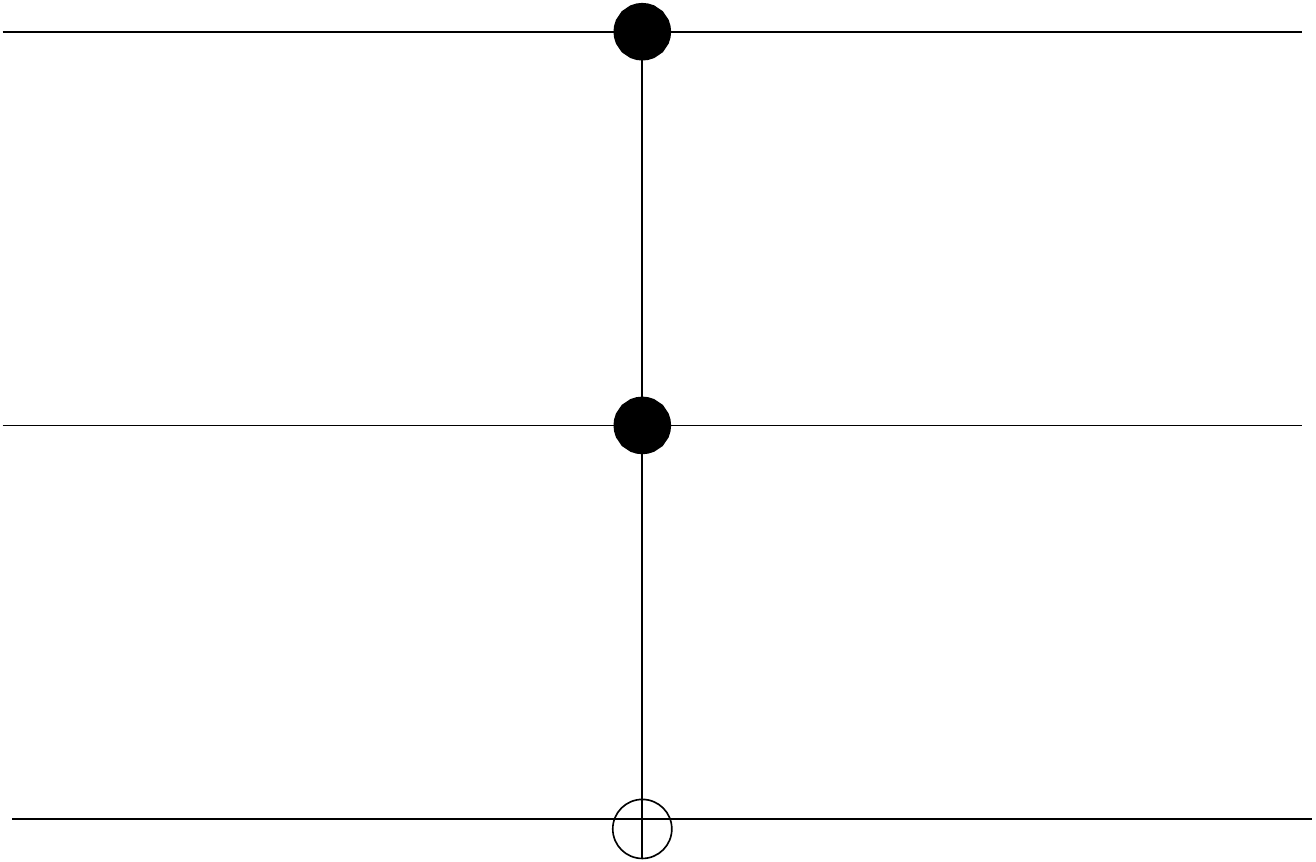}}
\caption{(a) Circuit diagram for the CNOT-gate. (b) Circuit diagram for the Toffoli-gate. }
\label{fig:cnot}
\end{center}
\end{figure}

In this paper, the controlled-not(also called CNOT-gate) and controlled-controlled-not gates (also known as Toffoli gate), are used to implement the translation operators. For 2 qubits  $|s_1\rangle$, $|s_2\rangle$, the controlled-not gate transforms $|s_1\rangle \mapsto |s_1\rangle$ and $|s_2\rangle \mapsto |s_1\oplus s_2 \; \mathrm{mod}\;2\rangle$; the corresponding quantum circuit is represented in Fig. \ref{fig:cnot}(a), where the ``plus dot'' represents the NOT-gate. These controlled gates can be generalized to any number of control qubits. 

The Toffoli gates transform 3 qubits $|s_1\rangle$, $|s_2\rangle$, $|s_3\rangle$ with $s_1$, $s_2$, $s_3$ in $\{0,1\}$ as follows: $|s_1\rangle \mapsto |s_{1}\rangle$, $|s_2\rangle \mapsto |s_2\rangle$ and $|s_3\rangle \mapsto |s_3\oplus(s_1 s_2) \; \mathrm{mod}\;2\rangle$ and is represented in Fig. \ref{fig:cnot}(b).

To sum-up, quantum computing works in the following way. First, the qubits forming the quantum register are initialized to a certain state, a vector in $\mathcal{H}_{\ell}$ of dimension $2^{\ell}$. Then, a sequence of unitary operations is applied to this quantum register, modifying its quantum state. The quantum circuits illustrate this procedure: each horizontal line represent a qubit, vertical lines represent control and boxes represent unitary operators. Finally, a measurement is performed where the state of the quantum system is observed.   

As discussed in the following sections, the most important operations of the classical algorithms considered in this paper are the changes of basis and the translations by upwinding. At the quantum level, these operations can be performed using rotation and translation operators. The latter can themselves be decomposed as elementary quantum gates (e.g. NOT,CNOT, Hadamard, Toffoli). The scaling of the number of fundamental quantum gates as a function of quantum resources yields the computational complexity of the algorithm. This will be discussed in Section \ref{sec:eff}.

\section{Reservoir method for symmetric linear first order hyperbolic systems}\label{sec:res_meth}

In this section, we recall the principle of the reservoir method for solving one-dimensional first order linear hyperbolic systems with constant matrices. This method was initially proposed in \cite{res0,res1,res3} and studied analytically in \cite{res2}. It basically consists of introducing flux-difference reservoirs and CFL counters in order to ensure a ``CFL=1'' or diffusion-less behavior of the approximate solution to hyperbolic systems of conservation laws at any time, on any characteristic field, and any location. 

We start by considering the one-dimensional case because this is the main building block to treat systems with many dimensions. Indeed, our strategy for the multi-dimensional case is the use of alternate direction iteration, whereas the multi-dimensional problem is reduced to a sequence of one-dimensional problems. 

\subsection{Reservoir method for one-dimensional linear systems}\label{sec:diagsys}

We first present some general remarks on finite volume discretization for first order hyperbolic systems in 1-D with constant matrices and then, describe how their discretization can be improved with the reservoir method. 

Specifically, we consider the following system:
\begin{eqnarray}\label{oneD1}
\left\{
\begin{array}{lcl}
\partial_t u + A\partial_x u & = & 0, \qquad (x,t) \in \R\times (0,T),\\
u(\cdot,0) & = & u_0, \qquad x \in \R,
\end{array}
\right.
\end{eqnarray}
where $u_0 :\R \mapsto \R^m$ is a given function in $\big(L^2(\R)\big)^m$ with compact support, so that $u_0 \in \big(L^1(\R)\big)^m$, and $A\in S_m(\R)$ has $m$ real eigenvalues ordered as $|\lambda_1| \leq \cdots \leq |\lambda_m|$. The fact that $u_0$ has compact support allows for avoiding boundary conditions issues on finite domain and for preserving the $L^{2}$-norm of the solution. The latter can be checked using an elementary calculation: in particular, it can be shown that $\partial_{t} \|u(t)\|_{L^2}=0$ if $A$ is symmetric and if $u(t)$ has compact support. As we will see later, the conservation of the $L^{2}$-norm is important to get simple quantum algorithms. 
We also assume that $\|u_0\|_{L^2}=1$. 

We denote by $\{s_k\}_{k=1,\cdots,m}$ the corresponding orthonormalized right-eigenvectors, and we denote by $S$=col($s_1,\cdots,s_m$) the corresponding transition matrix. To simplify the presentation, we will assume that the eigenvalues are all non-zero. Such a system is called hyperbolic \cite{serre,smoller,god1,god2}. Notice that in the basis of eigenvectors, the system is diagonal, and can then be rewritten as an uncoupled system
\begin{eqnarray*}
	\left\{
	\begin{array}{lcl}
		\partial_t w +\Lambda \partial_x w & =& 0, \qquad (x,t) \in \R\times (0,T),\\
		w(\cdot,0) & = & S^{-1}u_0, \qquad x \in \R,
	\end{array}
	\right.
\end{eqnarray*}  
where $\Lambda=$diag($\lambda_1,\cdots,\lambda_m$). The exact solution to this problem, for $k=1,\cdots,m$, is given by
\begin{eqnarray}
\label{eq:exact_1d}
w_{k}(x,t)= w_{k}(x-\lambda_{k}t,0).
\end{eqnarray}
This solution can be obtained by the method of characteristics and corresponds to a streaming with velocities $(\lambda_{k})_{k=1,\cdots,m}$.

We can now look how such systems are discretized on a grid to obtain a numerical scheme. For this purpose, we introduce a sequence of nodes $\{x_j\}_{j \in \Z}$ (resp. $\{x_{j \pm 1/2}\}_{j \in \Z}$), defined by $x_j = j\Delta x$ (resp. $x_{j \pm 1/2} = (j \pm 1/2)\Delta x$) for a given lattice spacing $\Delta x >0$. We also define finite volumes $\big\{\omega_j\big\}_{j \in \Z}$, where $\omega_j=(x_{j-1/2},x_{j+1/2})$. We introduce a sequence of times $\{t_n\}_{n\in \N}$ and time steps $\{\Delta t_n\}_{n \in \N}$ ulteriorly determined, as well as a sequence of vectors $\big\{U_j^n\big\}_{(j,n)\in \Z\times \N}$ of $\R^m$ approximating $u(x_j,t_n)$,  for any $j \in \Z$ and $n \in \N$. Initially, for any $j \in \Z$, the solution is projected on the mesh
\begin{eqnarray*}
	U_j^0 = \cfrac{1}{\Delta x} \int_{\omega_j}u_0(x)dx,
\end{eqnarray*}
corresponding to a finite volume formulation. For first order linear systems, the natural finite volume method consists of upwinding the solution on each characteristics field.  In other words, the corresponding finite volume scheme reads
\begin{eqnarray}\label{sch1}
U_j^{n+1} = U_j^n -\cfrac{\Delta t_n}{\Delta x}\big(\Phi_{j+1/2}^n-\Phi_{j-1/2}^n\big), \qquad j \in \Z, \, n\geq 0
\end{eqnarray}
where the interface flux is given by
\begin{eqnarray*}
	\Phi_{j-1/2}^{n} = \cfrac{1}{2}\Big\{A\big(U_j^n + U^n_{j-1}\big) - VA\big(U_j^n-U_{j-1}^n\big)\Big\},
\end{eqnarray*}
with $V=S\hbox{diag}_{k=1,\cdots,m}(\textrm{sgn}(\lambda_k))S^{-1}$ the so-called {\it sign matrix}.  Thus
\begin{eqnarray}\label{sch11}
U_j^{n+1} = U_j^n -\cfrac{\Delta t_n}{2 \Delta x} A\Big\{\big(I-V\big)(U_{j+1}^n-U_j^n) + \big(I+V\big)(U_j^n-U_{j-1}^n)\Big\} \, .
\end{eqnarray}
This is a straightforward generalization of the upwind scheme for transport equations.  This is equivalent in the basis of eigenvectors to solve
\begin{eqnarray*}
	W_j^{n+1} = W_j^n - \cfrac{\Delta t_n}{\Delta x}S^{-1}\big(\Phi_{j+1/2}-\Phi_{j-1/2}\big),
\end{eqnarray*}
where $W_j^n  =S^{-1}U_j^n$ with $W_j^n= \big(w_{1;j}^n,\cdots,w_{m;j}^n\big)^T$, which simply becomes per characteristic field
\begin{eqnarray}
\label{eq:diag_bas1}
		w_{k;j}^{n+1} & =& w_{k;j}^n - \cfrac{\lambda_k\Delta t_n}{\Delta x}(w^n_{k;j+1}-w^n_{k;j}), \qquad \lambda_k<0,\\
\label{eq:diag_bas2}
		w_{k;j}^{n+1} & =& w_{k;j}^n - \cfrac{\lambda_k\Delta t_n}{\Delta x}(w^n_{k;j}-w^n_{k;j-1}), \qquad \lambda_k>0 \, .
\end{eqnarray}
This scheme is stable iff the CFL-condition 
\begin{eqnarray*}
	\textrm{CFL}=\cfrac{\Delta t_n}{\Delta x}\max_{k=1,\cdots,m}|\lambda_k|=\cfrac{\Delta t_n}{\Delta x}|\lambda_m| \leq 1
\end{eqnarray*}
is satisfied. In practice, we choose CFL=1,  which allows for avoiding numerical diffusion on the $m$'th characteristic fields, but creating some on the other ones. However, if the eigenvalues have all the same modulus, the scheme \eqref{sch1} provides the exact solution on the grid even if it is only first order accurate. 
This is typically the case with the one-dimensional Dirac equation where eigenvalues are related to the speed of light (see \cite{cpc2012} for instance). The reservoir method \cite{res1,res2,res3} was precisely introduced as a tool to avoid numerical diffusion for all characteristics fields in first order finite volume schemes for hyperbolic systems of conservation laws. 

We now turn to the principle of the reservoir method. The main two ingredients of the reservoir techniques are i) the CFL counters and ii) the flux difference reservoirs at the finite volume interfaces $\{x_{j+1/2}\}_{j\in \Z}$. Indeed, when the matrix is non-constant $A(x)$ and the system is well-posed, the reservoir technique requires the introduction of $m$ time-dependent vectorial reservoirs at each interface $R_{1;j-1/2},\cdots,R_{m;j-1/2} \in \R^m$ and initially taken null, as well as $m$ scalar time-dependent counters $c_{1;j-1/2},\cdots,c_{m;j-1/2} \in [0,1)$ also initialized to $0$. In the specific cases discussed in this paper with a constant $A$ matrix, the reservoirs and counters are actually interface independent, greatly simplifying the notation$/$implementation of the scheme. Making this assumption allows us to consider only $m$ reservoirs  $R_{1},\cdots,R_m \in \R^m$ and $m$ counters $c_{1},\cdots,c_m$. Furthermore, we denote by $U^k_{\textrm{R}}(U,W)$ the solution of the Riemann problem with left (resp. right) state $U$ (resp. $W$), which lies between the $k$'th and $k+1$st wave, where we have set: $U^0_{\textrm{R}}(U,W)=U$ and $U^{m}_{\textrm{R}}(U,W)=W$. For linear systems, computing the solution of Riemann problems is almost straightforward (see \cite{god2,res1} for details). Additionally, we introduce a temporary variable:
%
\begin{eqnarray*}
	\mathcal{C}_k^{n+1}:=c_{k}^n + |\lambda_k|\cfrac{\Delta t_n}{\Delta x} \, .
\end{eqnarray*}
Then, we update the solution as follows
\begin{eqnarray}
\label{eq:update_reservoir}
U_j^{n+1} = U^n_j + \sum_{k=1}^{m}\big(\widetilde{U}_{k;j-1/2}^{n+1}+\widetilde{U}^{n+1}_{k;j+1/2}\big),
\end{eqnarray}
where we have (if $\lambda_k<0$)
\begin{eqnarray*}
	\left.
	\begin{array}{lcl}
		\left(
		\begin{array}{c}
			\widetilde{U}^{n+1}_{k;j+1/2} \\
			c_{k}^{n+1}\\
			R_{k}^{n+1}
		\end{array}
		\right) 
		& = & 
		\left\{
		\begin{array}{c}
			\left(
			\begin{array}{c}
				0 \\
				c_k^n+|\lambda_k|\cfrac{\Delta t_n}{\Delta x}\\
				R_{k}^n-\cfrac{\Delta t_n}{\Delta x}A\big(U^k_{\textrm{R}}(U^n_{j},U_{j+1}^n)-U^{k-1}_{\textrm{R}}(U^n_{j},U_{j+1}^n)\big)
			\end{array}
			\right),  \hbox{ when } \mathcal{C}^{n+1}_k < 1,\\
			\left(
			\begin{array}{c}
				R_{k}^n-\cfrac{\Delta t_n}{\Delta x}A\big(U^k_{\textrm{R}}(U^n_{j},U_{j+1}^n)-U^{k-1}_{\textrm{R}}(U^n_{j},U_{j+1}^n)\big)\\
				0\\
				0
			\end{array}
			\right),  \hbox{ when } \mathcal{C}^{n+1}_k =1
		\end{array}
		\right.
	\end{array}
	\right.
\end{eqnarray*}
and if $\lambda_k>0$, we have
\begin{eqnarray*}
	\left.
	\begin{array}{lcl}
		
		\left(
		\begin{array}{c}
			\widetilde{U}^{n+1}_{k;j-1/2} \\
			c_{k}^{n+1}\\
			R_{k}^{n+1}
		\end{array}
		\right) 
		
		& = & 
		
		\left\{
		\begin{array}{c}

			\left(
			\begin{array}{c}
				0 \\
				c_k^n+|\lambda_k|\cfrac{\Delta t_n}{\Delta x}\\
				R_{k}^n-\cfrac{\Delta t_n}{\Delta x}A\big(U^k_{\textrm{R}}(U^n_{j-1},U_{j}^n)-U^{k-1}_{\textrm{R}}(U^n_{j-1},U_{j}^n))\big)
			\end{array}
			\right),  \hbox{ when } \mathcal{C}^{n+1}_k < 1,\\

			\left(
			\begin{array}{c}
				R_{k}^n-\cfrac{\Delta t_n}{\Delta x}A\big(U^k_{\textrm{R}}(U^n_{j-1},U_{j}^n)-U^{k-1}_{\textrm{R}}(U^n_{j-1},U_{j}^n))\big)\\
				0\\
				0
			\end{array}
			\right),  \hbox{ when } \mathcal{C}^{n+1}_k =1 \, .

		\end{array}
		\right.

	\end{array}
	\right.
\end{eqnarray*}
The time steps are finally chosen as
\begin{eqnarray*}
	\Delta t_n = \min_{k=1,\cdots,m}\left[\big(1-c_{k}^n\big)\cfrac{\Delta x}{|\lambda_k|}\right] \, .
\end{eqnarray*}
Although, the scheme may look complicated, it simply consists in updating the $k$'th components of the solution in the basis of eigenvectors, when the corresponding local counter reaches $1$ or {\it any} prescribed value less or equal to $1$.  

The analysis of convergence is addressed in \cite{res2}. In particular, it was proven that the reservoir method provides exact solutions at the discrete level for linear hyperbolic systems with constant rational eigenvalues. This latter assumption was only technical, and the reservoir method is still applicable beyond this condition. More specifically, it is proven that at time say $T>0$, the reservoir method combined with an order 1 finite volume method provides a $L^1$-error $\epsilon$, which is bounded by the product of the maximal time step with the initial $L^1$-norm error ($L^1$-norm of the error between the exact initial data and its projection on the finite volume mesh):
\begin{eqnarray*}
	\epsilon = \norm{u(\cdot,T) - U^{n_T}}{L^{1}} < \norm{u(\cdot,t_{0}) - U^{0}}{L^{1}}   +  C \max_{k=1,\cdots,n_T}(\Delta t_{k}),
\end{eqnarray*}
where $C>0$ is a real constant.
As a consequence in one dimension, the error remains bounded for large $T$'s, unlike usual finite volume methods (including higher order ones, in general) for which the error grows linearly in $T$.

A priori, it is challenging to design a quantum algorithm that implements this numerical scheme. However, for the case considered here, where $A$ is constant, the reservoir method can be formulated in a very simple way, more suitable for a quantum implementation.
In the diagonal basis, the upwind scheme reads Eqs. \eqref{eq:diag_bas1} and \eqref{eq:diag_bas2}. Then, at each time step, the reservoir technique consists of freezing the components of the solution, until the corresponding CFL counter reaches $1$. In practice, we proceed as follows, assuming that we are at time $t_{n}$:
\begin{itemize}
	\item The time step is evaluated from
	\begin{eqnarray*}
		\Delta t_{n+1} = \min_{k=1,\cdots,m}\left[\big(1-c_{k}^n\big)\cfrac{\Delta x}{|\lambda_k|}\right] \, .
	\end{eqnarray*}
	\item Then,  the  CFL counter reaches 1 for some set of components $\mathcal{K}^{n} = \{ k^{*}_{1},\cdots, k_{a}^{*}\}$ with $k^{*}_{1},\cdots,k_{a}^{*} \in \{1,\cdots,m\}$, where $a$ is the number of components that needs to be updated. We also define a set which stores the sign of the corresponding eigenvalues, that is $\mathcal{Q}^{n} = \{\sigma(\lambda_{k^{*}_{1}}),\cdots, \sigma(\lambda_{k^{*}_{a}})\}$, where $\sigma$ is the sign function. Then, for all $k \in \mathcal{K}^{n}$ 
	\begin{eqnarray*}
		\left.
		\begin{array}{lcl}
			w_{k;j}^{n+1} & = & w^n_{k;j-1}, \qquad \hbox { if } \lambda_{k}>0,\\
			w_{k;j}^{n+1} & = &  w^n_{k;j+1}, \qquad \hbox { if } \lambda_{k}<0.
		\end{array}
		\right.
	\end{eqnarray*}
	This step corresponds to a simple translation of the solution, similar to the analytical solution given in \eqref{eq:exact_1d}. Therefore, the reservoir method reproduces the exact solution on the grid, even if the scheme is first order. Meanwhile, the other components are frozen, that is for any $k\notin \mathcal{K}^{n}$ 
	\begin{eqnarray*}
		w_{k;j}^{n+1} = w^n_{k;j} \, .
	\end{eqnarray*}
	\item The counters are updated as follow.  For $k \neq k^*$
	\begin{eqnarray*}
		c_{k}^{n+1} =  c_{k}^n + |\lambda_{k^*}|\cfrac{\Delta t_n}{\Delta x}, \, \hbox{ and } \, c_{k^*}^{n+1}=0.
	\end{eqnarray*}
	\item At the end of the calculation, we can use the transition matrix $S$ to obtain the approximate solution $U^{n_{T}}$. 
\end{itemize}

The reservoir method can be simply reformulated as a list of operations.
Let us denote by $n_T$ the total number of iterations, such that $\sum_{n=1}^{n_T} \Delta t_n= T$. Next, we denote by $\mathcal{I}^n,\mathcal{S}^{n}$ the ordered sets of indices and signs, respectively, corresponding to the characteristic fields which have been updated up to time $t_{n}$. Generally, we have $\mathcal{I}^{n}:=\big(\mathcal{K}^1,\cdots,\mathcal{K}^{n})$ and $\mathcal{S}^{n}:=\big(\mathcal{Q}^1,\cdots,\mathcal{Q}^{n})$, and we denote the $k$'th element of $\mathcal{I}^{n_T},\mathcal{S}^{n_T}$ with $k \leq n_T$, by $\mathcal{I}_k$ and $\mathcal{S}_k$, respectively. The only relevant information required by the quantum algorithm for linear systems with constant coefficients, is then the sets $\mathcal{I}^{n_T}$ and $\mathcal{S}^{n_T}$. In particular, there will be no need for explicitly  creating and updating reservoirs or even CFL counters. In practice, a classical algorithm can be run to determine the sets $\mathcal{I}^{n_T}$ and $\mathcal{S}^{n_T}$. 

\noindent{\bf Basic example.} We now propose as an illustration, a simple example to construct $\mathcal{I}^{n_T}$. We consider $A \in S_3(\R)$ with eigenvalues $\lambda_1=10^{-1}$, $\lambda_2=1$ and $\lambda_3=1+10^{-1}$. Numerically, we take $\Delta x=10^{-2}$, $T=10^{-1}$ and $n_T=23$. We then find
\begin{eqnarray}\label{Int1D}
\mathcal{I}^{n_T} & =& ( 3,     2,     3,    2,     3,     2,     3,     2,     3,     2,     3, 
   2,     3,     2,     3,     2,     3,     2,     3,     1,     3,     2,     3 ),
\end{eqnarray}
while the set $\mathcal{S}^{n_{T}}$ is trivial, containing only positive signs.
The first time steps are given by $\Delta t_{1}=9.09\times 10^{-3}$,  $\Delta t_{2}=9.09\times 10^{-4}$, $\Delta t_{3}=8.18\times 10^{-3}$, $\Delta t_{4}=1.82\times 10^{-3}, \cdots$.

\begin{figure}[!ht]
	\begin{center}
		\hspace*{1mm}\includegraphics[height=6cm, keepaspectratio]{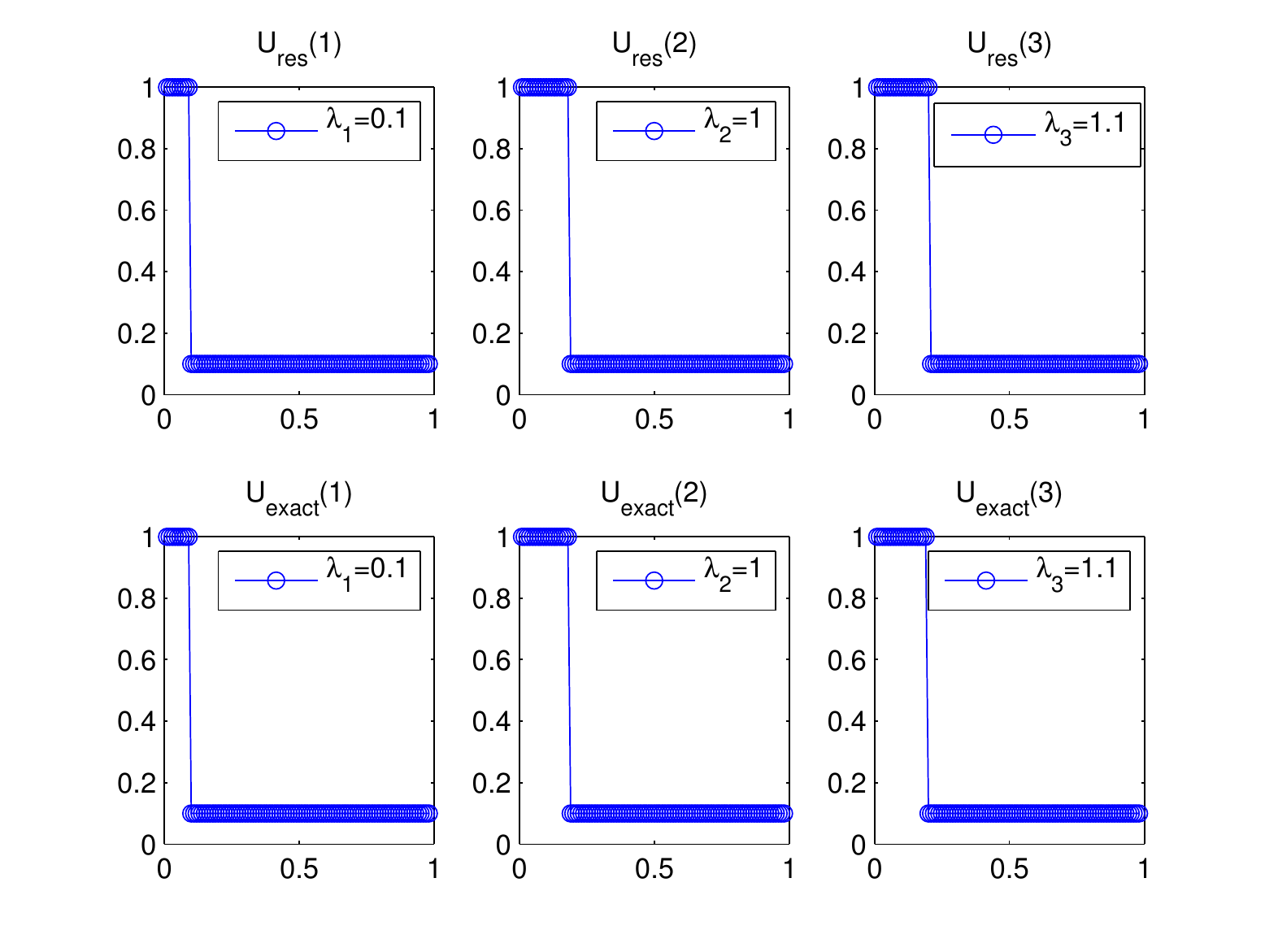}
		\hspace*{1mm}\includegraphics[height=6cm, keepaspectratio]{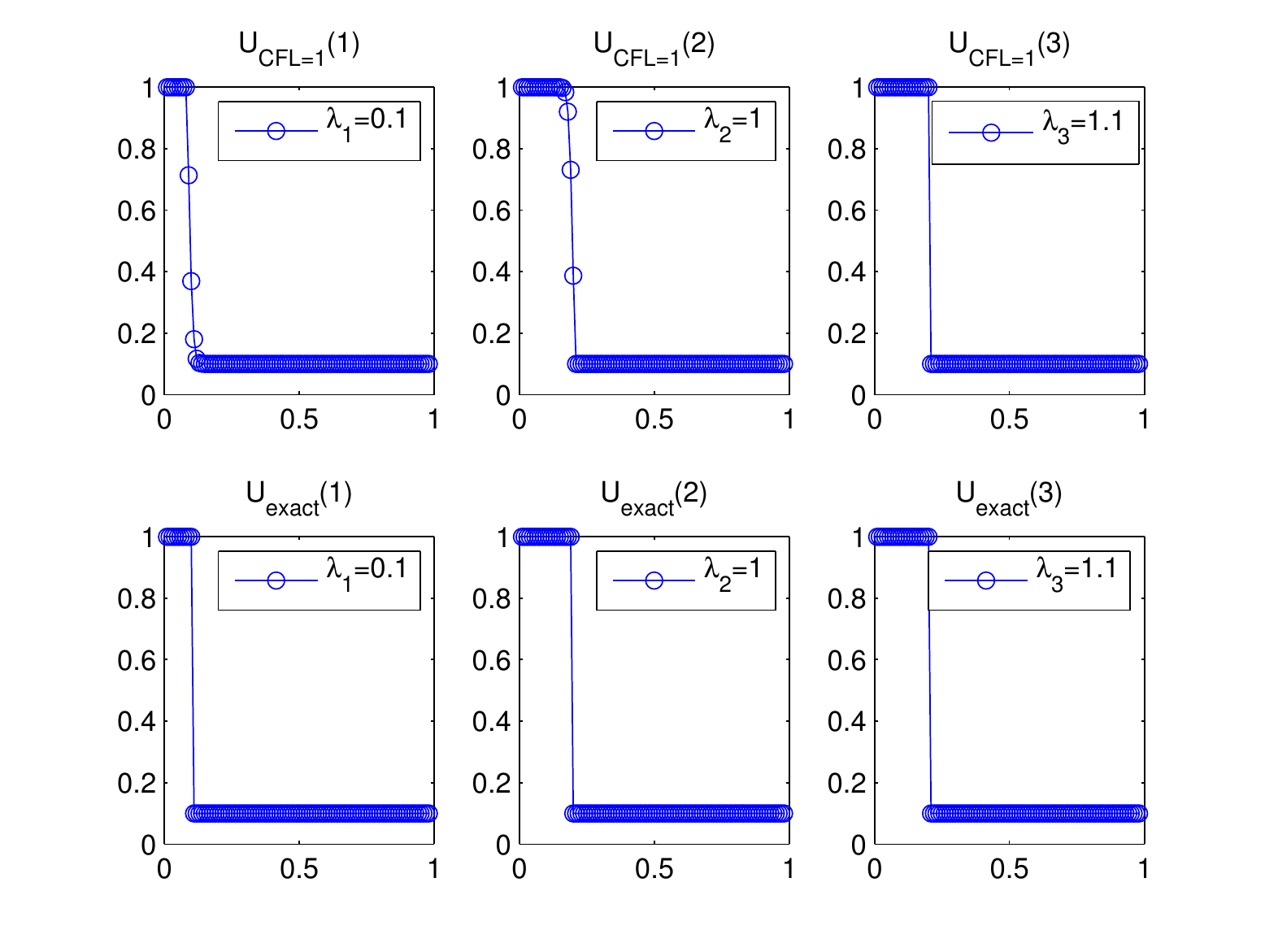}
		\caption{Reservoir method (on the left) and upwind scheme with CFL=1 (on the right) solutions at final time $T=0.1$.}
		\label{fig0}
	\end{center}
\end{figure}

For $A=\mathrm{diag}(\lambda_1,\lambda_2,\lambda_3)$ and with a domain given by $[0,1]$ (with Dirichlet boundary conditions), and an initial data $u_0(x)=1$ for $x\leq 0.1$ and $u_0(x)=0$, for $x \in (0.1,1]$, we report the solution components at the final time in Fig. \ref{fig0}, and compare to the ``CFL=1''-solution. As expected, the ``CFL=1''-solution displays some spurious diffusion for components with the lowest eigenvalues ($u_{1}$ and $u_{2}$), seen as a smoothing of the discontinuity. On the other hand, the component with the largest eigenvalue ($u_{3}$) does not show this effect as the CFL condition is exactly one for this particular component, in stark contrast with $u_{1}$ and $u_{2}$, for which the CFL condition is lower than 1, inducing diffusion in the numerical solution. The solution obtained from the reservoir technique is very few diffused because the CFL condition is 1 at all time and space points.

\subsection{Alternate direction iteration for multi-dimensional systems}
\label{sec:adi_mD}

We can now generalize the ideas presented in the previous section for multi-dimensional linear symmetric hyperbolic systems.  In particular, we consider, for $u=u(x,t)$ and $x = (x_1,\cdots,x_d)$, the following system
\begin{eqnarray}\label{multiD}
\left\{
\begin{array}{lcl}
\partial_tu + \sum_{j=1}^dA^{(j)}\partial_{x_{j}}u & = & 0, \qquad (x,t) \in \R^d\times (0,T),\\
u(\cdot,0) & = & u_0, \qquad x \in \R^d
\end{array}
\right.
\end{eqnarray}
where the matrices $A^{(j)} \in S_{m}(\mathbb{R})$ and such that for any $n=(n^{(1)},\cdots,n^{(d)})^T \in \R^d$, $\sum_{j=1}^dA^{(j)}n^{(j)}$ has real eigenvalues with linearly independent eigenvectors, and $u_0$ has compact support and $L^2$-norm equal to $1$. Again, the $L^{2}$-norm is conserved under these conditions. Then, for any $1 \leq j \leq d$,  the eigenvalues of $A^{(j)}$ are denoted $|\lambda^{(j)}_1| <\cdots <|\lambda^{(j)}_m|$. By symmetry-assumption, for any $j \in \{1,\cdots,d\}$ there exists $S^{(j)} \in U_n(\R)$ such that $A^{(j)}=S^{(j)}\textrm{diag}\big(\lambda^{(j)}_1,\cdots,\lambda^{(j)}_d\big)(S^{(j)})^T$. The corresponding reservoir-method for constant matrices and with directional splitting can be proven to be $L^2-$stable by construction. 

As mentioned earlier, it is convenient to use the alternate direction iteration method in order to apply the 1-D reservoir method for each dimension. The alternate direction iteration method proceeds as follows \cite{leveque2002finite}. From time $t_n$ to $t_{n+1}$, and assuming $u(\cdot,t_n)$ known, we successively solve
\begin{eqnarray}
\label{eq:splitting}
\left\{
\begin{array}{l}
\left\{
\begin{array}{lcll}
\partial_t u^{(1)} + A^{(1)}\partial_{x_{1}}u^{(1)} & = & 0, & (x,t) \in \R^d\times \in (t_n,t_{n_{1}}),\\
u^{(1)}(\cdot,t_n) & = & u(\cdot, t_n), &  x \in \R^d
\end{array}
\right.
\\
\\
\left\{
\begin{array}{lcll}
\partial_t u^{(2)} + A^{(2)}\partial_{x_{2}}u^{(2)} & = & 0, & (x,t) \in \R^d\times \in (t_{n_{1}},t_{n_2}),\\
u^{(2)}(\cdot,t_n) & = & u^{(1)}(\cdot, t_{n_1}), &  x \in \R^d
\end{array}
\right.
\\
\\
\cdots \\
\\
\cdots \\
\\
\left\{
\begin{array}{lcll}
\partial_t u^{(d)} + A^{(d)}\partial_{x_{d}}u^{(d)} & = & 0, & (x,t) \in \R^d\times \in (t_{n_{d-1}},t_{n+1}),\\
u^{(d)}(\cdot,t_n) & = & u^{(d-1)}(\cdot, t_{n_{d-1}}), &  x \in \R^d
\end{array}
\right.
\end{array}
\right.
\end{eqnarray}
Finally, we define the approximate solution at time $t_{n+1}$ by $u(\cdot,t_{n+1})=u^{(d)}(\cdot,t_{n+1})$.
This corresponds to a first order splitting scheme and therefore, has an error that scales like $O(\Delta t_{n}^{2})$. We note here that just like in the 1-D case, the set of all $(\Delta t_{n})_{n=1,\cdots,n_T}$ is still to be determined.

We can now discretize the solution on a space grid using finite volumes to obtain the full numerical scheme. This process is very similar to the 1-D case. For this purpose, we introduce, for each dimension $i=1,\cdots,d$, a sequence of nodes $\{x_{i;j}\}_{j \in \Z}$ (resp. $\{x_{i;j\pm1/2}\}_{j \in \Z}$), defined by $x_{i;j} = j\Delta x$ (resp. $x_{i;j\pm1/2} = (j\pm1/2)\Delta x$) for a given $\Delta x >0$. Then, we can define  $d$-dimensional cubic cells as $ \omega_{j_1,\cdots,j_d} = \otimes_{i=1}^d\omega_{i;j_{i}}$ where $\omega_{i;j}=(x_{i;j-1/2},x_{i;j+1/2})$. 
The discretization then proceeds as follows. Let us introduce a sequence of vectors $\big\{U_{j_1,\cdots,j_d}^n\big\}_{(j_1,\cdots,j_d,n)\in \Z^d\times\N}$ in $\R^m$, approximating the mean of $u(x,t_n)$ over $\omega_{i;k}$ for any $(j_1,\cdots,j_d) \in \Z^d$, $n \in \N$. The initial data is set to 
\begin{eqnarray*}
	U_{j_1,\cdots,j_d}^0 = \cfrac{1}{(\Delta x)^d} \int_{\omega_{j_1,\cdots,j_d}}u_0(x)dx_{1} \cdots dx_{d} \, .
\end{eqnarray*} 
Then, the update of the solution parallels the 1-D case. This is possible because the splitting has transformed the multi-dimensional system to a sequence of 1-D systems, using alternate direction iteration. Therefore, we can introduce reservoirs and CFL counters at the interface between the finite volumes to implement the reservoir method. However, it was demonstrated in Section \ref{sec:diagsys} that for constant matrices $A^{(j)}$, the reservoir are interface dependent, allowing for the introduction of $d \times m$ reservoirs $\{R_{i;k} \}$ and $d \times m$ counters $\{c_{i;k}\}$, for $i \in \{1,\cdots,d\}$ and $k\in \{1,\cdots,m\}$. However, the method amounts to the generation of a list of updated components where the reservoirs and CFL counters can be disregarded. This can be carried to the multi-dimensional case, except for one new ingredient: before each update, the system has to be diagonalized thanks to unitary operators $\{S_i\}_{1 \leq i \leq d} \in U_{m}(\R)$, where $i$ is the direction of the streaming.  Setting $w^{(i)}=S^{(i)T}u^{(i)}$ for any $i \in \{1,\cdots,d\}$, the equations in Eq. \eqref{eq:splitting} are transformed to uncoupled systems of transport equations in directions $x_{i}$ for $1 \leq i \leq d$, of the form
\begin{eqnarray*}
\partial_t w^{(i)} + \Lambda^{(i)}\partial_{x_{i}}w^{(i)} = 0,
\end{eqnarray*}
which can be solved as in the 1-D case. In particular, it is necessary to create the lists $\mathcal{I}^{n_T}$ and $\mathcal{S}^{n_{T}}$, similar to the ones defined in Section \ref{sec:diagsys}, which allows for sorting the translation operators. 
The numerical scheme at time $t_{n}$ reads as follows with $U_{j_1,\cdots,j_d}^n=\big(u_{1;j_1},\cdots,u_{m;j}^n\big)^T$ and $W_{j_1,\cdots,j_d}^n=\big(w_{1;j_1},\cdots,w_{m;j}^n\big)^T$.
\begin{itemize}
	\item The time step is evaluated from
	\begin{eqnarray*}
		\Delta t_{n+1} = \min_{k=1,\cdots,m}\min_{i=1,\cdots,d}\left[\big(1-c^{n}_{i;k}\big)\cfrac{\Delta x}{|\lambda^{(i)}_k|}\right] \,.
	\end{eqnarray*}
	\item Then, the CFL counter reaches 1 for some set of pairs $\mathcal{K}^{n} = \{ (k^{*}_{1},i^{*}_{1}),\cdots, (k_{a}^{*},i^{*}_{a})\}$ with components $k^{*}_{1},\cdots,k_{a}^{*} \in \{1,\cdots,m\}$ and dimensions $i^{*}_{1},\cdots,i_{a}^{*} \in \{1,\cdots,d\}$, where $a$ is the number of components to be updated. We also define $\mathcal{Q}^{n}$ as in the one-dimensional case. 
	Then for all pairs  $(k,i) \in \mathcal{K}^{n}$, the solution update proceeds in three steps:
	
	\begin{enumerate}
	\item We transform to the diagonal basis using $W_{j_1,\cdots,j_d}^n = (S^{(i)})^{T} U_{j_1,\cdots,j_d}^n$.
	
	\item  We evaluate the streaming step as
		\begin{eqnarray*}
			\left.
			\begin{array}{lcl}
				w_{k;j_1,\cdots,j_{i},\cdots j_d}^{n+1} & = & w^n_{k;j_1,\cdots,j_{i-1},j_{i}-1,j_{i+1},\cdots j_d}, \qquad \hbox {if } \lambda^{(i)}_{k}>0,\\
				w_{k;j_1,\cdots,j_{i},\cdots j_d}^{n+1} & = & w^n_{k;j_1,\cdots ,j_{i-1},j_{i}+1,j_{i+1},\cdots j_d}, \qquad \hbox {if } \lambda^{(i)}_{k}<0.
			\end{array}
			\right.
		\end{eqnarray*}
		Meanwhile, the other components are frozen, that is for any pairs for which  $k' \neq k$, we have
		\begin{eqnarray*}
			w_{k';j_1,\cdots,j_d}^{n+1} = w^n_{k';j_1,\cdots,j_d} \, .
		\end{eqnarray*}

	\item We transform back to the canonical basis using $U_{j_1,\cdots,j_d}^n = S^{(i)} W_{j_1,\cdots,j_d}^n$.
	\end{enumerate}

	\item Finally, the CFL counters are updated as follows.  For $k' \neq k$ with any $i  \in \{1,\cdots,d\}$, and for $k=k'$ with $i' \neq i$ 
	\begin{eqnarray*}
		c_{i';k'}^{n+1} =  c_{i';k'}^{n} + |\lambda^{(i)}_{k}|\cfrac{\Delta t_n}{\Delta x}, \, \hbox { and } \, c_{i;k}^{n+1}=0 \, .
	\end{eqnarray*}
\end{itemize}

The list $\mathcal{I}^n$ is now a set of pairs of indices, corresponding to the characteristic fields and dimension which have been updated. It is given by $\mathcal{I}^{n}=\big(\mathcal{K}^1,\cdots , \mathcal{K}^{n}\big)$. We also have the list $\mathcal{S}^{n}$ which stores the sign of the eigenvalues. With these two lists, it is possible to evolve the solution in time and this is equivalent to the reservoir method combined with alternate direction iteration.

\noindent{\bf Basic example.} We illustrate this approach with a two-dimensional test (another example can be found in Appendix \ref{app:ex_2d_diag}, for a diagonal system):
\begin{eqnarray*}
\left\{
\begin{array}{lcl}
\partial_tu + A^{(\textrm{x})}\partial_{x}u +  A^{(\textrm{y})}\partial_{y}u & = & 0, \qquad (x,y,t) \in [0,L]^2\times (0,T),\\
u(x,y,0) & = & u_0, \qquad (x,y) \in [0,L]^2
\end{array}
\right.
\end{eqnarray*}
The matrices are defined as $A^{(\textrm{x})} = S^{(\textrm{x})}\Lambda^{(\textrm{x})}S^{(\textrm{x})T}$, and $A^{(\textrm{y})} = S^{(\textrm{y})}\Lambda^{(\textrm{y})}S^{(\textrm{y})T}$ with $\lambda_1^{(1)}=1,\lambda_2^{(1)}=2$ and $\lambda_1^{(2)}=1,\lambda_2^{(2)}=4$ and with unitary transition matrices defined by
\begin{eqnarray*}
S^{(\textrm{x})} =
\cfrac{1}{\sqrt{2}}\left(
\begin{array}{cc}
1 & -1 \\
1 & 1
\end{array}
\right) , \qquad
S^{(\textrm{y})} =
\cfrac{1}{\sqrt{5}}\left(
\begin{array}{cc}
2 & -1 \\
-1 & 2
\end{array}
\right) \, . 
\end{eqnarray*}
Notice that in this case, the matrices $A^{(\textrm{x})}$ and $A^{(\textrm{y})}$ do not commute, but $A^{(\textrm{x})}A^{(\textrm{y})}-\big(A^{(\textrm{y})}A^{(\textrm{x})}\big)^T=0$ and $[S^{(\textrm{x})},S^{(\textrm{y})}]=0$.

\begin{figure}[!ht]
\begin{center}
\centering 
\subfloat[]{\includegraphics[height=6cm, keepaspectratio]{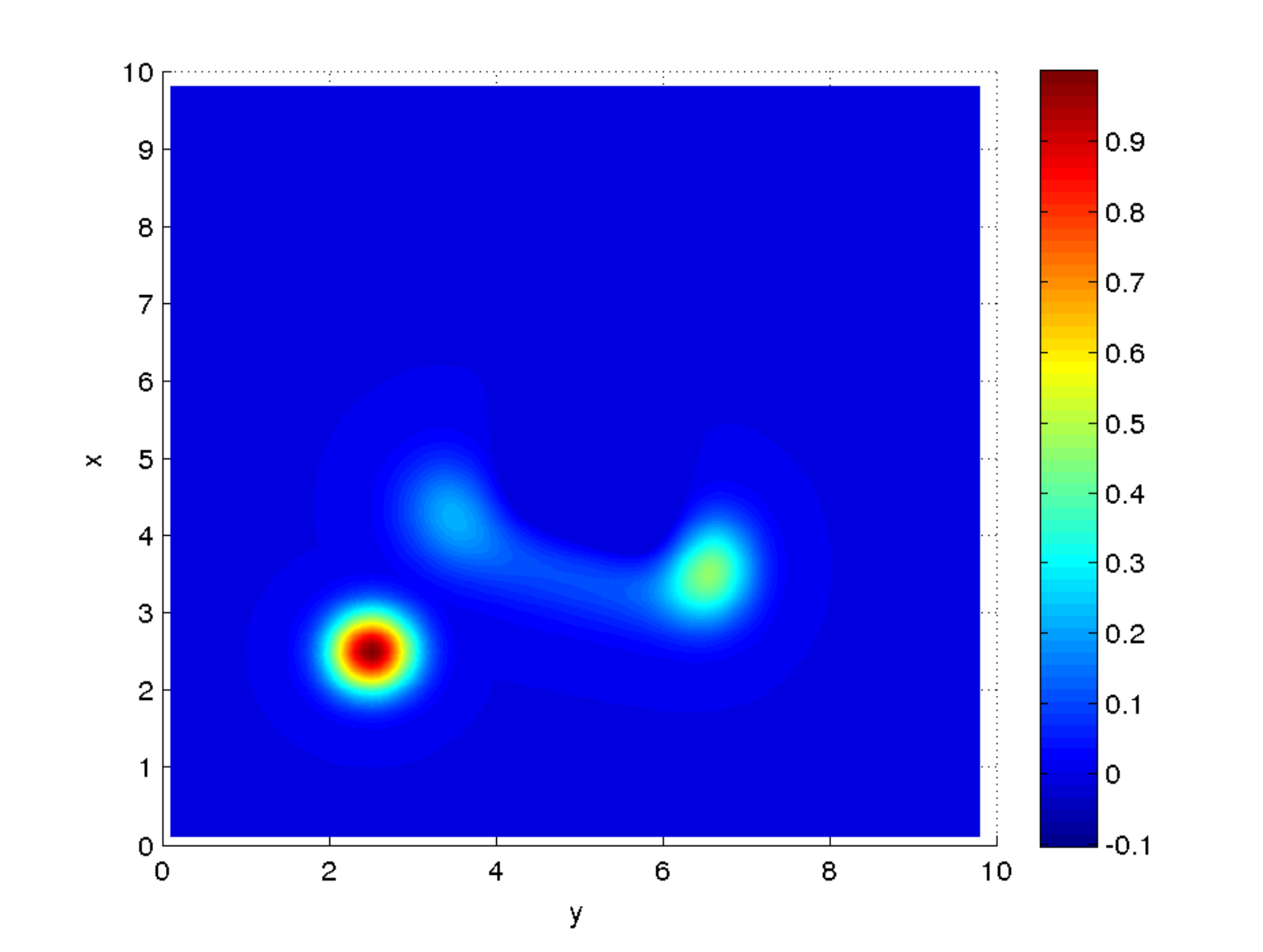}}
\subfloat[]{\includegraphics[height=6cm, keepaspectratio]{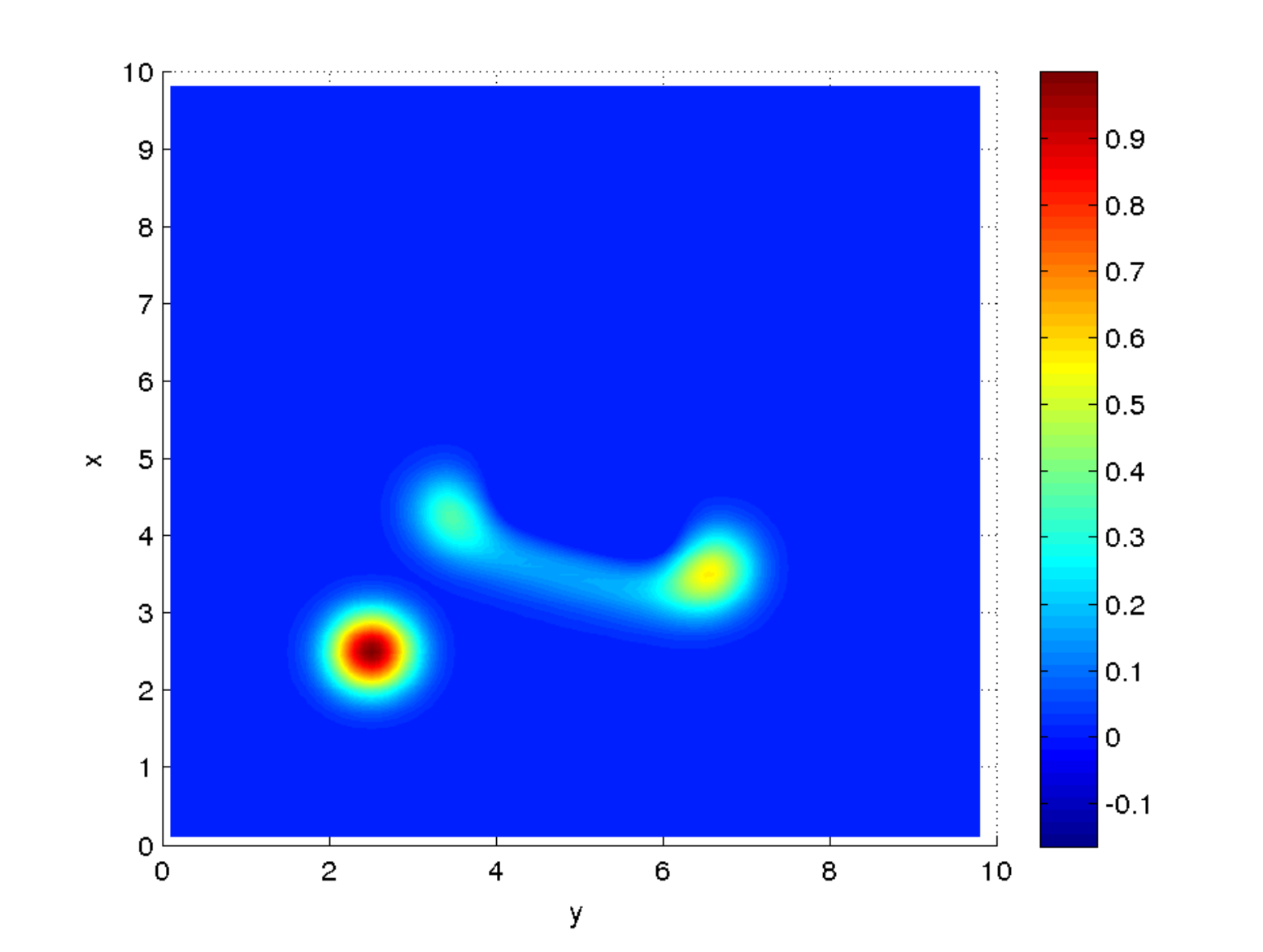}}\\
\subfloat[]{\includegraphics[height=6cm, keepaspectratio]{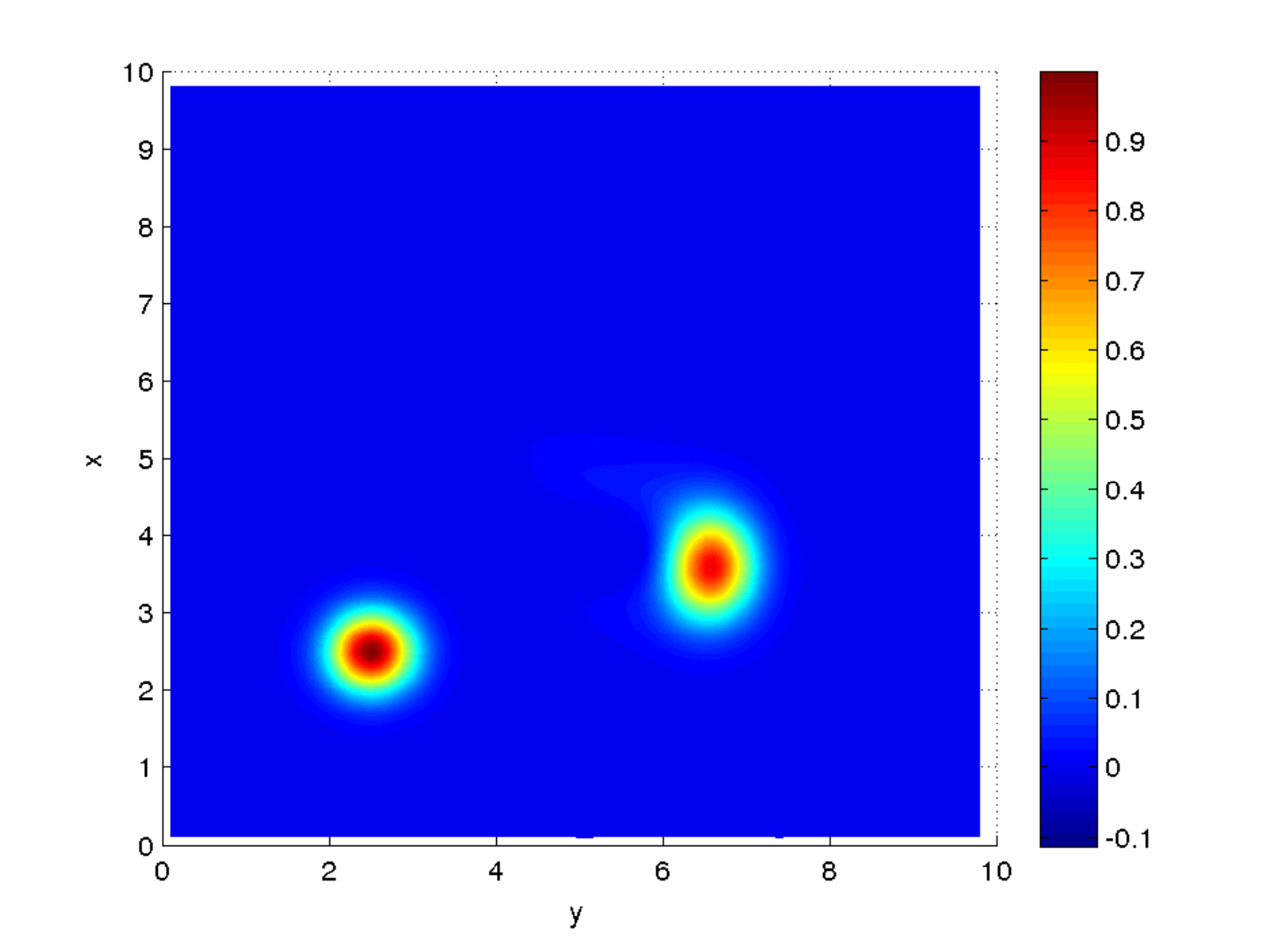}}
\subfloat[]{\includegraphics[height=6cm, keepaspectratio]{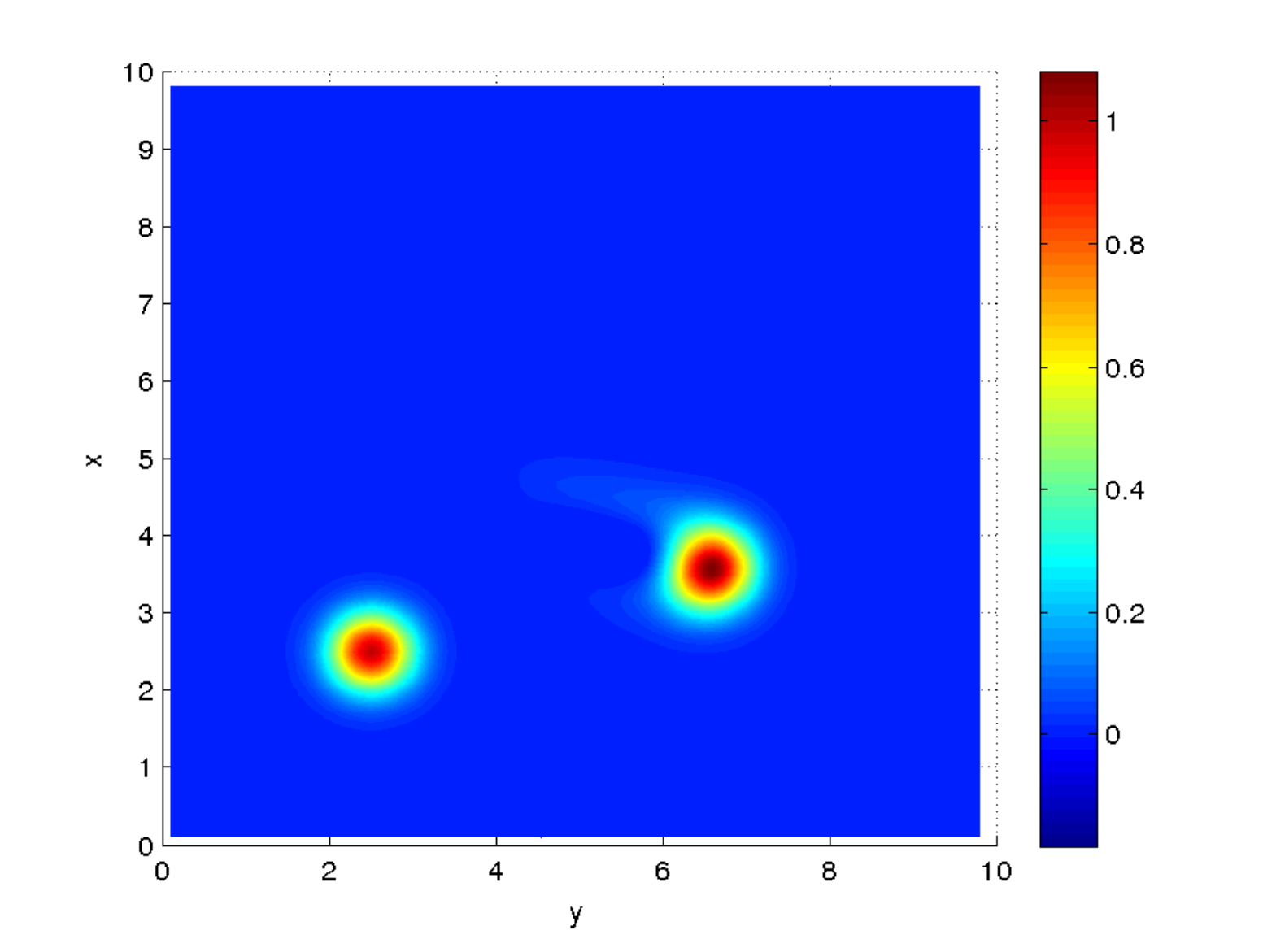}}
\caption{first (bottom-left corner) and second component solution at time $T=1$. (Left column) CFL=1 solution. (Right column) Reservoir method.}
\label{fig2}
\end{center}
\end{figure}

The computational domain is $[0,L=10]^2$ and $T=1$, $m=2$, $d=2$, $\Delta x_{1}=\Delta x_{2}=0.1$. The initial data is $u_{0,1}(x_{1},x_{2})=\exp\Big(-4\big((x_{1}-5/2)^2 + (x_{2}-5/2)^2\big)\Big)$, $u_{0,2}(x_{1},x_{2})=\exp\Big(-4\big((x_{1}-5/2)^2 + (x_{2}-5/2)^2\big)\Big)$. The reservoir solution components (first and second components at time $T=1$) are represented in Fig. \ref{fig2}, showing no numerical diffusion, unlike the ``CFL=1''-solutions also reported in Fig. \ref{fig2}.

\section{Quantum algorithm for linear first order hyperbolic systems}
\label{sec:general_case}

In this section, a quantum algorithm, i.e. an algorithm which can be implemented on a quantum computer, is developed for solving linear symmetric first order hyperbolic systems. The algorithm will be formulated as a quantum circuit which is equivalent to the reservoir method described in the previous section.  

\subsection{Amplitude encoding of the solution}\label{subsec:amp_encoding}

To develop a quantum algorithm and to benefit of the quantum nature of the computation, the first step is the encoding of the solution into the quantum register. Here, we use the same notations as in Ref. \cite{fillion}, and we map $U_{j_{1},\cdots,j_{d}}^{n}$  into the probability amplitudes. First, we assume that the quantum register is made of $\ell = p + \sum_{i=1}^{d}n_{i}$ qubits. Therefore, its state is given by Eq. \eqref{eq:qu_register}. However, for our purpose, it is more convenient to split the Hilbert space of the register in different parts as $\mathcal{H}_{\ell} = \mathcal{H}_p\otimes_{i=1}^d\mathcal{H}_{n_{x_{i}}}$ and to relabel the states such that
\begin{eqnarray*}
	|u_{\ell}\rangle = \sum_{k=1}^{m}\sum_{j_{1}=1}^{N_{x_{1}}} 
	\cdots \sum_{j_{d}=1}^{N_{x_{d}}}
	\alpha_{k,j_1,\cdots,j_d}|k\rangle \otimes |j_1,\cdots,j_d\rangle,
\end{eqnarray*}
where $m = 2^{p}$ and $N_{i}=2^{n_{i}}$ and where the relabeling is performed as $(k-1)_{10} = (s_1 \cdots s_p)_2$ and $(j_{i}-1)_{10} = (s_{p+ \sum_{l=1}^{i-1}n_{l}+1} \cdots s_{p+\sum_{l=1}^{i}n_{l}})_2 := (s_{i,1} \cdots s_{i,n_x})_2$. This means that the first $p$ qubits label the $m=2^{p}$ components of the solution, while the other $\tilde{n}:= \sum_{i=1}^{d}n_{i}$ qubits serves to label the coordinate space positions. Because this is a tensor product, there are $N:= \prod_{i=1}^{d}N_{x_{i}}= 2^{\tilde{n}}$ available amplitudes to store lattice  coordinates. 

Once the states of the quantum register are properly relabeled,  the mapping of the solution using amplitude encoding is straightforward:
\begin{eqnarray*}
	u_{k;j_1,\cdots,j_d}^{n} \mapsto \alpha_{k,j_1,\cdots,j_d} \, ,
\end{eqnarray*}
for a given time $n$.
In other words, the coefficients of the discretized solution are mapped into the probability amplitudes of the quantum register.

\subsection{Updating the solution in the quantum algorithm}

For the updating of the solution in the quantum algorithm, we are seeking a sequence of unitary transformations that are equivalent to the reservoir method described in the last section. Precisely, we are looking for a unitary transformation $\hat{V}$, such that, for all $k,j_1,\cdots,j_d$
\begin{eqnarray*}
u_{k;j_1,\cdots,j_d}^{n} \mapsto \alpha_{k,j_1,\cdots,j_d} \xrightarrow{\hat{V}|u_{\ell}\rangle}  \alpha'_{k,j_1,\cdots,j_d} \mapsto u_{k;j_1,\cdots,j_d}^{n+1}, 
\end{eqnarray*} 
where $u_{k;j_1,\cdots,j_d}^{n+1}$ is the same as the one obtained by the reservoir method. We will discuss the complexity of these mappings in the next section.  Notice that here and in the following, we denote unitary operations on the quantum register with the ``hat'' notation ($\hat{V}$). 

We saw in Section \ref{sec:adi_mD} that the reservoir method, for constant matrices, amounts to a sequence of three operations: transformation to the diagonal basis, streaming and transformation to the canonical basis. The components and dimension which are subjected to these transformations are given by the list $\mathcal{I}^{n_{T}}$ while the streaming direction is in $\mathcal{S}^{n_{T}}$. We would like to perform the same operations on the quantum register and therefore, we need to construct two types of operator: translation operators and rotation operators.

\subsubsection{Translation operators}
\begin{figure}
	\centering 
	\subfloat[]{\includegraphics[scale=0.7]{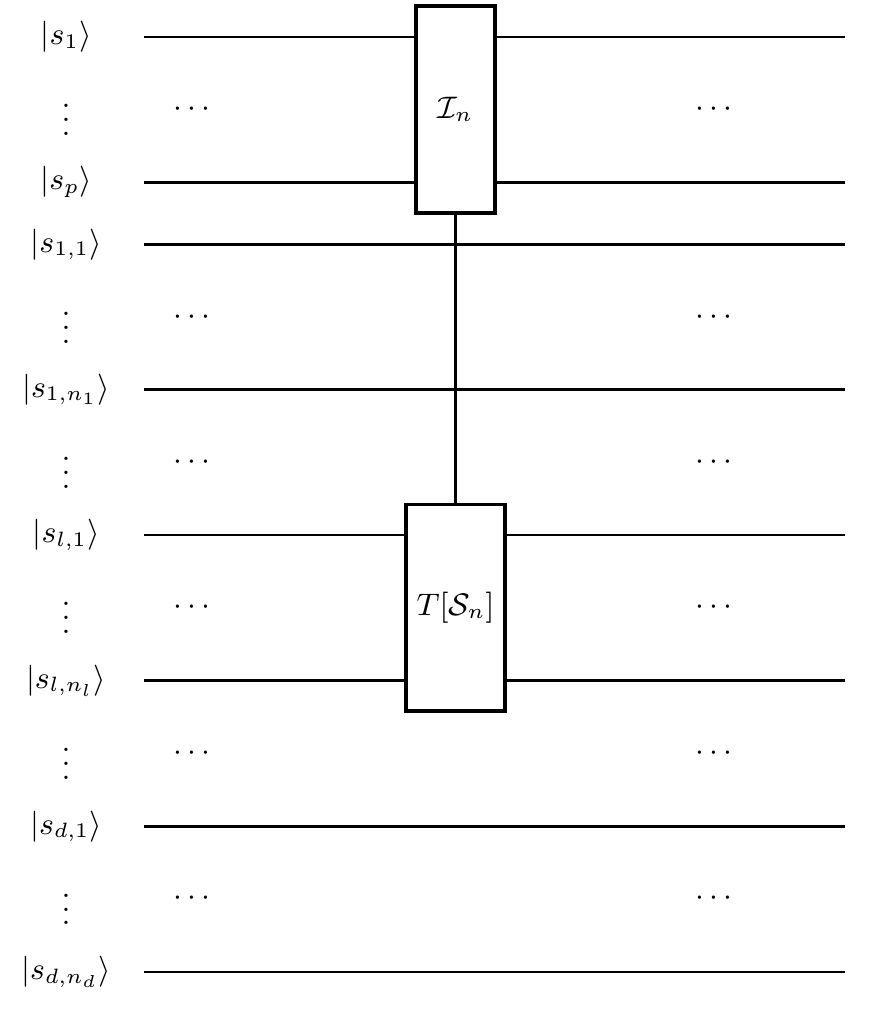} } 
	\hspace*{20mm}\subfloat[]{\includegraphics[scale=1.2]{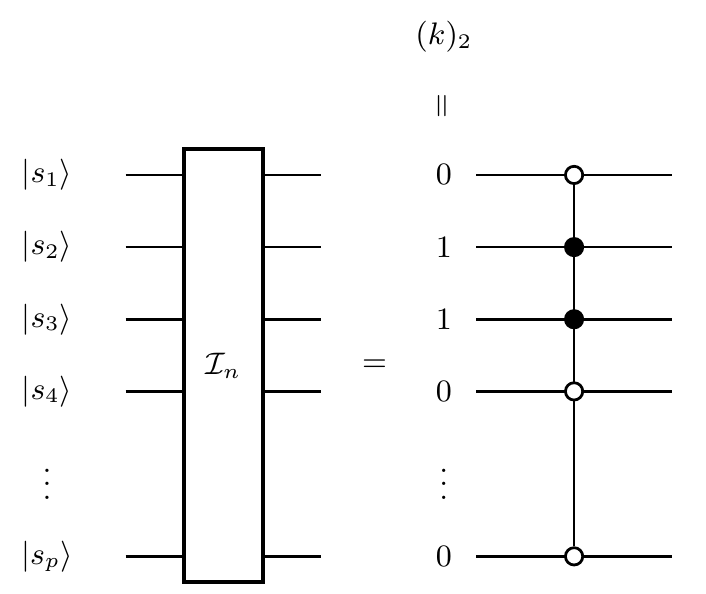}}
	\caption{(a) Circuit diagram for the implementation of the translation operator at time $t_{n}$, for the element in the list $\mathcal{I}_{n} = (k,l)$.  The box $T[\mathcal{S}_{n}]$ is the shift operator on the $n_{l}$-qubits (its decomposition is depicted in Fig. \ref{fig:incr}) and the box $\mathcal{I}_{n}$ is a control on the first $p$-qubits (its explicit circuit is depicted in (b)). (b) Explicit implementation of the controlled operations, where the control is set by the component value $k$ in the pair  $\mathcal{I}_{n}$, expressed as a binary string.  }
	\label{fig:translation_controlled}
\end{figure}

The translation operators perform the streaming operations on the quantum register. Therefore, they are defined, for any $( k,l)$ in $\mathcal{I}^{n_T}$, by
\begin{eqnarray*}
	\left.
	\begin{array}{clc}
		\hat{T}\big[(k,l)\big]|k\rangle\otimes |i_1,\cdots,i_d\rangle & = & |k\rangle\otimes |i_1,\cdots,i_{l-1},i_l\ominus 1 \; \textrm{mod}(N_{l}+1),i_{l+1},\cdots i_d\rangle, \qquad \hbox{ if } \lambda^{(l)}_k > 0,\\
		\hat{T}\big[(k,l)\big]|k\rangle\otimes |i_1,\cdots,i_d\rangle & = & |k\rangle\otimes |i_1,\cdots,i_{l-1},i_l\oplus 1 \; \textrm{mod}(N_{l}+1),i_{l+1},\cdots i_d \rangle, \qquad \hbox{ if } \lambda^{(l)}_k < 0.
	\end{array}
	\right.
\end{eqnarray*}
and when $k \neq k'$
\begin{eqnarray*}
	\left.
	\begin{array}{clc}
		\hat{T}\big[(k,l)\big]|k'\rangle\otimes |i_1,\cdots,i_d\rangle & = & |k'\rangle\otimes |i_1,\cdots, i_d\rangle.
	\end{array}
	\right.
\end{eqnarray*}
This is a unitary operation which can be represented by a quantum circuit that includes a shift operator $\hat{T}[\pm]$ controlled by some qubits. In particular, to translate a specific component $k$, the shift operator has to be controlled by the first quantum register $\mathcal{H}_{p}$. The quantum operation is a fully controlled gate on this register and the explicit controls are determined by the value $k$ expressed as a binary string with $p$ digits. In particular, we have $(k)_{10 \mapsto 2} = (s_{1},\cdots,s_{p})_{2}$. If the value of the digit $s_{i}=0$, the control is on state $|s_{i} \rangle = |0\rangle$ (white dot). Conversely, if the value of the digit $s_{i}=1$, the control is on state $|s_{i} \rangle = |1\rangle$ (black dot). These controlled operations allow for selecting and translating the proper component. Then, the sign of $\mathcal{S}_{n}$ determines which shift operator $\hat{T}[\pm]$ is used. The corresponding circuit is displayed in Fig. \ref{fig:translation_controlled}.

In practice, the computations are performed on finite domains. As $u_0$ has compact support, for finite time, the support of the solution to the considered system also has compact support. As a consequence, the boundary conditions imposed have no influence on the solution assuming the domain is large enough such that the solution is not scattered on the boundary. Here, we use periodic boundary conditions, which can easily be implemented within the quantum algorithms developed above, thus explaining the appearance of the $\textrm{mod}(N_{l}+1)$.

\begin{figure}
	\subfloat[]{\includegraphics[width=0.5\textwidth]{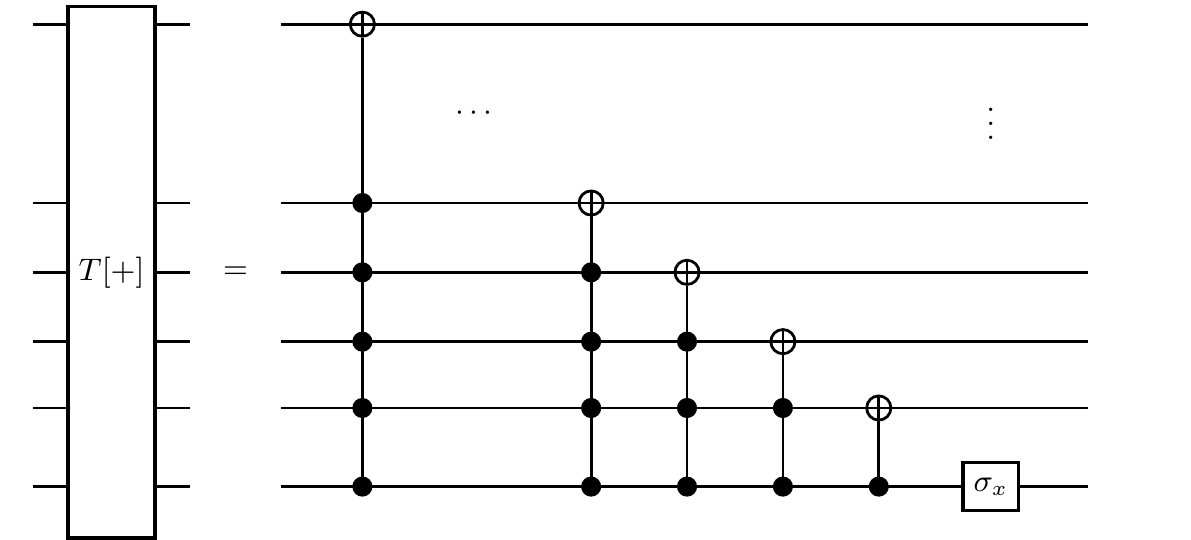}}
	\subfloat[]{\includegraphics[width=0.5\textwidth]{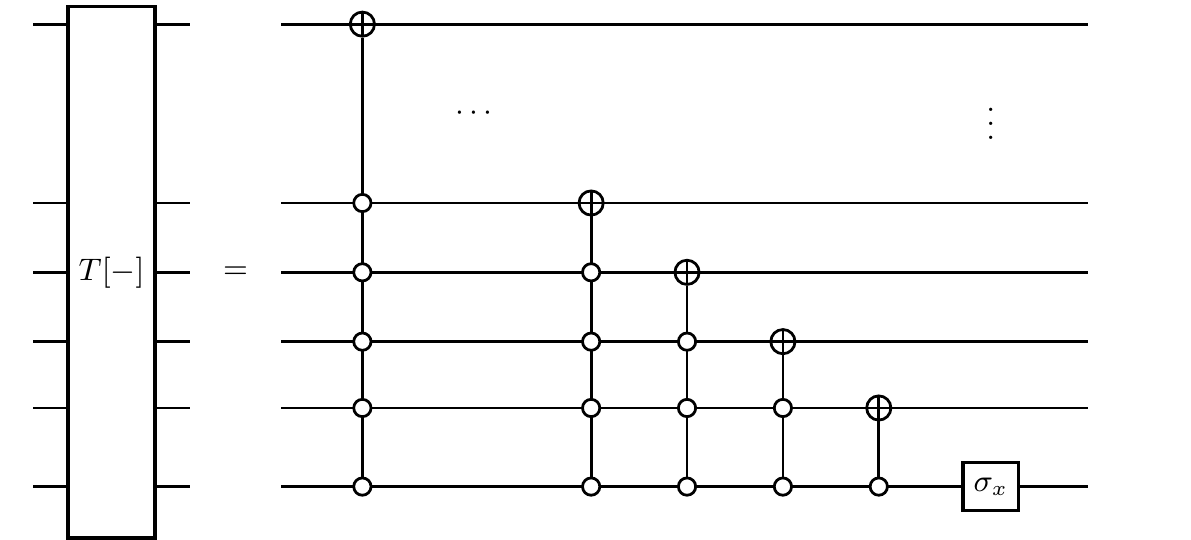}}
	\caption{Circuit diagram for the (a) increment and (b) decrement operator acting on qubits storing the space data of the solution (the set of $n_x$ qubits). This implementation was first considered in \cite{PhysRevA.79.052335}.}
	\label{fig:incr}
\end{figure}

The shift operators $\hat{T}[\pm]$ can be decomposed as a sequence of simpler unitary operations.  One possible decomposition is displayed in Fig. \ref{fig:incr} where controlled operations are used. The left (resp. right) quantum circuit corresponds to a shift from the left (resp. right) to the right (resp. left). This decomposition was first studied in \cite{PhysRevA.79.052335} and used for the solution of the Dirac equation in \cite{fillion}. It has a complexity which scales like $O(n_{l}^{2})$. 

\subsubsection{Rotation operators}\label{subsubsec:rot}

A unitary operation (diagonalization) is applied to transform the system from the canonical basis of eigenvectors of $A^{(l)}$ to the diagonal basis before and after any increment or decrement operator. In the classical algorithm, these unitary operations were denoted by $\{S^{(l)} \}_{l\in \{1,\cdots,d\}}$. We are now looking for the corresponding operators $\{\hat{S}[l] \}_{l\in \{1,\cdots,d\}}$ which implement the same operations on the quantum register. Because the transition operators  are unitary operations that rotates the characteristic fields, the quantum operations are simply unitaries applied in the Hilbert space $\mathcal{H}_{p}$, as $(\hat{S}[l]^T|k\rangle ) \otimes |i_1,\cdots,i_d\rangle$. Because of the tensor product, this operation is automatically carried over all grid points, i.e. at every grid points, the fields are transformed to the diagonal basis. The explicit value of the rotation operators in the computational basis is the same as  the classical $\{S^{(l)} \}_{l\in \{1,\cdots,d\}}$ and thus, they should be determined from the eigenvectors of $\{A^{(l)} \}_{l\in \{1,\cdots,d\}}$. 

The next fundamental issue is the decomposition of unitary operators $\{\hat{S}[l] \}_{l\in \{1,\cdots,d\}}$ in terms of quantum elementary gates. This question can be treated using existing techniques:  $m\times m$ unitary matrices can be  performed$/$simulated exactly using explicit quantum circuits with a quadratic complexity $O(m^2)$ using a Gray code approach \cite{EffDQG}. In this paper and for the sake of simplicity, we will present only examples where the unitary transition matrices are relatively simple, but existing decomposition techniques for more complex matrices can perfectly be coupled to the method presented. A simple example for a rotation operator is given in Appendix \ref{app:ex_rot_op}.

\subsubsection{Full quantum algorithm}

\begin{figure}
	\centering 
	\includegraphics[width=0.5\textwidth]{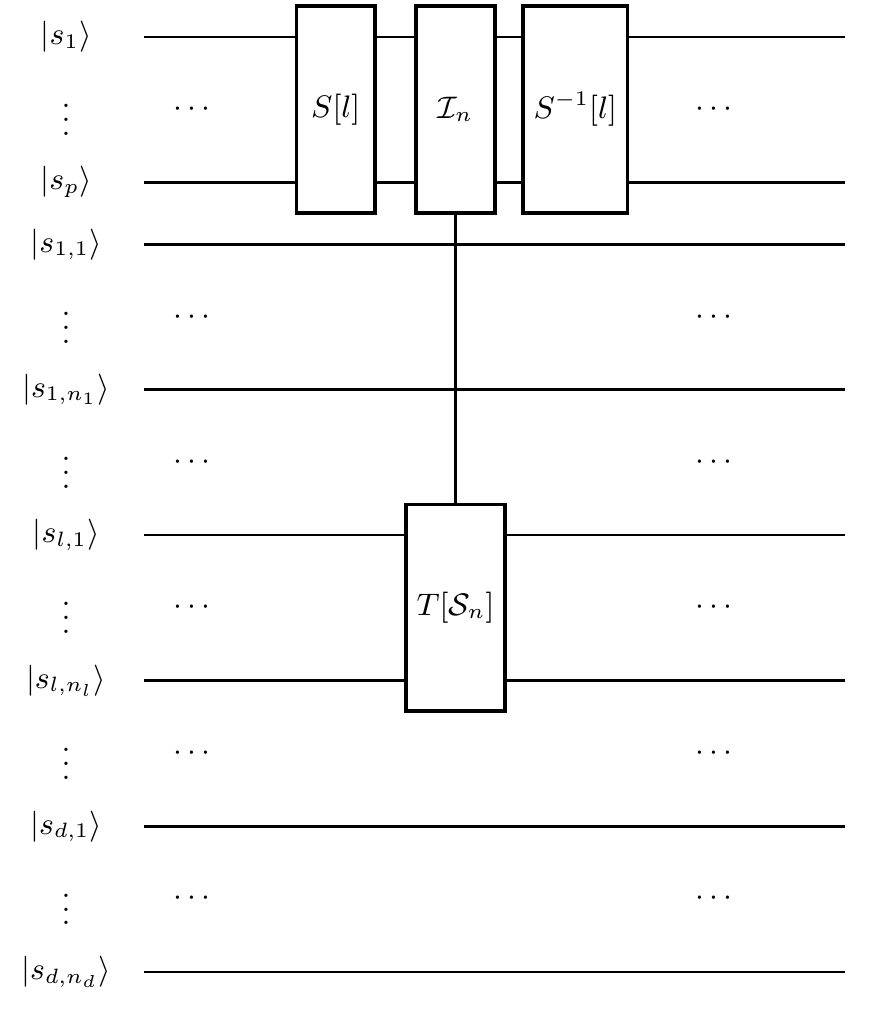}  
	\caption{Circuit diagram for the implementation of the algorithm for the $d$-dimensional linear system at time $t_{n}$ and where $\mathcal{I}_{n} = (k,l)$. Here, $S[l]$ is the unitary operation that implements the diagonalization of the system. The latter has to be decomposed into a set of fundamental gates.}
	\label{fig:scheme_multiD}
\end{figure}

The quantum circuit corresponding to the full quantum algorithm for one time step is displayed in Fig. \ref{fig:scheme_multiD}. The first $p$ qubits (from top) label the components of the solution, and the next qubits label the coordinate space positions in each of the $d$ dimensions. For $  \mathcal{I}^{n} = (k,l)$, i) we  apply a rotation $\hat{S}[l]^T$ (change of basis, to diagonalize $A^{(l)}$) to the system, followed ii) by a translation operator $\hat{T}[\pm k]$ of the $k$th component of the solution (upwinding), and finally iii) we apply $\hat{S}[l]$ (back to canonical basis).

The resulting quantum algorithm is relatively simple because it does not use sophisticated quantum techniques such as amplitude amplification, phase estimation or the quantum solution of linear systems \cite{lloyd}. This is possible because hyperbolic systems with constant symmetric matrices conserve the $L^{2}$-norm. Therefore, the algorithm relies on straightforward mappings from the classical unitary operators to the quantum operators.

\section{Efficiency and resource requirements}
\label{sec:eff}

In this section, the efficiency of the preceding algorithms is analyzed. This is performed by comparing the number of required operations on a quantum computer and on a classical computer, for the same numerical method, i.e. the reservoir scheme discussed in previous sections. In particular, we are interested in the quantum speedup defined by the asymptotic behavior of \cite{PhysRevA.88.022316,Ronnow420}
\begin{eqnarray*}
S \sim  \cfrac{N_{\mathrm{classical}}(\mathcal{N})}{N_{\mathrm{quantum}}(\mathcal{N})} \quad \mbox{as } \mathcal{N}\rightarrow \infty
\end{eqnarray*}
where $\mathcal{N}$ characterizes the system size, $N_{\mathrm{classical}}(\mathcal{N})$ is the computational cost (or complexity) of the classical algorithm and $N_{\mathrm{quantum}}(\mathcal{N})$ is the cost of the quantum algorithm. These costs are evaluated by counting the number of operations required to evolve the solution in time. 

\begin{prop}
	\label{prop:quant_speed}
Let us denote by $N$ the number of grid points in each dimension, $d$ the number of dimensions, $n_{T}$ the number of iterations, and $m$ the number of characteristic fields.
Let consider a hyperbolic system of equations with constant symmetric matrices. Then, the quantum speedup for the numerical resolution of this system using the reservoir method is
\begin{eqnarray*}
S =
 O\left( \cfrac{m^{2}}{\log^{2} m} \cfrac{N^{d}}{\log^{2} N} \right), & \;\; \mathrm{for}\; 
 O\left(\frac{m^{2}}{\log^{2} m} \left[1+\frac{d}{m} + \frac{dm^{2}}{n_{T}}  \right] \right) = O(\log^{2} N),
 %
\end{eqnarray*}
corresponding to an exponential quantum speedup, when $N$ and $n_{T}$ are much larger than $d$ and $m$. 
\end{prop}
\noindent{\bf Proof.} The number of operations can be evaluated as follows. On a classical computer, the shift operations defined in Fig.  \ref{fig:incr} requires $O(N^{d})$ swap operations to translate the solution along a given axis. Accessing the array for the component that is translated, given from the list $\mathcal{I}^{n_{T}}$,  requires $O(1)$ operations. On the other hand, the rotation operation is a matrix-vector multiplication at all grid points, thus requiring $O(m^{2}N^{d})$ operations. Finding the rotation operators is equivalent to solving an eigenvalue problem for each matrix $\{A^{(j)}\}_{j=1,\cdots,d}$, typically requiring $O(d m^{3})$ operations. Finally, constructing the list $\mathcal{I}^{n_{T}}$ requires a classical algorithm with a scaling of $O(n_{T}dm)$.  Then, the total number of operations for $n_{T}$ iterations can be written as
\begin{eqnarray*}
N_{\mathrm{classical}} &=& O\left(n_{T}m^{2}N^{d}  + d m^{3} + n_{T}dm \right).
%
\end{eqnarray*}  
For the quantum algorithm, the shift operator along an axis can be implemented using $O(\tilde{n}^{2}) = O(\log^{2} N )$ quantum logic gates \cite{fillion}, where $\tilde{n}$ is the number of qubits labeling the grid points in the given direction.  However, these gates are controlled by the register $p$: controlled gates need $O(p^{2}) = O(\log^{2} m)$ operations \cite{gates_quantum_computPRA95}. As discussed earlier, the rotation operator needs $O(m^{2})$ gates. If a classical algorithm is used to find the rotation matrices, the number of operations is also $O(d m^{3})$, as in the classical case. This however could be improved by using a quantum algorithm, such as the Abrams-Lloyd technique \cite{PhysRevLett.83.5162}.
Therefore, the total number of quantum logic gates after $n_{T}$ iterations is
\begin{eqnarray*}
N_{\mathrm{quantum}} &=& O\left( n_{T} \log^{2} m \log^{2} N+ n_{T}m^{2} + d m^{3} + n_{T}dm\right).
%
\end{eqnarray*} 
Finally, taking the ratio of $N_{\mathrm{classical}}$ and $N_{\mathrm{quantum}}$, and using the fact that $n_{T}=O(N)$, we can evaluate the quantum speedup which proves the proposition. $\Box$ 

The last proposition is an important result of this article, stating that in some regimes, for a large number of grid points, our algorithm which generates a state representing the solution of a hyperbolic system of equations using the quantum algorithm is much more efficient than its classical counterparts. In terms of computational complexity, the quantum implementation has an exponential speedup over the classical implementation for the time evolution of the solution.  

Here, we have neglected the initialization and reading phases of the quantum register. In particular, it is  assumed that the initialization of $U^{0}$ can be implemented in $\mathrm{polylog} (N)$ operations. However, this is not true in the general case. As a matter of fact, the initialization of the quantum register to a general initial state $U^{0}$ requires the implementation of diagonal unitary gates, which can be implemented with uniformly controlled gates having a computational cost scaling like $O(N^{2})$ \cite{PhysRevA.71.052330}. For a certain class of functions, which can be integrated analytically, this can be improved to $O(\log N)$ by using the algorithm described in \cite{Zalka08011998,kaye2004quantum,grover2002creating}, recovering the exponential quantum speedup. 

In addition, the reading of the solution is not more efficient in the quantum case in general because $U^{n_{T}}$ is stored in $O(N)$ quantum amplitudes. Reconstructing these amplitudes, a process called quantum tomography \cite{d2003quantum}, necessitates that the quantum algorithm is performed at least $O(N)$ times, an exponential number of operations. Therefore, to keep the efficiency of the algorithm, the final solution stored on the quantum register should either be post-processed with other efficient quantum algorithms or the measurement should be performed on some given observable $\langle u_{\ell} | \hat{O} | u_{\ell} \rangle$, where the operator $\hat{O}$ allows for extracting some relevant informations on the solution. In some particular cases, it may be possible to reconstruct the quantum state with some polynomial speedup using the method given in \cite{cramer2010efficient}. 

For a given error $\epsilon>0$, we estimate the computational resources and complexity necessary to implement the algorithm proposed above.  We consider a {\it $d$-dimensional} $m=2^p$-equation systems. We also assume that $u_0$ smooth with compact support, and the problem is set on a cubic domain $\Omega$ of size $L^d$.  For a $N^d$-point lattice, we define $\Delta x:=L/N$. The analysis of convergence for $d=1$ is provided in \cite{res2}, Theorem 2.4. We notice that the projection error of the initial data on the lattice $\|u_0-u^0_{h}\|_{\ell^1}$ is bounded by $\leq C\|\nabla u_0\|_{\infty}\Delta x^d$, for some $C>0$. In addition the directional splitting, when the matrices do not commute, creates an error in $O(\Delta t^2)$ per time iteration. For a total of $n_T$ time iterations and using that $\Delta t$ is proportional to $\Delta x$ (CFL or stability condition), there exists a constant $E=E(u_0,c,m,d,\Omega)>0$, $F>0$, such that
\begin{eqnarray}
\label{eq:error}
\|u(\cdot,t_{n_T})-u^{n_T}_h\|_1 & \leq & \epsilon_{\mathrm{reservoir}} + \epsilon_{\mathrm{splitting}} ,\\
\label{eq:error_exp}
& \leq & \|u_0-u^0_{h}\|_{\ell^1}+E\Delta x + F (n_T \Delta t) \Delta x ,
\end{eqnarray}
where $u$ (resp. $u^{n_T}_h$) denotes the exact (resp. approximate) solution, at $T=t_{n_T}$. 
The main feature of the reservoir method for linear systems with constant coefficients, is that the error remains bounded in time. Thus, the first term ($\epsilon_{\mathrm{reservoir}}$) on the right-and-side of \eqref{eq:error} and \eqref{eq:error_exp} is  independent of $n_{T}=T/\Delta t$. However, the splitting error $\epsilon_{\mathrm{splitting}}$ grows linearly with time.  As a consequence, we have the condition $r:=\epsilon_{\mathrm{reservoir}}/\epsilon_{\mathrm{splitting}} \ll 1$, i.e. the error due to reservoir is negligible compared to the splitting, for a long enough final time $T$, when $T = \Omega(L^{d-1}/N^{d-1})$.  Then, for a given error $\epsilon>0$, the necessary resources are such that
\begin{eqnarray*}
\epsilon = F (n_T \Delta t) \Delta x (1+r) = F T \frac{L}{N}(1+r) .
\end{eqnarray*}
As a consequence, we obtain that the number of grid points, for given error, final time and domain size should be
\begin{eqnarray}
\label{eq:N}
N = F(1+r) \frac{TL}{\epsilon} = O\left( \frac{TL}{\epsilon}\right). 
\end{eqnarray} 
As $\log_2(N) = \tilde{n}$, the total number of qubits necessary for representing the solution after $n_T$ iterations, with an error $\epsilon$ and for $T$ large enough, is given by
\begin{eqnarray}
\label{eq:tilde_n}
\tilde{n} = \log \left( F(1+r) \frac{TL}{\epsilon}\right) = O\left[ \log \left(  \frac{TL}{\epsilon}\right) \right].
\end{eqnarray}
Notice that in the case of diagonal matrices, $N$ is ``only'' a $O(L/\epsilon)$, due to the commutation of the matrices. 
Reporting these results for $N$ and $\tilde{n}$ in the cost of the algorithm, we can evaluate the speedup in terms of the problem parameters. We finally deduce the following proposition: 
\begin{prop}
	\label{prop:quant_speed_eps}
	Let us denote by $T$ the final time, $d$ the number of dimensions, $L$ the size of the domain in each dimension, $m$ the number of characteristic fields, and $\epsilon$ the numerical error.
	We consider a hyperbolic system of equations with constant symmetric matrices. Then, the quantum speedup for the numerical computation of this system using the reservoir method is
	\begin{eqnarray*}
		S &=& 
		O\left( \cfrac{m^{2}}{\log^{2} m} \cfrac{T^{d}L^{d}}{\epsilon^{d}} \cfrac{1}{\log^{2} \left(  \frac{TL}{\epsilon}\right)} \right) \, , 
	\end{eqnarray*}
	which corresponds to an exponential quantum speedup.
\end{prop}
The proof essentially follows the same logic as for Proposition \ref{prop:quant_speed}, but replacing $N$ and $\tilde{n}$ by \eqref{eq:N} and \eqref{eq:tilde_n}, respectively.

The previous results considered the asymptotic computational complexity of the classical and quantum algorithms. However, it is interesting to look at minimal examples to verify if a proof-of-principle calculation could be performed on actual quantum computers. Two such examples are described in Appendix \ref{app:quipper} where explicit gate decompositions are carried out with {\tt Quipper}.

\section{Some possible extensions of the proposed quantum algorithms}\label{sec:extension}
In this section, we propose some ideas to extend the algorithms proposed above. First, we discuss the extension of the above quantum algorithms to linear hyperbolic equations with space-dependent velocity. In the second part of this section,  we discuss method-of-line based quantum algorithms for linear hyperbolic systems.

\subsection{Reservoir method for linear hyperbolic equations with non-constant velocity}\label{sec:quantvar}

As introduction to this problem, we consider the following one-dimensional transport equation
\begin{eqnarray}\label{oneD1d}
\left\{
\begin{array}{lcl}
\partial_tu + \partial_x\big(A(x)u\big) &  = & 0, \qquad (x,t) \in \R\times (0,T),\\
u(\cdot,0) & = & u_0, \qquad x \in \R
\end{array}
\right.
\end{eqnarray}
where $A$ is assumed smooth, with a derivative denoted $a(x):=\partial_{x} A(x)$. We also assume that $u_0$ is assumed smooth with compact support. For the sake of simplicity of the presentation and notation, we will assume that $a(x)>0$.The $L^2-$norm of the solution to \eqref{oneD1d} is not preserved in general, instead: 
\begin{eqnarray*}
\cfrac{d}{dt}\int_{\R}A(x)|u(x,t)|^2dx=0.
\end{eqnarray*}
However, it is still possible to implement a quantum algorithm preserving the $\ell^2(\Z)-$norm of the quantum register. This problem can be reduced to a constant velocity transport equation by using a change of variable $y = f(x) = \int^{x}a^{-1}(x')dx'$, allowing for an analytical solution when $f$ can be obtained in analytical form. If this is not available, one should resort to a numerical approach like the reservoir method. In this case however, the reservoirs and counters are space-dependent. Quantum algorithms are based on unitary transformations. By default, the reservoir method for non-constant velocity on uniform mesh, is a priori not based on unitary operations. We then propose a version of the reservoir method on non-uniform mesh. More specifically, we define a sequence grid points $\big\{x_{j+1/2}\big\}_{j \in \Z}$, and one-dimensional volumes $\{\omega_j\}_{j \in \Z}$, with $\omega_j:=(x_{j-1/2},x_{j+1/2})$ of lengths $\Delta x_j:=x_{j+1/2}-x_{j-1/2}$. We denote, for any $j \in \Z$
\begin{eqnarray*}
u_j^0 := \cfrac{1}{\Delta x_j}\int_{\omega_j}u_0(x)dx
\end{eqnarray*}
and we denote $\{u_j^n\}_{(j,n)\in \Z \times \N}$ the sequences approximating 
\begin{eqnarray*}
\Big\{\cfrac{\Delta t_n}{\Delta x_j}\int_{\omega_j}u(x,t)dx\Big\}_{(j,n) \in \Z \times \N} \, .
\end{eqnarray*}
We also denote for $j\in \Z$, $a_{j-1/2}:=a(x_{j-1/2})$. 
For $a(x)$ assumed fixed (and positive), it is convenient to consider a non-uniform mesh as follows: $\omega_j = (x_{j-1/2},x_{j+1/2})$ with non-constant $\Delta x_j=x_{j+1/2}-x_{j-1/2}$ such that 
\begin{eqnarray*}
\left.
\begin{array}{lcl}
\Delta x_j  & = & \left\{
\begin{array}{cc}
\cfrac{a_{j-1/2}}{1+\lfloor a_{j-1/2}\rfloor}, & \hbox { if } \, a_{j-1/2} >1, \\
a_{j-1/2}, & \hbox { if } \, a_{j-1/2} \leq  1 \, .
\end{array}
\right.
\end{array}
\right.
\end{eqnarray*}
Notice that by construction, for any $j \in \Z$
\begin{eqnarray*}
\ell_j := \cfrac{a_{j-1/2}}{\Delta x_j} \in \N^* \, .
\end{eqnarray*}
Thus, the reservoir method can then be rewritten
\begin{eqnarray*}
\left.
\begin{array}{lcl}

\left(
\begin{array}{c}
u^{n+1}_{j} \\
c_{j-1/2}^{n+1}\\
r_{j-1/2}^{n+1}
\end{array}
\right) 

& = & 

\left\{
\begin{array}{c}

\left(
\begin{array}{c}
0 \\
c_{j-1/2}^n+\ell_j\Delta t_n\\
r_{j-1/2}^n-\ell_j\Delta t_n(u^n_j-u^n_{j-1})
\end{array}
\right),  \hbox{ when } c^n_{j-1/2} + \ell_j\Delta t_n<1, \\
\left(
\begin{array}{c}
u_j^n+r_{j-1/2}^n-\Delta t_n \ell_j(u^n_j-u_{j-1}^n)\\
0\\
0
\end{array}
\right),  \hbox{ when } c^n_{j-1/2} + \ell_j\Delta t_n=1 \, .
\end{array}
\right.

\end{array}
\right.
\end{eqnarray*}
The counters can then be defined by $c^n_{j-1/2}=\ell_j\sum_{k=p_j}^{n-1}\Delta t_k$, for some $0 \leq p_j \leq n$ and 
\begin{eqnarray*}
\Delta t_n = \min_j \left[(1-c_{j-1/2}^n)\cfrac{1}{\ell_j}\right] \, .
\end{eqnarray*}
Notice that if $a$ is small enough, we can simplify even more the algorithm, and we can determine a priori the time steps and the list of space indices to be updated per iteration, $\mathcal{I}^{n_T}$ as in the case of constant velocity. In order to achieve this procedure, we construct the grid nodes $\{x_{j+1/2}\}_{j\in \Z}$ such that for all $j\in \Z$, $a_{j-1/2}/\Delta x_j = \ell$ with $\ell$ constant.  Initially, the counters and reservoirs are as usual, empty. 
The counters are of the form $c_{j-1/2}^n = \ell\sum_{k=p}^{n-1}\Delta t_k$, for some $0 \leq p \leq n$ (for $p=n$, the counters are null). In other words, the counters as well as the time steps are spatially independent:
\begin{eqnarray*}
\Delta t_n = \cfrac{1}{\ell}\Big(1-\ell\sum_{k=p}^{n-1}\Delta t_k\Big) \, .
\end{eqnarray*}
In practice, we then have for any $n \in \N$ :
\begin{eqnarray*}
\Delta t_n = \cfrac{1}{\ell}, \qquad \hbox{ and } \,  u_{j}^{n+1} = u_{j-1}^n \, .
\end{eqnarray*}
At the quantum level for $a(x)>0$, the algorithm is simply given by $T_x|i\rangle = |i\ominus1 \; \textrm{mod}(N)\rangle$. Recall though that this diffusion-less approach only works for very specific velocities, and is a priori not conservative.  

This algorithm should also be efficient on a quantum computer, as it is based on the application of the shift operator, as in the constant velocity cases, and therefore, scales like $O(n_{T}\log^2 N)$. However, finding the grid point positions and the time step may require an exponential amount of resources. A naive classical algorithm for this task requires $O(N)$ operations. Therefore, the method is more efficient on a quantum computer as long as it is possible to find a quantum algorithm that can evaluate these grid positions more efficiently than $O(N)$.

\begin{figure}[!ht]
\begin{center}
\hspace*{1mm}\includegraphics[height=6cm, keepaspectratio]{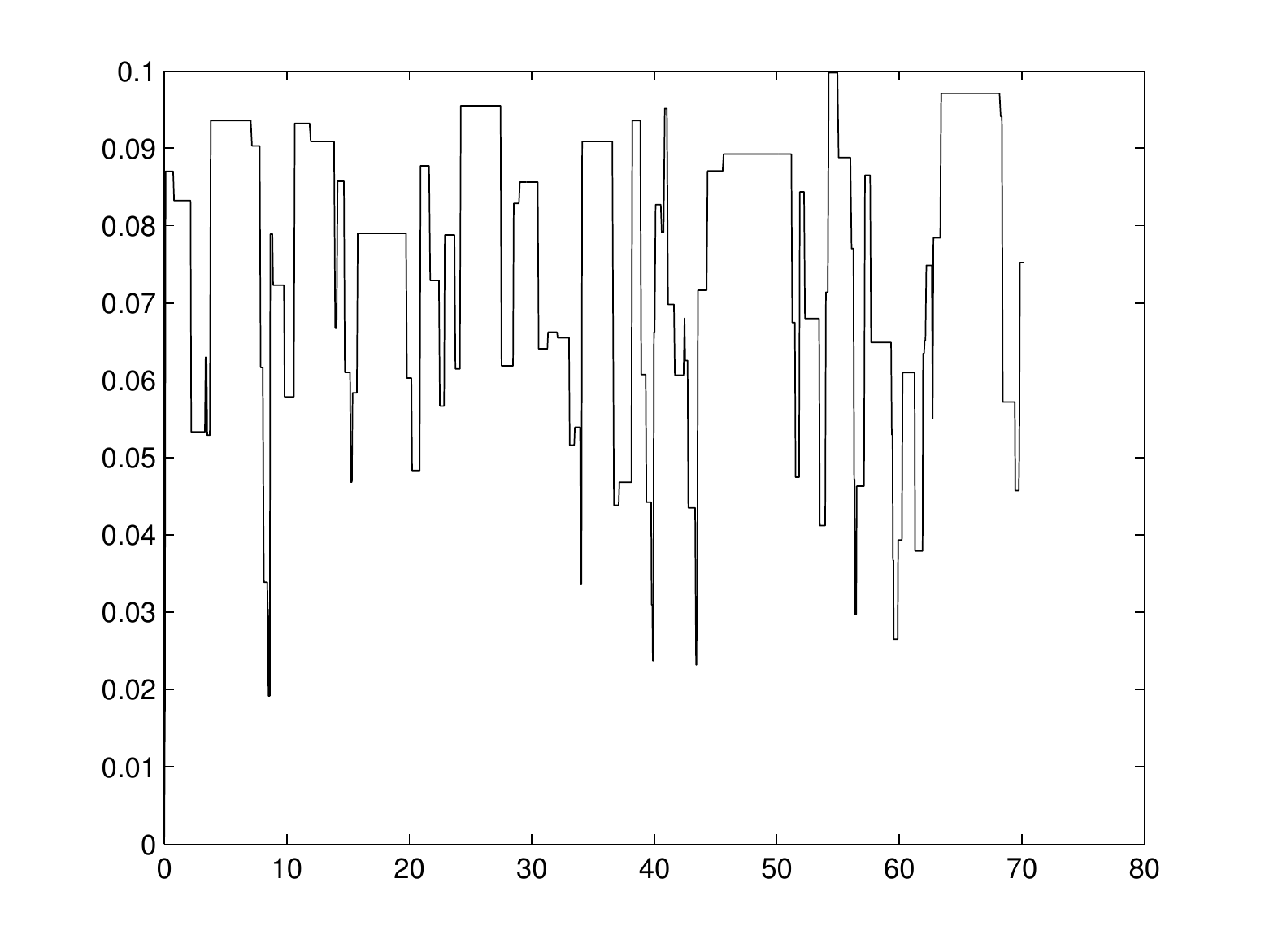}
\hspace*{1mm}\includegraphics[height=6cm, keepaspectratio]{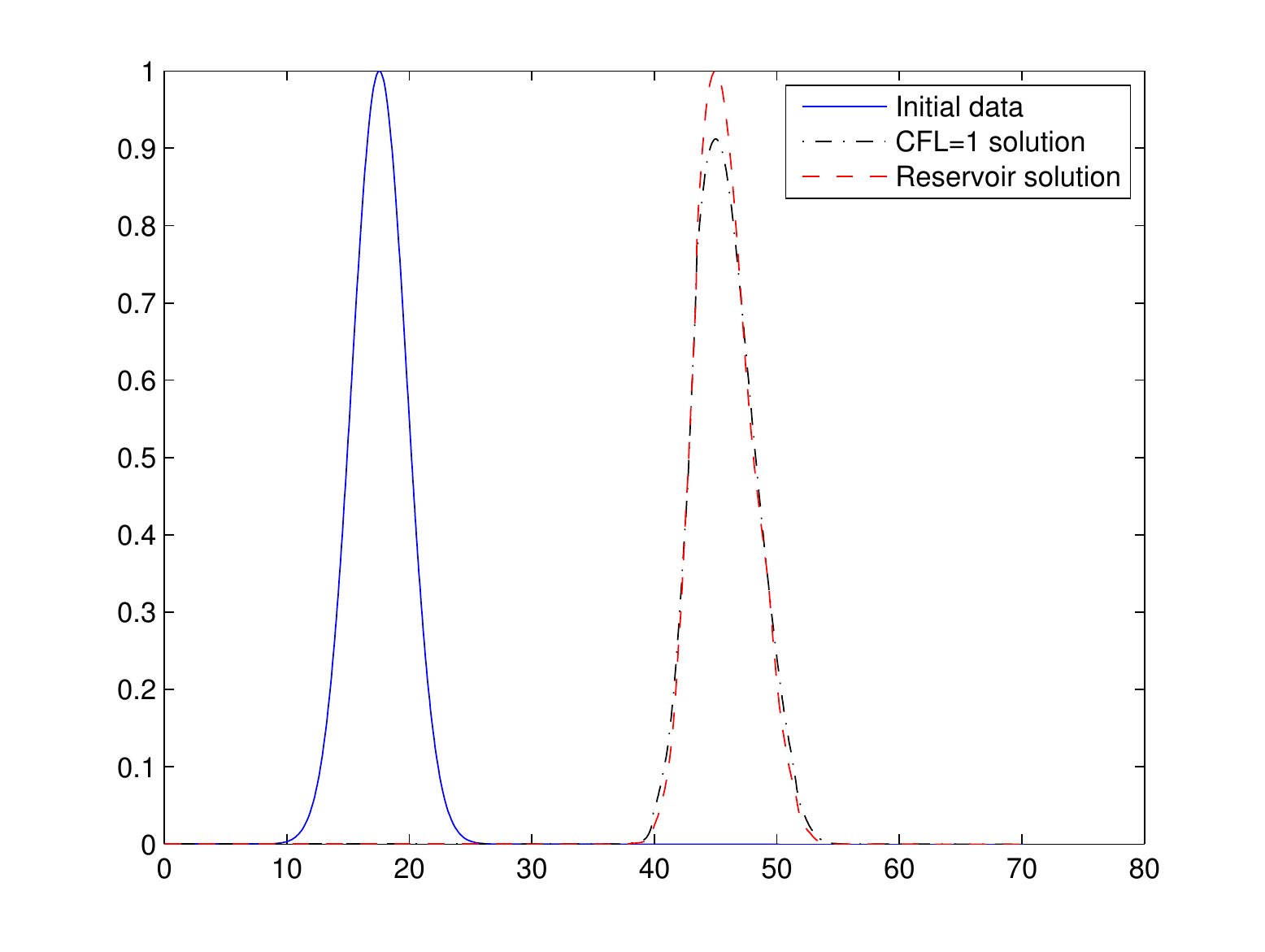}
\caption{(Left) Non-constant velocity $\{(x,a(x)), \, x \in (0,70)\}$. (Right) CFL=1 (on uniform mesh) and reservoir (CFL=1 on non-uniform mesh) solutions. }
\label{fig5}
\end{center}
\end{figure}

\noindent{\bf Example.} We illustrate the above approach with the following simple example.
\begin{eqnarray*}
\partial_t u + a(x)\partial_x u = 0, \, (x,t) \in (0,70)\times (0,T)
\end{eqnarray*} 
where a velocity is randomly constructed, using a uniform probability density function, $\mathcal{U}(0,1)$. It is defined $\{a_{j-1/2}\}_j$ with $a_{j-1/2} = a(x_{j-1/2})$ see Fig. \ref{fig5} (Left), and where 
$x_{j+1}=x_j+\Delta x_j$ with $\Delta x_j = a_{j-1/2}$. In particular $a_{j-1/2}/\Delta x_j=\ell$ with $\ell=1$ in the following, and $T=400$.
We report in Fig. \ref{fig5} (Right), the solution reservoir solution\footnote{this terminology is here a bit abusive, as this is nothing but a "CFL=1"-solution on a non-uniform mesh.} $\{u_j^n\}_{(j,n)\in \Z\times \N}$
\begin{eqnarray*}
u_{j}^{n+1} =  u_{j-1}^n
\end{eqnarray*}
The $\ell^2(\Z)-$ norm of $\{u_j^n\}_j$ is trivially well-preserved, while $\ell^2(\Delta x\Z)$ is not.  The CFL=1 solution $\{v_j^n\}_{(j,n) \in \Z\in\N}$ is constructed on uniform mesh $\Delta x_j=\Delta x$, given by (as the $a(x)>0$):
\begin{eqnarray*}
v^{n+1}_{j} = v^n_{j} - a_{j-1/2}\cfrac{\Delta t}{\Delta x}(v^{n+1}_j-v_{j}^n)
\end{eqnarray*}
at \textrm{CFL}=1, corresponding to a time step satisfying
\begin{eqnarray*}
\Delta t  = \cfrac{\Delta x}{\max_j |a_{j-1/2}|} \, .
\end{eqnarray*}
Notice that using a directional splitting, it is easy to extend this idea to $d$-dimensional transport equations for $u:=u(x_1,\cdots,x_d,t)$, of the form 
\begin{eqnarray*}
\partial_t u + \sum_{i=1}^da_i(x_{i})\partial_{x_{i}}u  =0, \, (x_{1}\cdots,x_{d},t) \in \R^d\times \R_+
\end{eqnarray*}
where $\{a_i(x_{i})\}_{1 \leq i \leq d}$ is a sequence of given functions. In this case, the space steps are simply defined directionally as follows: $\{\Delta x_{i;j}\}_{j \in \Z}=\{a_i(x_{i;j+1/2})\}_{j \in \Z}$.

\subsection{Generalization to other numerical scheme: Method of lines}\label{subsec:MoL}
Other numerical schemes could also be considered. For instance, the method of lines is particularly interesting \cite{1751-8121-47-10-105301}. In this case, the hyperbolic system is discretized in space only, and can be written as
\begin{eqnarray}
\cfrac{dU}{dt} = \mathcal{A} U,
\end{eqnarray}
where $\mathcal{A}$ is a sparse matrix obtained from the space discretization. When the matrix $\mathcal{A}$ is skew-symmetric the solution to the initial value problem is given by an orthogonal transformation as
\begin{eqnarray}\label{syseq2}
U^{n_{T}} = e^{\mathcal{A} T}U^{0}.
\end{eqnarray} 
%
The orthogonal operation $e^{\mathcal{A}T}$ can then be implemented efficiently for sparse matrices on a quantum computer, see for instance \cite{Berry2007,aharonov2003adiabatic}. However, the construction of the matrix $\mathcal{A}$ may be efficient only for a subclass of all matrices. This promising avenue will be also explored in a forthcoming paper.

\section{Conclusion}\label{sec:conclusion}
In this paper, we have proposed an efficient quantum scheme for the solution of first order linear hyperbolic systems by combining the diffusion-less reservoir method, and operator splitting. We demonstrated that in some cases, namely for a large class of multi-dimensional hyperbolic systems with constant symmetric matrices and for one-dimensional scalar hyperbolic equations with non-constant velocity, the reservoir method, along with alternate direction iteration, can be simplified significantly and becomes a set of streaming steps $\mathcal{I}^{n_{T}}$. These steps can be implemented efficiently on a quantum computer. We have also shown that the combination of these techniques yields a speedup over the classical implementation of the same numerical scheme, promising interesting performance. However, as with other similar quantum algorithms, the measurement of the solution, encoded in the probability amplitudes, requires $O(N)$ operations which is not more efficient than a classical algorithm. To be useful, the algorithm has to be combined with some efficient post-processing procedure that either computes some observables or that yields some general properties of the solution. 

The generalization of this quantum algorithm to more general hyperbolic systems is certainly harder to obtain. First, for other types of matrices (non-constant and non-symmetric), the $L^{2}$-norm is not preserved, implying that non-unitary operations has to be implemented on the quantum computer. It  is possible in principle to use non-unitary operations by using projective measurements at every time steps \cite{doi:10.1063/1.4917056} but this is certainly more challenging to implement. Second, the space-dependent matrices has to be diagonalized at every grid points, which requires $O(N^{d}m^{3})$ operations for a classical algorithm, instead of $O(dm^{3})$ for constant matrices. It is unclear how efficient this operation can be on a quantum computer. Finally, when the matrices depends on space, the reservoir and CFL counters has to be considered explicitly at each volume interface and updated according to \eqref{eq:update_reservoir}. Again, it is unclear if an efficient quantum algorithm can be formulated for this purpose.

The techniques and ideas developed in this paper, can be generalized to the derivation of efficient algorithms solving other types of linear differential systems. However, the required key element is the use of an efficient implementation of translation operators which allow for  an efficient overall algorithm. We are currently investigating  the extension of some of the ideas presented in this paper to first order nonlinear hyperbolic equations. In principle, it is possible to derive explicit quantum versions of simple first order finite difference$/$volume schemes for any linear partial differential equation; however in addition to the poor accuracy, deriving efficient quantum versions of these classical algorithms, that is having exponential speedups w.r.t. classical algorithms, is far from trivial and is even the source of many open problems, which could certainly be investigated by numerical analysts.

\bibliographystyle{plain}
\bibliography{biblio}

\begin{thebibliography}{10}

\bibitem{PhysRevLett.83.5162}
Daniel~S. Abrams and Seth Lloyd.
\newblock Quantum algorithm providing exponential speed increase for finding
  eigenvalues and eigenvectors.
\newblock {\em Phys. Rev. Lett.}, 83:5162--5165, Dec 1999.

\bibitem{aharonov2003adiabatic}
Dorit Aharonov and Amnon Ta-Shma.
\newblock Adiabatic quantum state generation and statistical zero knowledge.
\newblock In {\em Proceedings of the thirty-fifth annual ACM symposium on
  Theory of computing}, pages 20--29. ACM, 2003.

\bibitem{res0}
F.~Alouges, F.~De~Vuyst, G.~Le~Coq, and E.~Lorin.
\newblock A process of reduction of the numerical diffusion of usual order one
  flux difference schemes for nonlinear hyperbolic systems [un proc\'ed\'e de
  r\'eduction de la diffusion num\'erique des sch\'emas \`a diff\'erence de
  flux d'ordre un pour les syst\`emes hyperboliques non lin\'eaires].
\newblock {\em Comptes Rendus Mathematique}, 335(7):627--632, 2002.

\bibitem{res4}
F.~Alouges, F.~De~Vuyst, G.~Le~Coq, and E.~Lorin.
\newblock The reservoir scheme for systems of conservation laws.
\newblock In {\em Finite volumes for complex applications, {III}
  ({P}orquerolles, 2002)}, pages 247--254. Hermes Sci. Publ., Paris, 2002.

\bibitem{res1}
F.~Alouges, F.~De~Vuyst, G.~Le~Coq, and E.~Lorin.
\newblock The reservoir technique: a way to make godunov-type schemes zero or
  very low diffuse. application to {C}olella-{G}laz solver.
\newblock {\em European Journal of Mechanics, B/Fluids}, 27(6):643--664, 2008.

\bibitem{res3}
F.~Alouges, G.~Le~Coq, and E.~Lorin.
\newblock Two-dimensional extension of the reservoir technique for some linear
  advection systems.
\newblock {\em J. of Sc. Comput.}, 31(3):419--458, 2007.

\bibitem{1751-8121-47-46-465302}
Pablo Arrighi, Vincent Nesme, and Marcelo Forets.
\newblock The {D}irac equation as a quantum walk: higher dimensions,
  observational convergence.
\newblock {\em Journal of Physics A: Mathematical and Theoretical},
  47(46):465302, 2014.

\bibitem{Aspuru-Guzik1704}
Al{\'a}n Aspuru-Guzik, Anthony~D. Dutoi, Peter~J. Love, and Martin Head-Gordon.
\newblock Simulated quantum computation of molecular energies.
\newblock {\em Science}, 309(5741):1704--1707, 2005.

\bibitem{gates_quantum_computPRA95}
A.~Barenco, C.H. Bennett, R.~Cleve, D.P. Divincenzo, N.~Margolus, P.~Shor,
  T.~Sleator, J.A. Smolin, and H.~Weinfurter.
\newblock Elementary gates for quantum computation.
\newblock {\em Physical Review A}, 52(5), 1995.

\bibitem{barends2015digital}
R~Barends, L~Lamata, J~Kelly, L~Garc{\'\i}a-{\'A}lvarez, AG~Fowler, A~Megrant,
  E~Jeffrey, TC~White, D~Sank, JY~Mutus, et~al.
\newblock Digital quantum simulation of fermionic models with a superconducting
  circuit.
\newblock {\em Nature communications}, 6, 2015.

\bibitem{barends2016digitized}
Rami Barends, Alireza Shabani, Lucas Lamata, Julian Kelly, Antonio Mezzacapo,
  Urtzi Las~Heras, Ryan Babbush, AG~Fowler, Brooks Campbell, Yu~Chen, et~al.
\newblock Digitized adiabatic quantum computing with a superconducting circuit.
\newblock {\em Nature}, 534(7606):222--226, 2016.

\bibitem{Strini2008}
Giuliano Benenti and Giuliano Strini.
\newblock Quantum simulation of the single-particle schroedinger equation.
\newblock {\em American Journal of Physics}, 76(7):657--662, 2008.

\bibitem{PhysRevA.71.052330}
Ville Bergholm, Juha~J. Vartiainen, Mikko M\"ott\"onen, and Martti~M. Salomaa.
\newblock Quantum circuits with uniformly controlled one-qubit gates.
\newblock {\em Phys. Rev. A}, 71:052330, May 2005.

\bibitem{1751-8121-47-10-105301}
Dominic~W Berry.
\newblock High-order quantum algorithm for solving linear differential
  equations.
\newblock {\em Journal of Physics A: Mathematical and Theoretical},
  47(10):105301, 2014.

\bibitem{Berry2007}
Dominic~W. Berry, Graeme Ahokas, Richard Cleve, and Barry~C. Sanders.
\newblock Efficient quantum algorithms for simulating sparse hamiltonians.
\newblock {\em Communications in Mathematical Physics}, 270(2):359--371, 2007.

\bibitem{doi:10.1063/1.4917056}
Andreas Blass and Yuri Gurevich.
\newblock Ancilla-approximable quantum state transformations.
\newblock {\em Journal of Mathematical Physics}, 56(4):042201, 2015.

\bibitem{boghosian1998simulating}
Bruce~M Boghosian and Washington Taylor.
\newblock Simulating quantum mechanics on a quantum computer.
\newblock {\em Physica D: Nonlinear Phenomena}, 120(1):30--42, 1998.

\bibitem{e12112268}
Katherine~L. Brown, William~J. Munro, and Vivien~M. Kendon.
\newblock Using quantum computers for quantum simulation.
\newblock {\em Entropy}, 12(11):2268, 2010.

\bibitem{1367-2630-15-1-013021}
Yudong Cao, Anargyros Papageorgiou, Iasonas Petras, Joseph Traub, and Sabre
  Kais.
\newblock Quantum algorithm and circuit design solving the poisson equation.
\newblock {\em New Journal of Physics}, 15(1):013021, 2013.

\bibitem{cramer2010efficient}
Marcus Cramer, Martin~B Plenio, Steven~T Flammia, Rolando Somma, David Gross,
  Stephen~D Bartlett, Olivier Landon-Cardinal, David Poulin, and Yi-Kai Liu.
\newblock Efficient quantum state tomography.
\newblock {\em Nature Communications}, 1:149, 2010.

\bibitem{d2003quantum}
G~Mauro D'Ariano, Matteo~GA Paris, and Massimiliano~F Sacchi.
\newblock Quantum tomography.
\newblock {\em Advances in Imaging and Electron Physics}, 128:206--309, 2003.

\bibitem{Deutsch97}
D.~Deutsch.
\newblock Quantum theory, the church-turing principle and the universal quantum
  computer.
\newblock {\em Proceedings of the Royal Society of London A: Mathematical,
  Physical and Engineering Sciences}, 400(1818):97--117, 1985.

\bibitem{PhysRevA.79.052335}
B.~L. Douglas and J.~B. Wang.
\newblock Efficient quantum circuit implementation of quantum walks.
\newblock {\em Phys. Rev. A}, 79:052335, May 2009.

\bibitem{feynman1982simulating}
Richard~P Feynman.
\newblock Simulating physics with computers.
\newblock {\em International journal of theoretical physics}, 21(6):467--488,
  1982.

\bibitem{cpc2012}
F.~Fillion-Gourdeau, E.~Lorin, and A.~D. Bandrauk.
\newblock Numerical solution of the time-dependent {D}irac equation in
  coordinate space without fermion-doubling.
\newblock {\em Comput. Phys. Comm.}, 183(7):1403 -- 1415, 2012.

\bibitem{schwinger}
F.~Fillion-Gourdeau, E.~Lorin, and A.D. Bandrauk.
\newblock Resonantly enhanced pair production in a simple diatomic model.
\newblock {\em Phys. Rev. Lett.}, 110(1), 2013.

\bibitem{fillion}
Fran\c{c}ois Fillion-Gourdeau, Steve MacLean, and Raymond Laflamme.
\newblock Algorithm for the solution of the dirac equation on digital quantum
  computers.
\newblock {\em Phys. Rev. A}, 95:042343, Apr 2017.

\bibitem{RevModPhys.86.153}
I.~M. Georgescu, S.~Ashhab, and Franco Nori.
\newblock Quantum simulation.
\newblock {\em Rev. Mod. Phys.}, 86:153--185, Mar 2014.

\bibitem{god1}
E.~Godlewski and P.-A. Raviart.
\newblock {\em Hyperbolic systems of conservation laws}, volume 3/4 of {\em
  Math\'ematiques \& Applications (Paris) [Mathematics and Applications]}.
\newblock Ellipses, Paris, 1991.

\bibitem{god2}
E.~Godlewski and P.-A. Raviart.
\newblock {\em Numerical approximation of hyperbolic systems of conservation
  laws}, volume 118 of {\em Applied Mathematical Sciences}.
\newblock Springer-Verlag, New York, 1996.

\bibitem{quipper}
A.S. Green, P.L. Lumsdaine, N.J. Ross, P.~Selinger, and B.~Valiron.
\newblock An introduction to quantum programming in quipper.
\newblock {\em Lecture Notes in Computer Science (including subseries Lecture
  Notes in Artificial Intelligence and Lecture Notes in Bioinformatics)}, 7948
  LNCS:110--124, 2013.

\bibitem{quipper2}
A.S. Green, P.L. Lumsdaine, N.J. Ross, P.~Selinger, and B.~Valiron.
\newblock Quipper: A scalable quantum programming language.
\newblock pages 333--342, 2013.

\bibitem{grover2002creating}
Lov Grover and Terry Rudolph.
\newblock Creating superpositions that correspond to efficiently integrable
  probability distributions.
\newblock {\em arXiv preprint quant-ph/0208112}, 2002.

\bibitem{lloyd}
A.~W. Harrow, A.~Hassidim, and S.~Lloyd.
\newblock Quantum algorithm for linear systems of equations.
\newblock {\em Phys. Rev. Lett.}, 103(15):150502, 4, 2009.

\bibitem{jordan2012quantum}
Stephen~P Jordan, Keith~SM Lee, and John Preskill.
\newblock Quantum algorithms for quantum field theories.
\newblock {\em Science}, 336(6085):1130--1133, 2012.

\bibitem{kassal2008polynomial}
Ivan Kassal, Stephen~P Jordan, Peter~J Love, Masoud Mohseni, and Al{\'a}n
  Aspuru-Guzik.
\newblock Polynomial-time quantum algorithm for the simulation of chemical
  dynamics.
\newblock {\em Proceedings of the National Academy of Sciences},
  105(48):18681--18686, 2008.

\bibitem{PMID:21166541}
Ivan Kassal, James~D Whitfield, Alejandro Perdomo-Ortiz, Man-Hong Yung, and
  Alán Aspuru-Guzik.
\newblock Simulating chemistry using quantum computers.
\newblock {\em Annual review of physical chemistry}, 62:185—207, 2011.

\bibitem{kaye2004quantum}
Phillip Kaye and Michele Mosca.
\newblock Quantum networks for generating arbitrary quantum states.
\newblock {\em arXiv preprint quant-ph/0407102}, 2004.

\bibitem{kelly2015state}
Julian Kelly, R~Barends, AG~Fowler, A~Megrant, E~Jeffrey, TC~White, D~Sank,
  JY~Mutus, B~Campbell, Yu~Chen, et~al.
\newblock State preservation by repetitive error detection in a superconducting
  quantum circuit.
\newblock {\em Nature}, 519(7541):66--69, 2015.

\bibitem{res2}
S.~Labb\'e and E.~Lorin.
\newblock On the reservoir technique convergence for nonlinear hyperbolic
  conservation laws. {I}.
\newblock {\em J. Math. Anal. Appl.}, 356(2):477--497, 2009.

\bibitem{Lanyon57}
B.~P. Lanyon, C.~Hempel, D.~Nigg, M.~M{\"u}ller, R.~Gerritsma,
  F.~Z{\"a}hringer, P.~Schindler, J.~T. Barreiro, M.~Rambach, G.~Kirchmair,
  M.~Hennrich, P.~Zoller, R.~Blatt, and C.~F. Roos.
\newblock Universal digital quantum simulation with trapped ions.
\newblock {\em Science}, 334(6052):57--61, 2011.

\bibitem{leveque2002finite}
Randall~J LeVeque.
\newblock {\em Finite volume methods for hyperbolic problems}, volume~31.
\newblock Cambridge university press, 2002.

\bibitem{SLoyd}
S.~{Lloyd}.
\newblock {Universal Quantum Simulators}.
\newblock {\em Science}, 273:1073--1078, August 1996.

\bibitem{Meyer395}
David~A. Meyer.
\newblock Quantum computing classical physics.
\newblock {\em Philosophical Transactions of the Royal Society of London A:
  Mathematical, Physical and Engineering Sciences}, 360(1792):395--405, 2002.

\bibitem{mezzacapo2015quantum}
A~Mezzacapo, M~Sanz, L~Lamata, IL~Egusquiza, S~Succi, and E~Solano.
\newblock Quantum simulator for transport phenomena in fluid flows.
\newblock {\em Scientific reports}, 5, 2015.

\bibitem{PhysRevLett.96.170501}
C.~Negrevergne, T.~S. Mahesh, C.~A. Ryan, M.~Ditty, F.~Cyr-Racine, W.~Power,
  N.~Boulant, T.~Havel, D.~G. Cory, and R.~Laflamme.
\newblock Benchmarking quantum control methods on a 12-qubit system.
\newblock {\em Phys. Rev. Lett.}, 96:170501, May 2006.

\bibitem{nielsen2010quantum}
Michael~A Nielsen and Isaac~L Chuang.
\newblock {\em Quantum computation and quantum information}.
\newblock Cambridge university press, 2010.

\bibitem{PhysRevA.88.022316}
Anargyros Papageorgiou and Joseph~F. Traub.
\newblock Measures of quantum computing speedup.
\newblock {\em Phys. Rev. A}, 88:022316, Aug 2013.

\bibitem{Ronnow420}
Troels~F. R{\o}nnow, Zhihui Wang, Joshua Job, Sergio Boixo, Sergei~V. Isakov,
  David Wecker, John~M. Martinis, Daniel~A. Lidar, and Matthias Troyer.
\newblock Defining and detecting quantum speedup.
\newblock {\em Science}, 345(6195):420--424, 2014.

\bibitem{PhysRevX.5.021027}
Y.~Salath\'e, M.~Mondal, M.~Oppliger, J.~Heinsoo, P.~Kurpiers,
  A.~Poto\ifmmode~\check{c}\else \v{c}\fi{}nik, A.~Mezzacapo, U.~Las~Heras,
  L.~Lamata, E.~Solano, S.~Filipp, and A.~Wallraff.
\newblock Digital quantum simulation of spin models with circuit quantum
  electrodynamics.
\newblock {\em Phys. Rev. X}, 5:021027, Jun 2015.

\bibitem{serre}
D.~Serre.
\newblock {\em Syst\`emes de lois de conservation. {I}}.
\newblock Fondations. [Foundations]. Diderot Editeur, Paris, 1996.
\newblock Hyperbolicit\'e, entropies, ondes de choc. [Hyperbolicity, entropies,
  shock waves].

\bibitem{shor1998}
P.~Shor.
\newblock Polynomial-time algorithms for prime factorization and discrete
  logarithms on a quantum computer.
\newblock {\em SIAM Journal on Computing}, 26(5):1484--1509, 1997.

\bibitem{Sinha2010}
Siddhartha Sinha and Peter Russer.
\newblock Quantum computing algorithm for electromagnetic field simulation.
\newblock {\em Quantum Information Processing}, 9(3):385--404, 2010.

\bibitem{smoller}
J.~Smoller.
\newblock {\em Shock waves and reaction-diffusion equations}, volume 258 of
  {\em Grundlehren der Mathematischen Wissenschaften [Fundamental Principles of
  Mathematical Science]}.
\newblock Springer-Verlag, New York-Berlin, 1983.

\bibitem{PhysRevA.65.042323}
R.~Somma, G.~Ortiz, J.~E. Gubernatis, E.~Knill, and R.~Laflamme.
\newblock Simulating physical phenomena by quantum networks.
\newblock {\em Phys. Rev. A}, 65:042323, Apr 2002.

\bibitem{0034-4885-61-2-002}
Andrew Steane.
\newblock Quantum computing.
\newblock {\em Reports on Progress in Physics}, 61(2):117, 1998.

\bibitem{strikwerda}
J.~C. Strikwerda.
\newblock {\em Finite Difference Schemes and Partial Differential Equations}.
\newblock Society for Industrial and Applied Mathematics (SIAM), Philadelphia,
  PA, second edition, 2004.

\bibitem{EffDQG}
J.J. Vartiainen, M.~Mötiönen, and M.M. Salomaa.
\newblock Efficient decomposition of quantum gates.
\newblock {\em Physical Review Letters}, 92(17):177902--1, 2004.

\bibitem{PhysRevLett.117.210502}
Xi-Lin Wang, Luo-Kan Chen, W.~Li, H.-L. Huang, C.~Liu, C.~Chen, Y.-H. Luo,
  Z.-E. Su, D.~Wu, Z.-D. Li, H.~Lu, Y.~Hu, X.~Jiang, C.-Z. Peng, L.~Li, N.-L.
  Liu, Yu-Ao Chen, Chao-Yang Lu, and Jian-Wei Pan.
\newblock Experimental ten-photon entanglement.
\newblock {\em Phys. Rev. Lett.}, 117:210502, Nov 2016.

\bibitem{1751-8121-43-6-065203}
Nathan Wiebe, Dominic Berry, Peter Høyer, and Barry~C Sanders.
\newblock Higher order decompositions of ordered operator exponentials.
\newblock {\em Journal of Physics A: Mathematical and Theoretical},
  43(6):065203, 2010.

\bibitem{wiesner1996simulations}
Stephen Wiesner.
\newblock Simulations of many-body quantum systems by a quantum computer.
\newblock {\em arXiv preprint quant-ph/9603028}.

\bibitem{PhysRevA.82.060302}
Man-Hong Yung, Daniel Nagaj, James~D. Whitfield, and Al\'an Aspuru-Guzik.
\newblock Simulation of classical thermal states on a quantum computer: A
  transfer-matrix approach.
\newblock {\em Phys. Rev. A}, 82:060302, Dec 2010.

\bibitem{yung2014}
Man-Hong Yung, James~D. Whitfield, Sergio Boixo, David~G. Tempel, and Alán
  Aspuru-Guzik.
\newblock {\em Introduction to Quantum Algorithms for Physics and Chemistry},
  pages 67--106.
\newblock John Wiley \& Sons, Inc., 2014.

\bibitem{OPPROP:PR877}
Christof Zalka.
\newblock Efficient simulation of quantum systems by quantum computers.
\newblock {\em Fortschritte der Physik}, 46(6-8):877--879, 1998.

\bibitem{Zalka08011998}
Christof Zalka.
\newblock Simulating quantum systems on a quantum computer.
\newblock {\em Proceedings of the Royal Society of London A: Mathematical,
  Physical and Engineering Sciences}, 454(1969):313--322, 1998.

\end{thebibliography}

\appendix

\section{Numerical example for the reservoir method with diagonal system}
\label{app:ex_2d_diag}

\begin{figure}[!ht]
	\begin{center}
		\hspace*{1mm}\includegraphics[height=4cm, keepaspectratio]{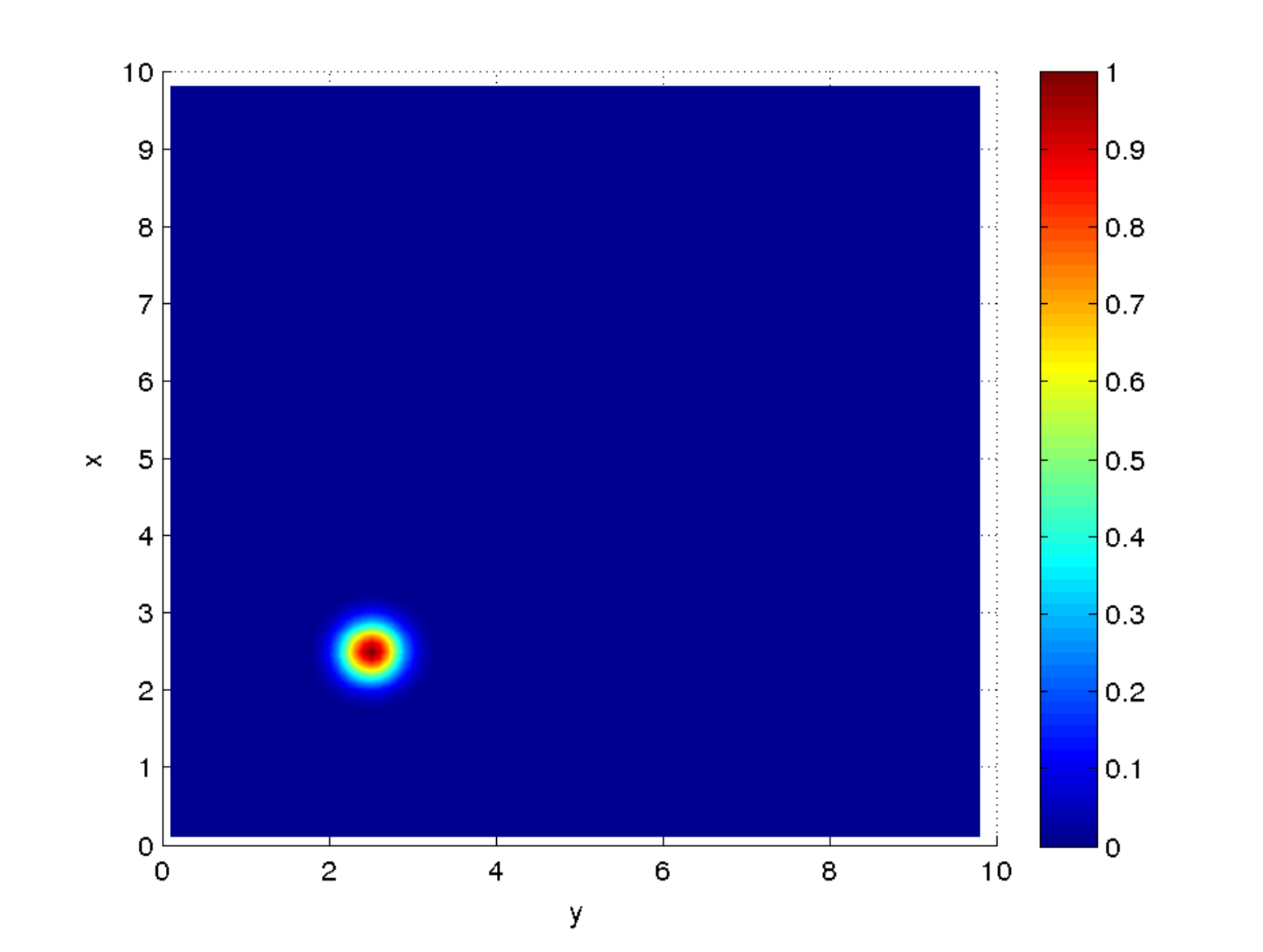}
		\hspace*{1mm}\includegraphics[height=4cm, keepaspectratio]{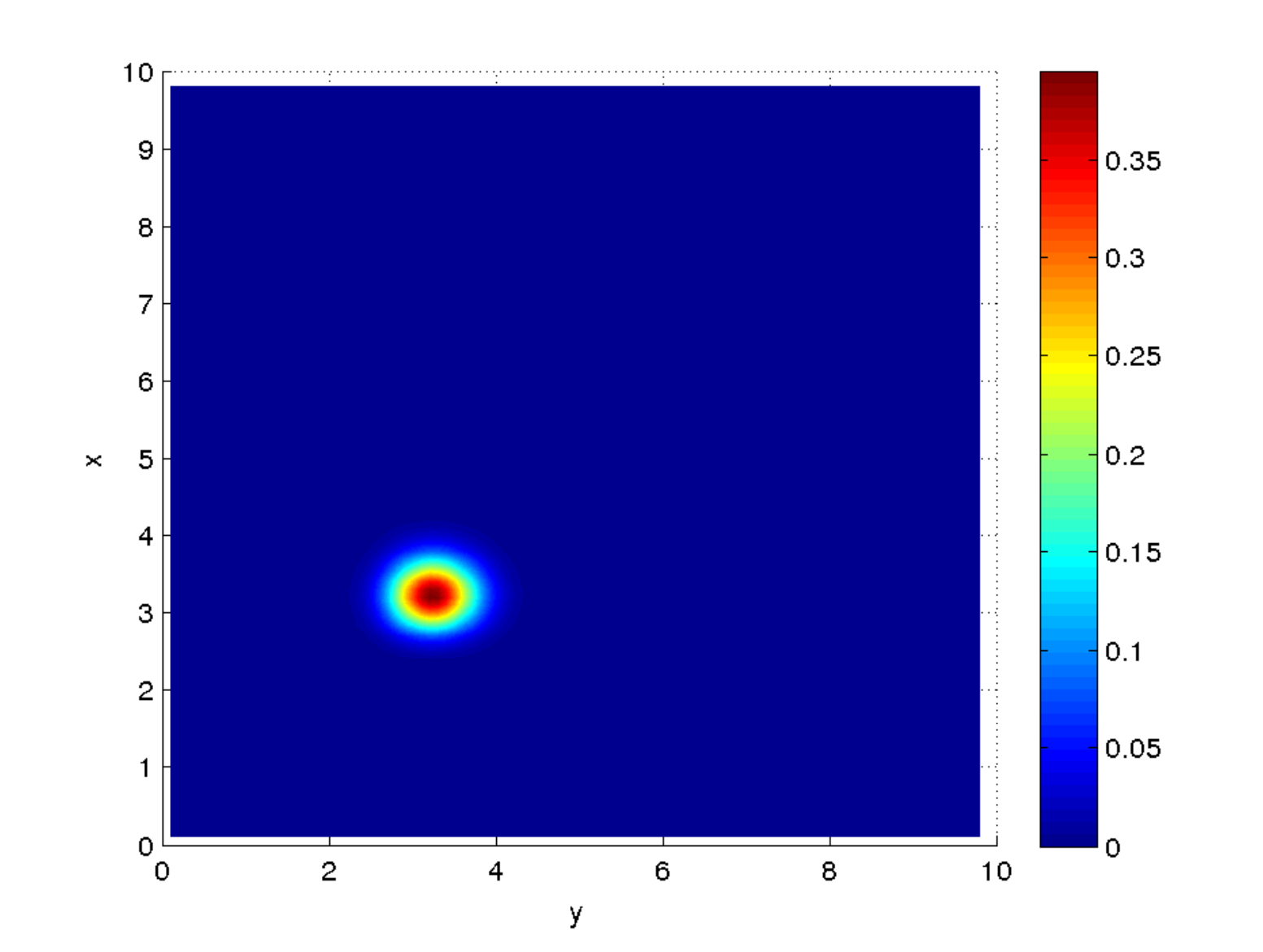}
		\hspace*{1mm}\includegraphics[height=4cm, keepaspectratio]{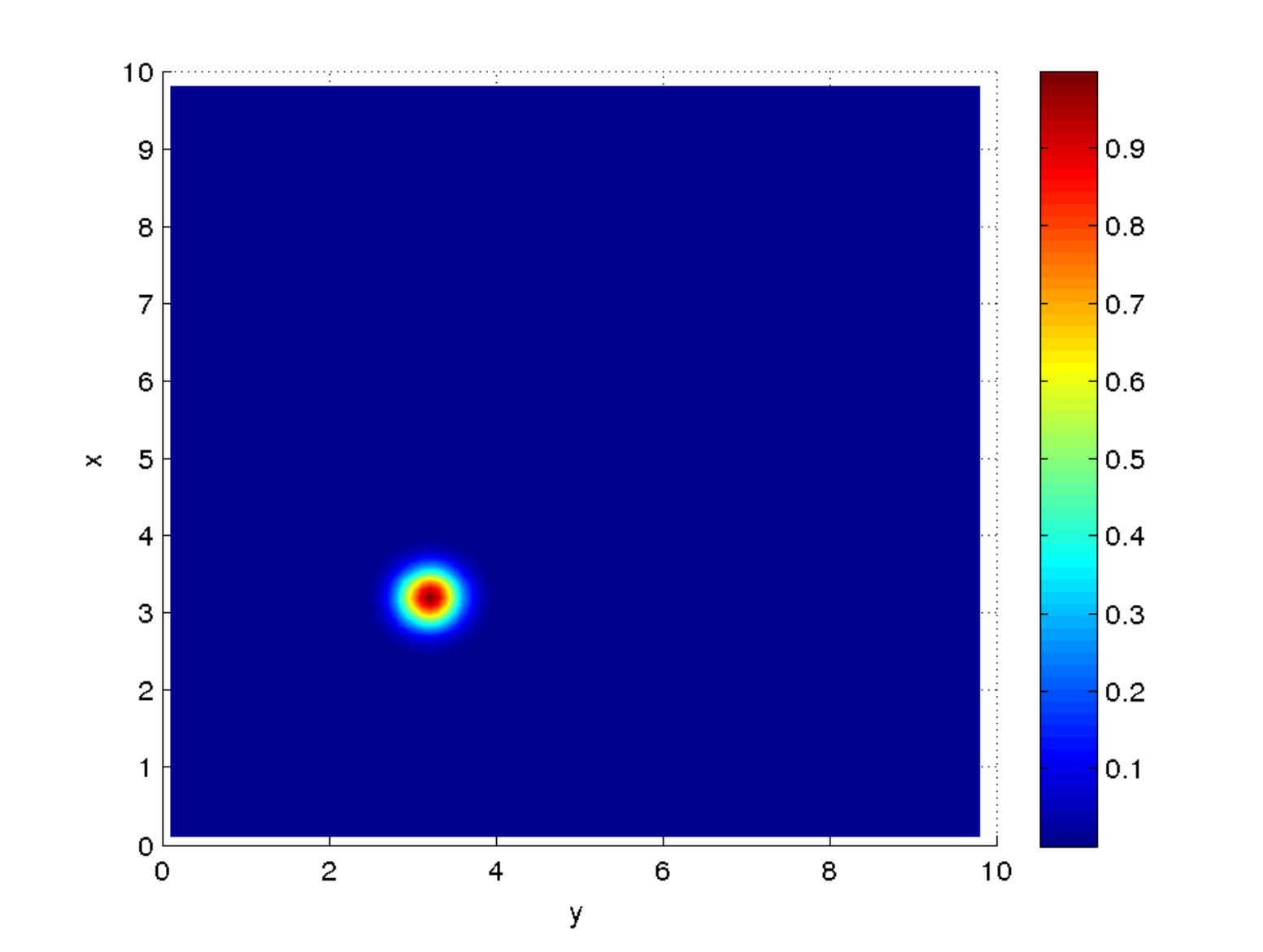}
		\hspace*{1mm}\includegraphics[height=4cm, keepaspectratio]{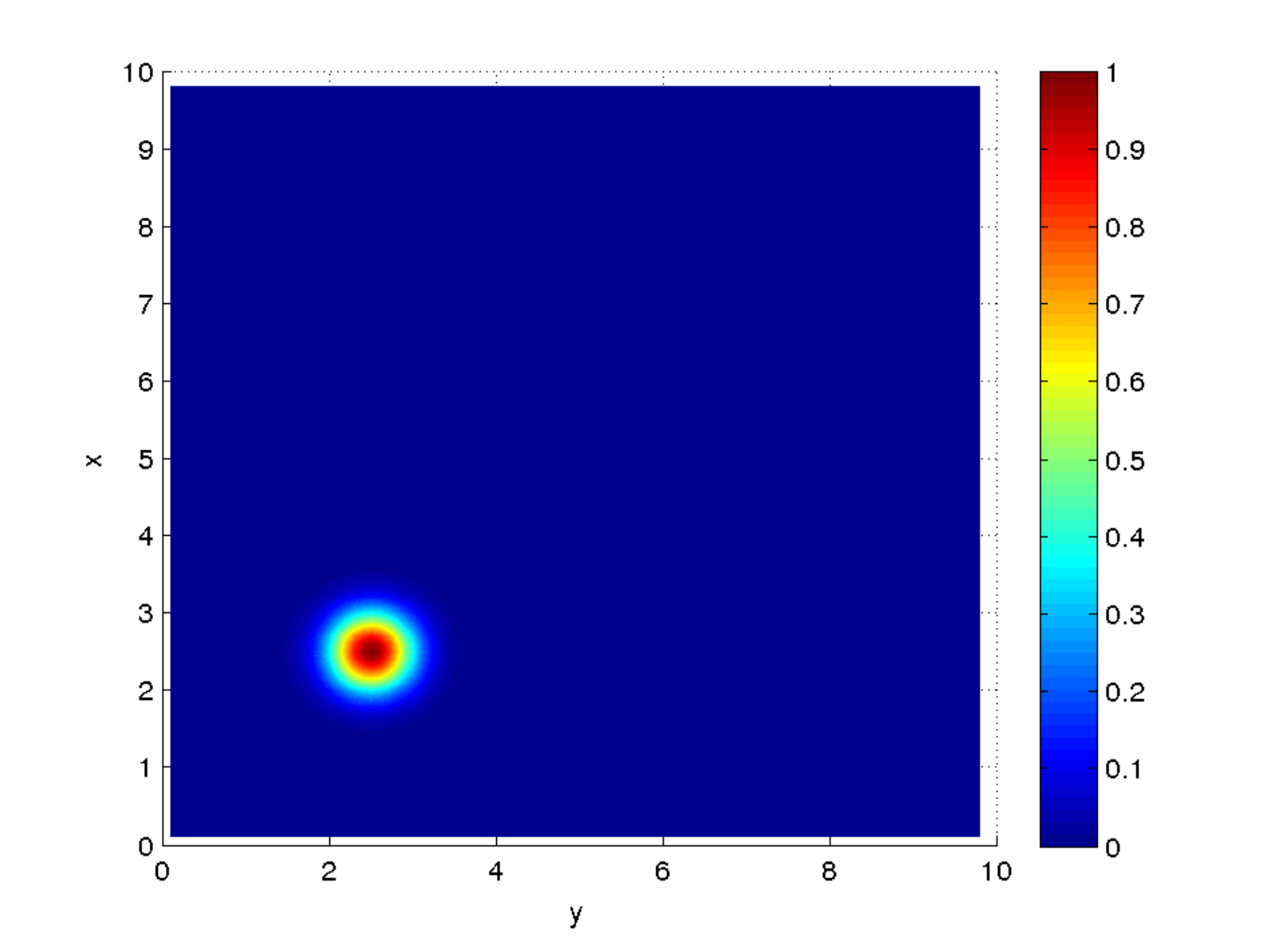}
		\hspace*{1mm}\includegraphics[height=4cm, keepaspectratio]{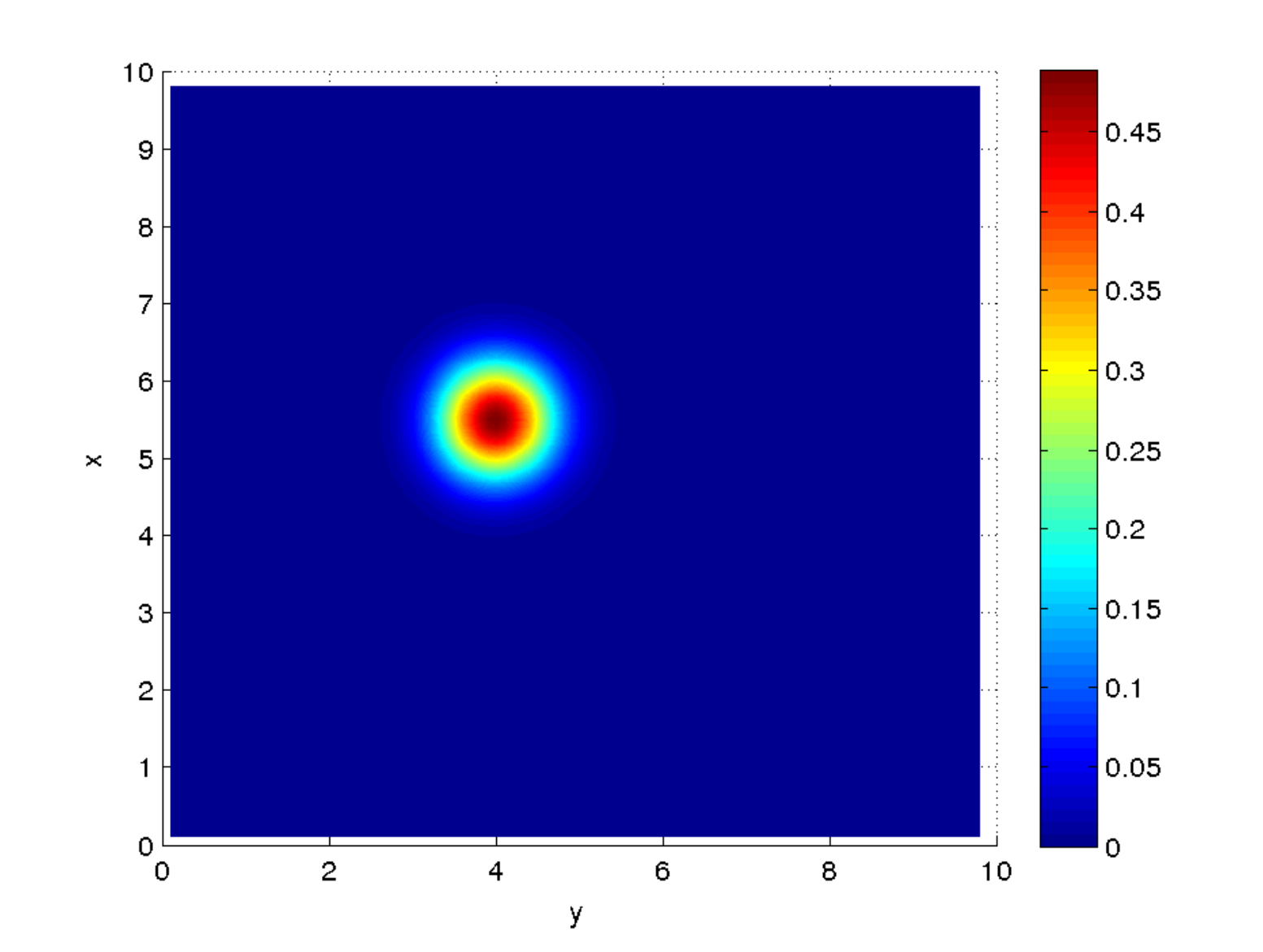}
		\hspace*{1mm}\includegraphics[height=4cm, keepaspectratio]{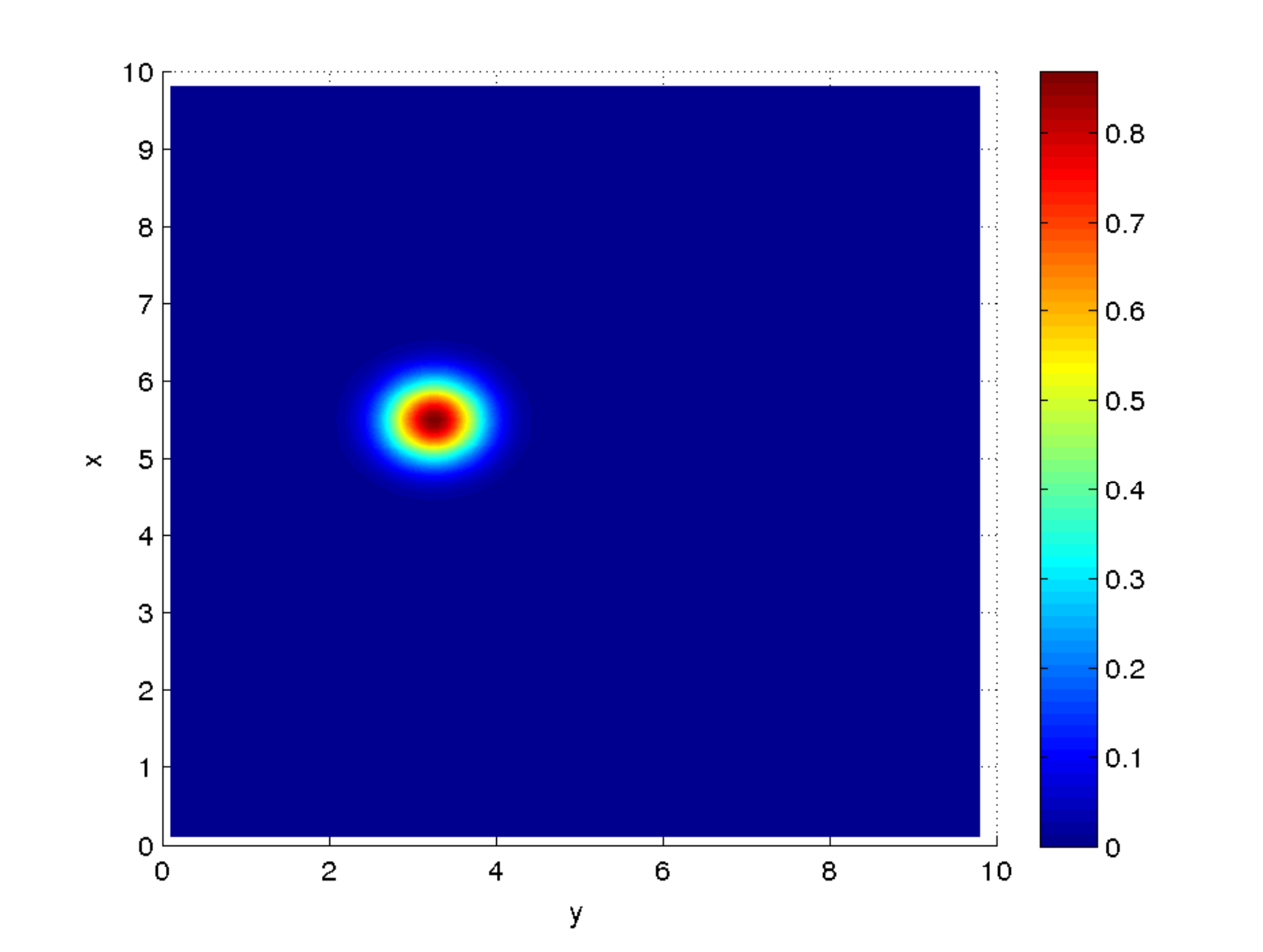}
		\hspace*{1mm}\includegraphics[height=4cm, keepaspectratio]{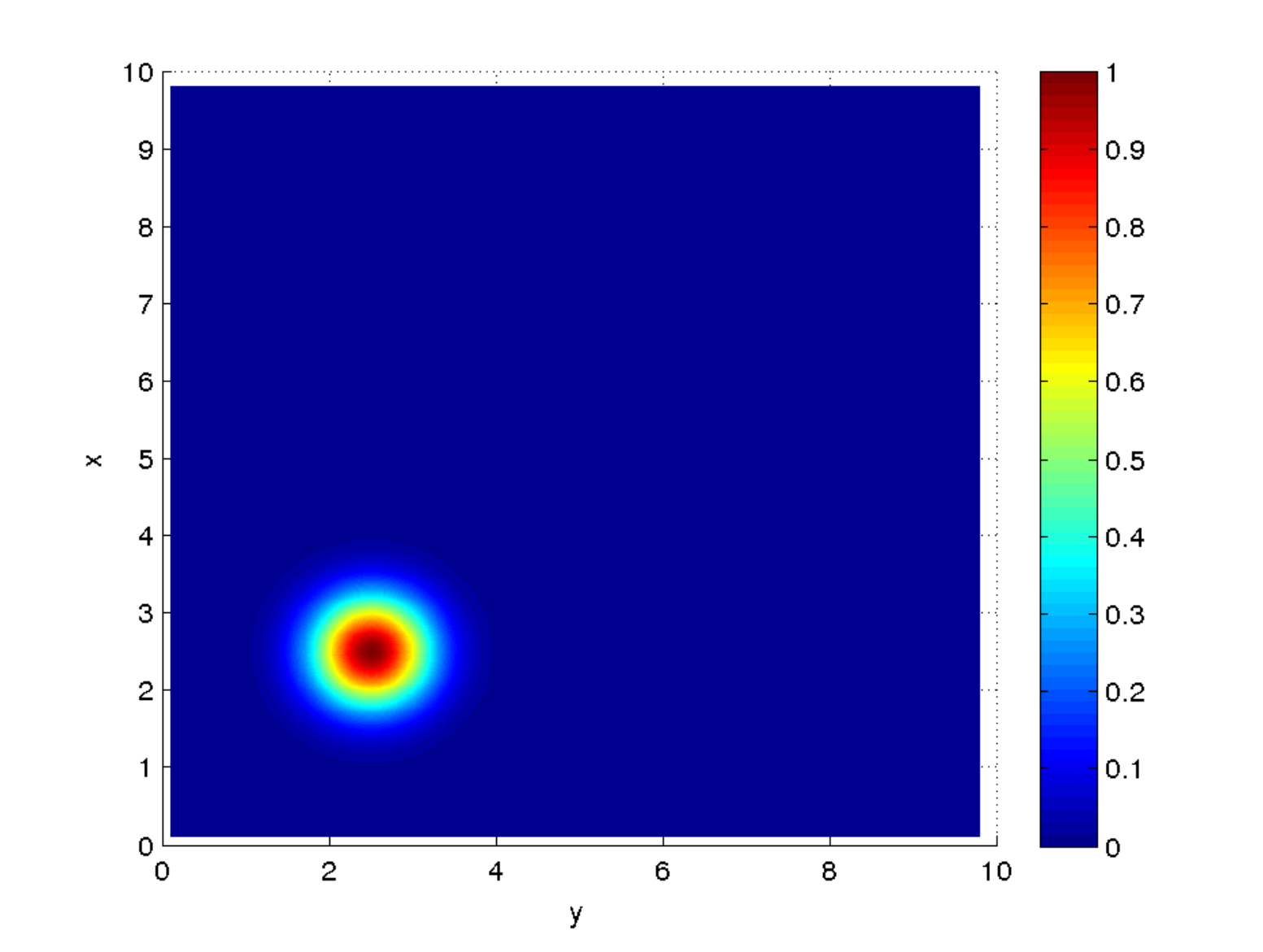}
		\hspace*{1mm}\includegraphics[height=4cm, keepaspectratio]{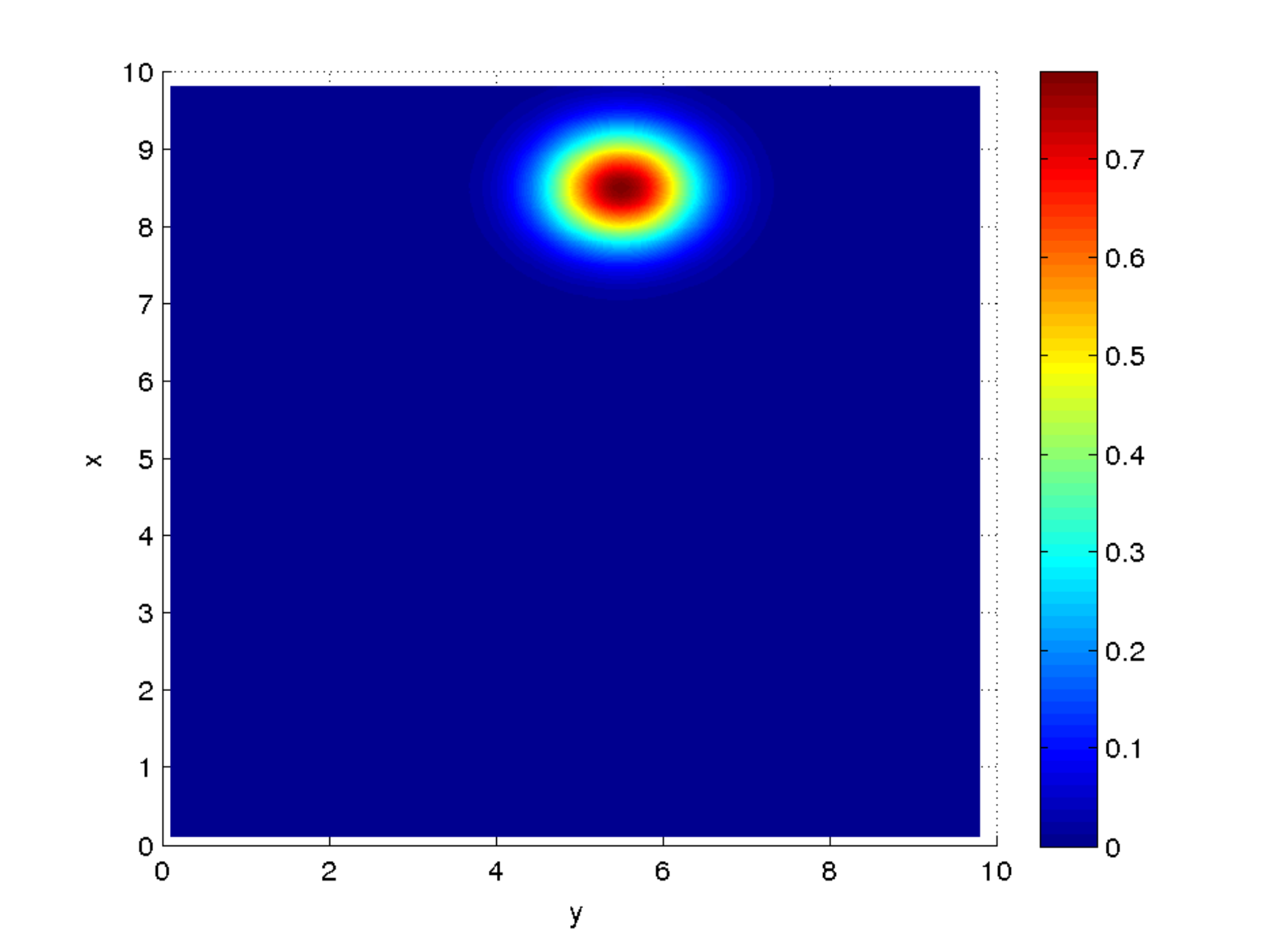}
		\hspace*{1mm}\includegraphics[height=4cm, keepaspectratio]{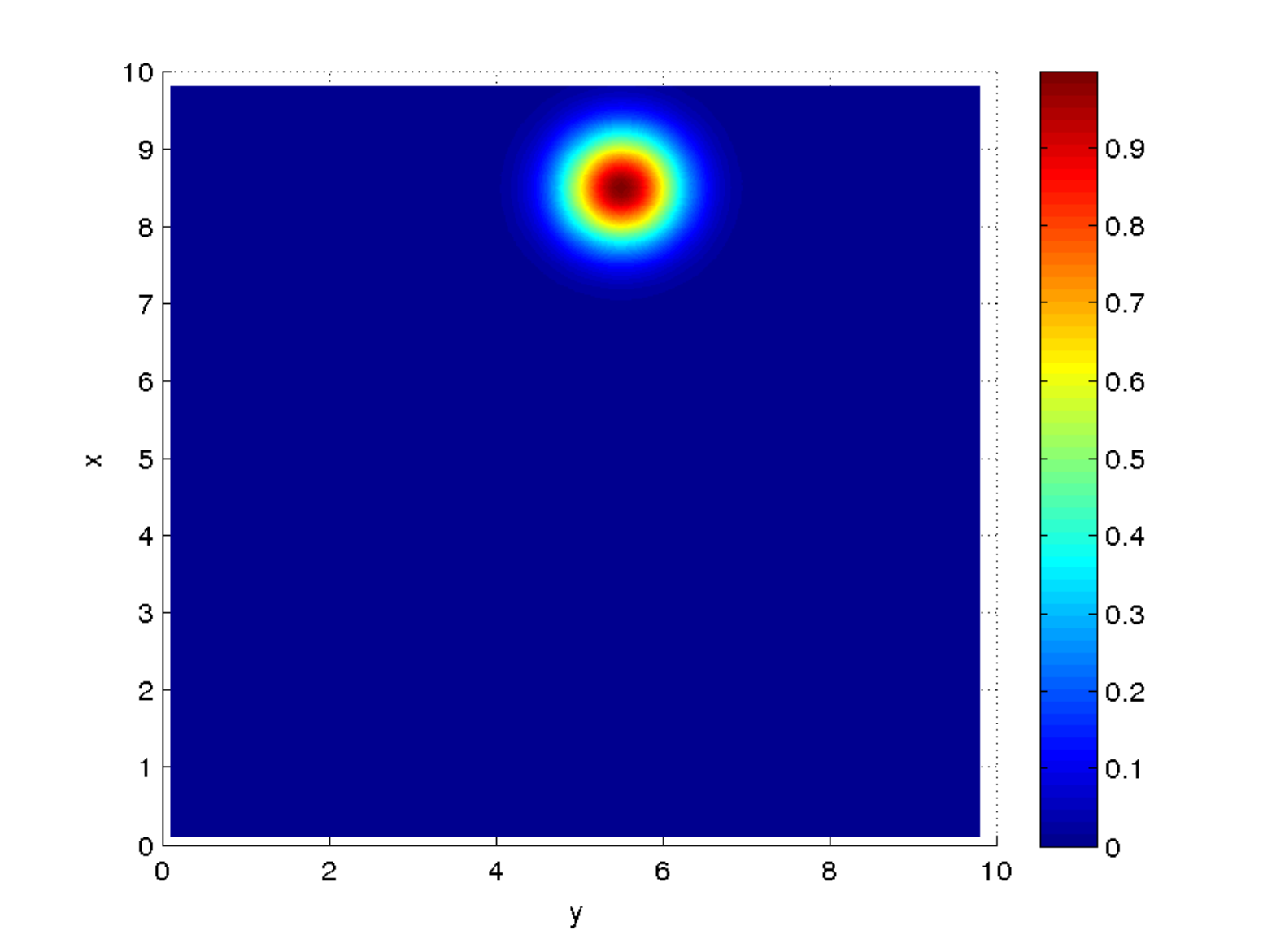}
		\caption{Solution components (rows $1$ to $3$): initial data (1st column), CFL=1 solution (2nd column), and reservoir method solution (3rd column), at final time $T=0.75$.}
		\label{fig1}
	\end{center}
\end{figure}

We now illustrate this approach with a simple classical diagonal two-dimensional test, with $\lambda_1^{(1)}=1,\lambda_2^{(1)}=4,\lambda_3^{(1)}=8$ and $\lambda_1^{(2)}=1,\lambda_2^{(2)}=2,\lambda_3^{(2)}=4$. The computational domain is $[0,10]^2$, and $T=0.75$, $m=3$, $d=2$, $\Delta x_{1}=\Delta x_{2}=0.1$. The time step is given by $\Delta t_{n}=0.012488$ for all $n \leq n_T=60$.  The initial data is $U_{0;1}(x_{1},x_{2})=\exp\Big(-2\big((x_{1}-5/2)^2 + (x_{2}-5/2)^2\big)\Big)$, $U_{0;2}(x_{1},x_{2})=\exp\Big(-4\big((x_{1}-5/2)^2 + (x_{2}-5/2)^2\big)\Big)$, $U_{0;3}(x_{1},x_{2})=\exp\Big(-8\big((x_{1}-5/2)^2 + (x_{2}-5/2)^2\big)\Big)$. The reservoir solution components are represented in Fig. \ref{fig1} (2nd column), showing no numerical diffusion whatsoever, unlike the ``CFL=1''-solutions also represented in Fig. \ref{fig1} (3rd column).

\section{Example for a rotation operator}
\label{app:ex_rot_op}

A simple explicit example for a rotation operator can easily be constructed, inspired from \cite{fillion}. We consider for $d=3$, $m=2^2$ the following system
\begin{eqnarray}\label{syst3d}
\left\{
\begin{array}{lcl}
\partial_tu + A^{(\textrm{x})}\partial_{x}u  + A^{(\textrm{y})}\partial_{y}u  +A^{(\textrm{z})}\partial_{z}u& = & 0, \qquad (x,y,z,t) \in \R^3\times (0,T),\\
u(\cdot,0) & = & u_0, \qquad (x,y,z) \in \R^3
\end{array}
\right.
\end{eqnarray}
with $A^{(\textrm{x})}=S^{(\textrm{x})}\Lambda^{(\textrm{x})}S^{(\textrm{x}T)}$ (resp. $A^{(\textrm{y})}=S^{(\textrm{y})}\Lambda^{(\textrm{y})}S^{(\textrm{y})T}$, $A^{(\textrm{z})}=S^{(\textrm{z})}\Lambda^{(\textrm{z})}S^{(\textrm{z})T}$), where $S^{(\gamma)}$ with $\gamma=x,y,z$, are defined by 
\begin{eqnarray}\label{dirac}
\left.
\begin{array}{clc}
S^{(\gamma)} & = & \cfrac{1}{\sqrt{2}}
\left(
\begin{array}{cc}
\textrm{I}_2 & \sigma_{\gamma}\\
\sigma_{\gamma} & -\textrm{I}_2
\end{array}
\right)
\end{array}
\right.
\end{eqnarray}
and where $\big\{\sigma_{\gamma}\big\}_{\gamma=x,y,z}$ are the Pauli matrices.  From a quantum algorithm point of view, these rotations operators can be decomposed \cite{fillion} as 
\begin{eqnarray*}
S[\gamma] = \textrm{C}(\sigma_{\gamma})\textrm{H}\otimes \textrm{I}_2 \textrm{C}(\sigma_{\gamma})
\end{eqnarray*}
where $C(\sigma_{\gamma})$ (resp. $\textrm{H}$) is $\sigma_{\gamma}$-controlled (resp. Hadamard) gate, see Fig. 1 in \cite{fillion}.
In this case the construction of the quantum circuit can directly be deduced from \cite{fillion}, thanks to a simple decomposition in elementary quantum gates of $S^{(\gamma)}$. More elaborated cases corresponding to more complex $S^{(\gamma)}$ can be considered as discussed above, but then necessitate more complex quantum circuits.

\section{Determining minimal resource requirements of the quantum algorithm with {\tt Quipper}}\label{app:quipper}
{\tt Quipper} is a {\tt Haskell}-based embedded functional language whose purpose is to emulate {\it the implementation of quantum algorithms on realistic quantum computers} by providing explicit gate decompositions of quantum algorithms \cite{quipper,quipper2}. In particular, functions for transforming complex quantum circuits into  elementary gates (Hadamard, CNOT, Clifford,...) are included in {\tt Quipper}, along with many other functionalities allowing for circuit assembly and for resource requirement diagnosis. In this section, we will implement some of the algorithms presented above.

Throughout, it is assumed that all logical operations are implemented without error. In a real quantum device, the quantum system interacts with its environment, generating some noise and error in each operation. In this sense, the gate counts given below represent a lower bound estimates for the ``true'' algorithm, which may require error-correcting steps.

The algorithm for the solution of hyperbolic systems was formulated in the abstract Hilbert space of the quantum register. Therefore, it is independent of the computer architecture and thus, is amenable to any digital quantum devices. For example, quantum computers based on superconducting circuits \cite{kelly2015state,barends2015digital,barends2016digitized}, trapped ions \cite{Lanyon57} and cavity quantum electrodynamics \cite{PhysRevX.5.021027}, have been used with some success for other problems and could be used in principle to implement our algorithm. The main limitations however are i) the number of available qubits in current register and ii) the coherence time of these devices that restricts the number of logic quantum gates. State-of-the-art quantum computations on actual digital computers reach $\approx 1000$ quantum logic gates on $\approx 9$ qubits \cite{barends2016digitized}. 

In all the examples considered in the following, minimal requirements are studied in order to assess the feasibility of simulations on these real quantum devices. In particular, we count the number of quantum gates (circuit depth) and the total number of qubits (circuit width) required to evolve the initial condition data to a given final time. Due to physical limitations in terms of the number of qubits and the coherence time, only systems with overly small meshes are investigated. As emphasized in \cite{fillion}, this may be enough for proof-of-principle calculations but is far from outperforming classical computations. Nevertheless, given the amount of effort and resources devoted to the development of these quantum devices, these numbers will likely be improved in the future.

\subsection{One-dimensional hyperbolic system}
\begin{figure}
\begin{center}
\subfloat[]{\includegraphics[scale=1.5]{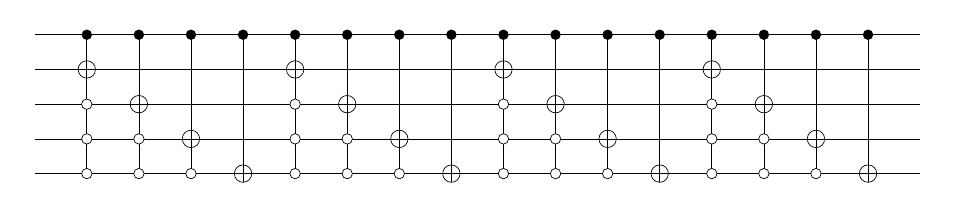}} \\
\subfloat[]{\includegraphics[scale=1.5]{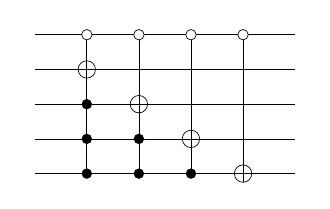}}
\caption{Circuit diagram with $N_x=16$. (a) first 4 iterations of the quantum reservoir method and first component. (b) Second component.}
\label{fig:first4}
\end{center}
\end{figure}

We consider a one-dimensional system with a computational domain as $[0,1] \times [0,T]$:
\begin{eqnarray*}
\partial_t u + \Lambda\partial_x u = 0, \qquad (x,t) \in [0,1] \times [0,T]
\end{eqnarray*}
where $A=\Lambda$ is a diagonal matrix in $M_2(\R)$, with eigenvalues $\lambda_1=1,\lambda_2 = -3$. If $A$ were not diagonal, it would be necessary to add a transition operator at the beginning and the end of the circuit. For $T=10^{-1}$ and $n_{T}=40$ we can determine the sets
\begin{eqnarray*}
\mathcal{I}^{n_T} & =& (  2,    2, 2,     1,    2,    2,   2,     1,    2,    2,    2,     1, 
    2,  2,    2,   1,    2,    2,  2,     1,    2,  2,    2,     1, \nonumber \\
    && 
  2,    2,    2,     1,   2,    2,  2,     1,   2,    2,  2,  1,  
    2,    2,    2,     1   ) \, , 
 \\
\mathcal{S}^{n_T} & =& (  -,    -, -,     +,    -,    -,    -,     +,    -,    -,    -,     +,  
   -,   -,    -,   +,    -,    -,  -,     +,    -,  -,    -,     +,  \nonumber \\
   &&
  -,    -,    -,     +,   -,    -,  -,     +,   -,    -,  -,  +,  
    -,    -,    -,     +,   ) \, . 
\end{eqnarray*}
We then implement the quantum algorithm described in Subsection \ref{sec:quantvar}, with {\tt Quipper}. In this case, it is only necessary to implement the decrement and increment operators, thanks to the preliminarily established list $\mathcal{I}^{n_T}$ of characteristics to be updated. Following Table I from \cite{fillion}, we report the minimal requirements for simulating the one-dimensional system described above, for proof-of-principle calculations. More specifically using {\tt Quipper}, we can evaluate precisely the number of elementary gates necessary to implement the quantum algorithm. Say for respectively $n_x=2$, $n_x=4$, and $n_x=8$, the circuit depth computed by {\tt Quipper} is $780$, $5520$, $27960$, and the circuit  width is respectively $3$, $7$, $15$. For instance for $n_x=4$ that is $N_x=16$, the respective number of Hadamard, Clifford, Toffoli, CNOT gates, is found to be  $540$, $1890$, $1350$, $1560$.  We report in Fig. \ref{fig:first4} the quantum circuit for the first 4 iterations with $N_x=16$. This quantum circuit has a width equal to $5$, that is $4$ qubits for labeling the coordinate space positions, and $1$ qubit for the labeling of the component. The circuits which are represented corresponds to the first 4 time iterations. We notice that the circuit for the first component (top) has $4$ times the same pattern (elementary circuit), corresponding to $4$ translations from the left to the right, while the circuit for the second component (bottom) corresponds to only $1$ iteration from the right to the left, for the same lapse of time. This is due to the fact that $\lambda_2=-3 \times \lambda_1$, so that after $4$ iterations, $4$ translations to the right are applied to the first component, while the second component is only translated once to the left. The quantum circuit for 1 iteration is generated by {\tt Quipper}.

\begin{figure}
\begin{center} 
\includegraphics[width=1.0\textwidth]{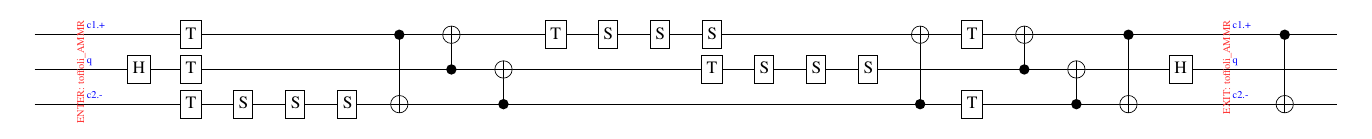}
\caption{Circuit diagram obtained from the gate decomposition with $N_x=4$ in 1-D for one iteration.}
\label{fig:firstdecomp}
\end{center}
\end{figure}

As an illustration for the same test, but with $N_x=4$, we also report in Fig. \ref{fig:firstdecomp} the quantum circuit for the corresponding gate decomposition and which is still generated by {\tt Quipper}.  This time we only have $3$ qubits, $1$ for the component index, and $2$ for the positions.


The case with the smallest circuit depth, for the lower number of lattice points ($n_{x}=2,N_{x}=4$), could possibly be implemented on actual device for a proof-of-principle calculation, as long as the initialization phase requires less than approximately $300$ gates. However, the results obtained from the gate decomposition demonstrate that it would be a challenging task to perform quantum simulations of hyperbolic systems on actual quantum device with larger lattice size. Moreover, performing a relevant quantum simulation outperforming classical computation demands much improvement from both the coherence time and the number of qubits.

\subsection{Three-dimensional hyperbolic system}
\begin{figure}
\begin{center}
\includegraphics[width=1.0\textwidth]{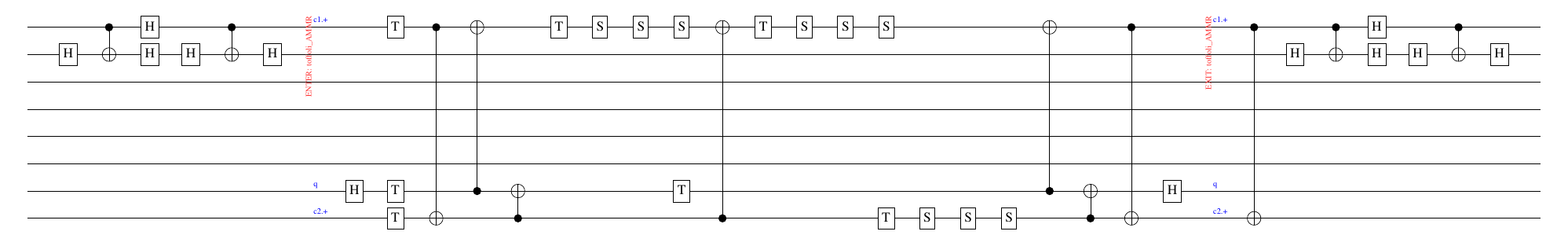}
\caption{Circuit diagram obtained from the gate decomposition with $N_x=16$ in 3-D for the first iteration.}
\label{fig:first1iter3d}
\end{center}
\end{figure}

In this subsection, we implement with {\tt Quipper} the quantum version of the reservoir method for a three-dimensional linear hyperbolic system \eqref{syst3d} with $\lambda_1^{(\textrm{x})}=\lambda_2^{(\textrm{x})}=-1$, $\lambda_3^{(\textrm{x})}=\lambda_4^{(\textrm{x})}=\sqrt{2}$, $\lambda_1^{(\textrm{y})}=\lambda_2^{(\textrm{y})}=-2$, $\lambda_3^{(\textrm{y})}=\lambda_4^{(\textrm{y})}=2\sqrt{2}$, $\lambda_1^{(\textrm{z})}=\lambda_2^{(\textrm{z})}=-4$, $\lambda_3^{(\textrm{z})}=\lambda_4^{(\textrm{z})}=4\sqrt{2}$. Notice that we have associated the following upper indices: $(\textrm{x}) \leftrightarrow (1)$, $(\textrm{y}) \leftrightarrow (2)$, $(\textrm{z}) \leftrightarrow (3)$. We then select $S^{(\gamma)}$ as in \eqref{dirac}, where we recall that Pauli's matrices are defined by:
\begin{eqnarray}
\sigma_{x} = 
\begin{bmatrix}
0 & 1 \\ 1 & 0  
\end{bmatrix}
\;\; \mbox{,} \;\;
\sigma_{y} = 
\begin{bmatrix}
0 & -{\tt i} \\ {\tt i} & 0 
\end{bmatrix}
\;\; \mbox{and} \;\;
\sigma_{z} = 
\begin{bmatrix}
1 & 0 \\ 0 & -1 
\end{bmatrix}\,.
\end{eqnarray}
The motivation for considering such ``simple'' transition matrices is to design simple quantum circuit portion for diagonalization. As mentioned in Section \ref{subsubsec:rot}, more complex transition unitary matrices could be considered, but would then require additional work for decomposing in elementary quantum gates. The quantum algorithm is applied from time $0$ to $10^{-2}$, with $\Delta x=10^{-2}$, corresponding to $n_T=30$ iterations. In addition, $\mathcal{I}^{n_T}$ and  $\mathcal{S}^{n_T}$ are such that:
\begin{eqnarray*}
\mathcal{I}^{n_T} & =& (  (3,3), (4,3), (1,3), (2,3), (3,2), (3,3), (4,2), (4,3), (1,2), (1,3),  \nonumber \\
&&
  (2,2), (2,3), (3,3),(4,3),(3,1),(3,2), (3,3),(4,1),(4,2), (4,3), \nonumber \\
  &&
  (1,3),(2,3),(3,3),(4,3),(1,1),(1,2),(1,3),(2,1),(2,2),(2,3)    )
\\
\mathcal{S}^{n_T} & =& (   +, +, -, -, +, +, +, +, -, -,  \nonumber \\
&&
 -, -, +,+,+,+, +,+,+, +, \nonumber \\
 &&
  -,-,+,+,-,-,-,-,-,-  )
\end{eqnarray*}
The corresponding time steps take the value $0.0177,0.0073,0.0104,0.0030,0.0177,0.0043,\cdots$.   We then encode $\mathcal{I}^{n_T}$ in the quantum algorithm in order to implement the quantum reservoir method.  This list provides the directional characteristic field to update. Say for respectively $n_x=n_y=n_z=2$, $n_x=n_y=n_z=4$, and $n_x=n_y=n_z=8$, the circuit depth computed by {\tt Quipper} is $672$, $3042$, $14262$, and the circuit  width is respectively $8$, $16$, $32$. For instance for $n_x=n_y=n_z=4$ that is $N_x=N_y=N_z=16$, the respective numbers of Hadamard, Clifford, Toffoli, CNOT gates, are found to be  $400$, $1037$, $675$, $840$.  We report in Fig. \ref{fig:first1iter3d} the quantum circuit for the first iteration with $N_x=N_y=N_z=4$, and a circuit width equal to $8$. The quantum circuit which is generated by {\tt Quipper} is much more complex (Toffoli, Hadamard, Clifford quantum gates, etc), as it also includes the changes of basis (rotations), and the translations.

The same conclusion as in the 1-D case can be reached from these results, i.e. a proof-of-principle calculation could possibly be performed with the smaller systems, but relevant calculations would require improvements in quantum technologies.

\end{document}